\documentclass[a4paper,12pt]{article}
\pdfoutput=1 

\usepackage{jheppub} 

\usepackage[T1]{fontenc} 
\usepackage{amsmath}
\usepackage{amsthm}
\usepackage{float}
\usepackage{braket}
\theoremstyle{definition}
\newtheorem{thm}{Theorem}[section]
\newtheorem*{thm*}{Theorem}

\newtheorem*{defn*}{Definition}

\newtheorem*{lem*}{Lemma}
\newtheorem{rem}[thm]{Remark}
\newtheorem*{rem*}{Remark}

\newtheorem*{con*}{Conjecture}

\newtheorem*{cor*}{Corollary}

\newtheorem*{prop*}{Proposition}

\newtheorem*{hypoth*}{Hypothesis}

\newtheorem*{claim*}{Claim}
\newtheorem*{prf}{Proof}

\title{\Large{Universal Computation with Quantum Fields}}

\author{Kazuki Ikeda}

\affiliation{Department of Physics, Osaka University,\\Toyonaka, Osaka 560-0043, Japan}
\date{\empty}
\arxivnumber{\empty}

\emailAdd{kazuki7131@gmail.com}

\abstract{We explore a way of universal quantum computation with particles which cannot occupy the same position simultaneously and are symmetric under exchange of particle labels. Therefore the associated creation and annihilation operators are neither bosonic nor fermionic. In this work we first show universality of our method and numerically address several examples. We demonstrate dynamics of a Bloch electron system from a viewpoint of adiabatic quantum computation. In addition we provide a novel Majorana fermion system and analyze phase transitions with spin-coherent states and the time average of the OTOC (out-of-time-order correlator). We report that a first-order phase transition is avoided when it evolves in a non-stoquastic manner and the time average of the OTOC diagnoses the phase transitions successfully. }

\begin{document} 
\maketitle 
\section{Yet Another Imitation Game}
Physics is a study on the computational principles of nature and on the algorithms implemented by an as-yet-unknown way. Future developments in quantum devises will help us explore many body systems. Technically there are two ways to simulate quantum physical dynamics of particles with quantum computers. An orthodox one is to prepare circuits exactly like the way real things work. The second is to create circuits so that the desired outcomes should be obtained. Physicists prefer the former idea and want to fingerprint their models or to find novel physical aspects which may be testable by experiments. Programmers prefer the latter. Instead of going to tiny theoretical details, they feel happy as long as their programs work without bugs. Programmers are unlikely to find new physics, but could find efficient ways to reproduce precise results. Some of the key criteria for accepting a theoretical model of physics are whether its descriptions and predictions are consistent with experimental results. We now ask the question, "Can a technically valid algorithm be scientifically sound?" An algorithm which is capable of accurately reproducing results would have a chance of predicting phenomena in a way that they seem consistent with experiment, even if real things do not actually obey it.

The new form of the problem can be described in terms of the 'imitation game \cite{turing2009computing}." It is played with three people, an experimental physicist (A), a programmer (B), and an interrogator (C) who is a theoretical physicist and know them by labels X and Y. Suppose both of A and B have sufficient skills in their own fields, but may be less familiar with the other fields. The experimenter is allowed to use any experimental device and material, and the programmer is allowed to access the internet and any computing devise. The interrogator stays in a room apart from the other two and put questions about physics to A and B. The object of the game for the interrogator is to determine which of the other two is the experimental physicist and which is the programmer. Though the programmer is not necessarily an expert of physics, he/she might manage to find the best algorithm which approximates experimental results well. At the end of the game, A and B submit their results, based on which the interrogator says either "X is A and Y is B" or "X is B and Y is A." If the interrogator cannot reliably tell the programmer from the experimenter, the programmer is said to have succeeded in "coding nature".

\section{Quantum Field Computation}\label{sec:qfc}
\subsection{Preliminaries}
A quantum computer consists of a many particle system, thereby allows us to study more about the advanced properties of materials and molecules, and also to explore fundamental physics. Quantum field theory (QFT) is a powerful tool to address countably or uncountably many particles.  Indeed it has had a major impact in condensed matter, cosmology, high energy physics and even pure mathematics. The question whether QFTs can be efficiently simulated by quantum computers was raised by Feynman \cite{Feynman1986}. According to Deutsch \cite{doi:10.1098/rspa.1985.0070}, the quantum version of Church-Turing thesis \cite{church1936unsolvable, doi:10.1112/plms/s2-42.1.230} states that "Every finitely realizable physical system can be perfectly simulated by a universal model computing machine operating by finite means". Here a finitely realizable physical system includes any experimentally testable physical object. Usually a quantum computer works on a graph consisting of a fixed number of particles. Quantum algorithms which simulate the dynamics of quantum lattice systems with a fixed number of particles have been considered by many authors \cite{PhysRevLett.79.2586,berry2007efficient,1998RSPSA.454..313Z}. Although theories on a connected space and a discrete space look very different, some QFT on a connected space can be well approximated by quantum mechanics on a discrete system. For instance, scalar field theory can be precisely approximated with finitely many qubits by discretization of space via a lattice \cite{Jordan:2011ne}. Moreover scattering in scalar field theory is also known to be BQP complete \cite{Jordan2018bqpcompletenessof}. (A problems is in BQP if it is solvable in polynomial time with a quantum computer and it is BQP-hard if it is at least as hard as the hardest problems in BQP. A problem which is both BQP-hard and contained in BQP is called BQP-complete.) Those references imply that quantum algorithms are nice technological substitutes for QFT on a connected space. So this motivates us to consider QFT from a viewpoint of quantum computation. However, from a viewpoint of QFT, it is believed that a theory on a fixed-particle-number Hilbert space is not powerful to describe high energy physics including cosmology and quantum gravity, where relativistic effects are dominant. 

In this work we propose a universal quantum computation with unfixed number of particles. What is new to us is that our model is based on the dynamics of somewhat unphysical particles that are neither bosonic nor fermionic. Those particles cannot occupy the same state and are symmetric under exchange of particle labels. They have the simplest representation and sufficient properties for universal computation. In practice, only two independent states are enough to do computation and generic linear combinations of them are essentially important to establish quantum supremacy. So particle statistics is not a main concern. Commutation relations of creation and annihilation operators would help one concentrate on coding. This is an advantage of our particles. Yet another question we should ask is whether such particles exist in a laboratory or nature. In later part of this note we will reproduce fermion dynamics with our particles. The fact that they exists in computer language leads us into our original, the "imitation game" played by physicists and programmers. In addition, our computational method is based on the idea of adiabatic quantum computation (AQC) \cite{2000quant.ph..1106F,RevModPhys.90.015002}. In general, AQC is as powerful as universal quantum computation when non-stoquastic Hamiltonian is used. In practice, however, the adiabatic condition is rarely satisfied and it remains open as to when such quasi-adiabatic dynamics can provide a computational advantage. One of the important aspects of AQC is quantum annealing \cite{PhysRevE.58.5355}, which is a metaheuristic algorithm to solve combinatorial optimization problems by changing the parameters adiabatically or even non-adiabatically. Measurements on the current quantum annealer \cite{Johnson2011,Troels14} are done only in the standard computational basis, thereby the Hamiltonian is stoquastic. Although it is speculated that quantum speedup is realized with stoquastic Hamiltonian, quantum annealing attracts interests of many authors and many applications have been developed \cite{10.3389/fphy.2014.00005,ikeda2019NSP}.

The rest of this piece is orchestrated as follows. Section \ref{sec:qfc} includes a short review on quantum computation and describes the universality of a computational method by means of a particle that cannot occupy the same state simultaneously and are symmetric under exchange of particle labels. In Sec. \ref{sec:ex}, we numerically investigate some basic properties of our quantum computation. There we carefully address a Bloch electron system with open boundary condition and study the probability of finding the eigenstates of the tight-binding Hamiltonian. Moreover in Sec. \ref{sec:PT}, we address phase transition associated with our quantum algorithm. We provide several models which experience first-order and second-order phase transitions and show quantum speedup is realized when non-stoquastic Hamiltonian is used. Finally in Sec. \ref{sec:conc} we present several future directions of research.

\subsection{Computation with Elemental Fields}
We first formulate quantum field theoretical computation on a lattice. In the standard physics literature, particles are generally classified into two categories: bosonic or fermionic. Bosons are particles that obey the Bose-Einstein statistics, can occupy the same state simultaneously and are symmetric under exchange of particle labels\footnote{$[a_i,a^\dagger_j]=\delta_{ij}, [a_i,a_j]=[a^\dagger_i,a^\dagger_j]=0$ for bosons}. On the other hand, fermions obey the Fermi-Dirac statistics, cannot occupy the same state simultaneously and are anti-symmetric under exchange of particle labels\footnote{$\{a_i,a^\dagger_j\}=\delta_{ij}, \{a_i,a_j\}=\{a^\dagger_i,a^\dagger_j\}=0$ for fermions}. When it comes to coding, fermionic operators are not good since they produce negative signs, which trigger coding errors. Bosonic operators can solve this problem, but they occupy the same state simultaneously, hence there exist some inactive particles that consume memory. Therefore, a simple question may come to our mind: is there any particles that cannot occupy the same state simultaneously and are symmetric under exchange of particle labels? A simple way to construct such particles is as follows. Let $a^\dagger$ and $a$ be the 0-dimensional fermionic creation and annihilation operators whose representations are 
\begin{equation}
    a=\begin{pmatrix}
    0&1\\
    0&0
    \end{pmatrix}~~
    a^\dagger=\begin{pmatrix}
    0&0\\
    1&0
    \end{pmatrix}. 
\end{equation}
Then $a^\dagger+a$ corresponds to the Pauli $X$ operator and $[a,a^\dagger]$ corresponds to the Pauli $Z$ operator. Let $a_i$ and $a^\dagger_i$ be the annihilation and creation operators acting on the Hilbert space of $i$-th particle:
\begin{align}
    a_i:&=1 \otimes\cdots \otimes 1\otimes a\otimes 1\otimes \cdots\otimes 1\\
    a^\dagger_i:&=1 \otimes\cdots \otimes 1\otimes a^\dagger\otimes 1\otimes \cdots\otimes 1,
\end{align}
which obey $[a_j,a^\dagger_i]=Z_i\delta_{ij}$, $[a^\dagger_i,a^\dagger_j]=[a_i,a_j]=0$ and  $a^\dagger_ia^\dagger_i=a_ia_i=0$ for all $i,j$. Note that $\{a_i,a^\dagger_j\}\neq \delta_{ij}$ for different $i,j$. We denote by $n_i=a^\dagger_i a_i$ the number operator. So a general state in a system with size $L$ is a superposition of  
\begin{equation}
    \ket{n_1\cdots n_L} 
\end{equation}
and the vacuum state is $\ket{\emptyset}=\ket{0\cdots0}$ which vanishes $a_i\ket\emptyset=0$ by any $a_i$. A particle $a^\dagger_i\ket{\emptyset}=\ket{1_i}$ created at $i$ disappears when $a^\dagger_i$ acts on the same state again due to $a^{\dagger 2}_i=0$. Regarding a two particle state $\ket{1_i,1_j}=a^\dagger_ia^\dagger_j\ket{\emptyset}$, one cannot tell one particle from another because of the commutation relation $a^\dagger_ia^\dagger_j=a^\dagger_ja^\dagger_i$ at different sites $(i\neq j)$. Therefore the creation and annihilation operators describe particles which cannot occupy the same position simultaneously and particles labels are indistinguishable. A common interpretation of those operators is that $a_i,a^\dagger_i$ annihilates/creates the $z$-spin at $i$, respectively. One may wonder if multiple fermions can be addressed with $a_i,a^\dagger_i$, but any operation can be reconstructed by them as we discuss below. 

\subsection{Universality}
Universality of quantum computation can be described in two ways. In a strong sense it means one can obtain any unitary operation, and in a weak sense it means one can get any desired probability distribution. Since wave functions are not physical observables, the latter is adequate for practical use. We first show our model is as powerful as universal computation in the following sense. 
\begin{thm}\label{thm:universal}
Let $\{x_i\}_{i=1}^M$ and $\{x'_j\}_{j=1}^N$ be coordinates of $M$ and $N$ particles on a discrete system. Any operator $\mathcal{O}$ on a discrete system can be written with $a_x,a^\dagger_x$ in such a way that
\begin{equation}
    \mathcal{O}=\sum_{M=0,N=0}\sum_{x_i,x'_j} A_{MN}(x_1,\cdots,x_M,x'_1,\cdots, x'_N)a^\dagger_{x_1}\cdots a^\dagger_{x_M}a_{x'_1}\cdots a_{x'_N}. 
\end{equation}
\end{thm}
\begin{prf}
Proof can be done by induction. Let $\ket{\psi_{{x_1\cdots x_M}}}=\prod_{i=1}^M a^\dagger_{x_i}\ket{\emptyset}$ be a $M$-particle state. The zero particle state corresponds to the vacuum $\ket{\psi_0}=\ket{\emptyset}$. Clearly it is always possible to assign $\bra{\emptyset}\mathcal{O}\ket{\emptyset}$ with any value, by choosing some $A_{00}$. Suppose the same things are true for matrix elements $\bra{\psi_{x_1\dots\cdots x_M}}\mathcal{O}\ket{\psi_{x'_1\dots\cdots x'_N}}$ of all $M$ and $N$ particle states satisfying $M<K, N\le L$ or $M\le K, N< L$. Then we obtain
\begin{align}
    \begin{aligned}
    \bra{\psi_{x_1\cdots x_K}}\mathcal{O}\ket{\psi_{x'_1\cdots x'_L}}=&L!K!A_{KL}(x_1,\cdots, x_K,x'_1,\cdots, x'_L)\\
    &+\text{terms with $A_{MN}$ for $M<K, N\le L$ or $M\le K, N<L$}. 
    \end{aligned}
\end{align}
By choosing appropriate $A_{KL}$, one can assign the right hand side with any desired value. Therefore, with an appropriate set of coefficients $\{A_{MN}\}$, any operation $\mathcal{O}$ can be approximated by some combinations of creation and annihilation operators. 
\end{prf}

Thanks to this theorem, one can realize any Hamiltonian thereby approximate any result of quantum computation. More simply, it is also possible to construct a universal gate set. For example, the CNOT operator acting on different $i,j$ cites can be represented by
\begin{align}
\begin{aligned}
    \text{CNOT}=&\frac{1}{2}(1+[a_i,a^\dagger_i])+\frac{1}{2}(1-[a_i,a^\dagger_i])(a_j+a^\dagger_j)\\
=&1-n_i+n_i(a_j+a^\dagger_j)    
\end{aligned}
\end{align}
Since any $n$-qubit unitary gate can be well simulated by CNOT and a single qubit gates, we can create a universal gate set by $\{a_i,a^\dagger_i\}$. Any unitary operator can be created by some algebraic operation to $a$ defined over $\mathbb{C}$ and, in this sense, $a$ is the primary operator of quantum computation. 

In what follows we give another representation of universal computation with $\{a_i,a^\dagger_i\}$. To this end, we begin with several operators \cite{RevModPhys.90.015002}:
\begin{enumerate}
    \item $H_\text{init}=H_{\text{clock init}}+H_\text{input}+H_\text{clock}$
    \item $H_\text{final}=H_\text{prop}+H_\text{input}+H_\text{clock}$
    \item $H_\text{prop}=\frac{1}{2}\sum_{\tau=1}^LH_\tau$
\end{enumerate}

And consider the total Hamiltonian defined by
\begin{equation}
    H(t)=(1-t)H_\text{init}+tH_\text{final},~~t\in[0,1]. 
\end{equation}
The clock Hamiltonian $H_\text{clock}$ 
\begin{equation}
    H_\text{clock}=\sum_{\tau=1}^{L-1}\ket{0_\tau1_{\tau+1}}_c\bra{0_\tau1_{\tau+1}}
\end{equation}
has the correct clock state as its ground state. Here $\ket{0_\tau 1_{\tau+1}}_c$ denotes Feynman's clock register \cite{Feynman:85}. The clock register can be more simply written as $\ket{\tau}\bra{\tau-1}=\ket{1_{\tau-1}1_\tau 0_{\tau+1}}_c\bra{1_{\tau-1}0_\tau0_{\tau+1}}$, for example. Since the initial clock state is 0 at $\tau=0$,   
\begin{equation}
    H_{\text{clock init}}=\ket{1_1}_c\bra{1_1}
\end{equation}
assigns $0$ to the correct initial state, otherwise 1. Moreover, at the initial time, 
\begin{equation}
    H_\text{input}=\sum_{i=1}^N\ket{1_i}\bra{1_i}\otimes\ket{0_1}_c\bra{0_1}
\end{equation}
gives 0 if all qubits used for computation are 0, otherwise 1. So the ground states of $H_\text{init}$ has 0 as the corresponding eigenvalue. A family of unitary operations $\{U_\tau\}_{\tau=1}^L$ can be accommodated into $\{H_\tau\}_{\tau=1}^L$ in such a way that \begin{align}
\begin{aligned}
    H_1=&1\otimes \ket{0_10_2}_c\bra{0_10_2}-U_1\otimes \ket{1_10_2}_c\bra{0_10_2}-U^\dagger_1\otimes\ket{0_10_2}_c\bra{1_10_2}+1\otimes\ket{1_10_2}_c\bra{1_10_2}\\
    H_\tau=&1\otimes \ket{1_{\tau-1}0_l0_{\tau+1}}_c\bra{1_{\tau-1}0_\tau0_{\tau+1}}-U_\tau\otimes\ket{1_{\tau-1}1_\tau0_{\tau+1}}_c\bra{1_{\tau-1}0_\tau0_{\tau+1}}\\
    &-U^\dagger_\tau\otimes\ket{1_{\tau-1}0_\tau0_{\tau+1}}_c\bra{1_{\tau-1}1_\tau0_{\tau+1}}+1\otimes\ket{1_{\tau-1}1_\tau0_{\tau+1}}_c\bra{1_{\tau-1}1_\tau0_{\tau+1}}\\
    H_L=&1\otimes\ket{1_{L-1}0_L}_c\bra{1_{L-1}0_L}-U_L\otimes\ket{1_{L-1}1_L}_c\bra{1_{L-1}0_L}\\ &-U^\dagger_L\otimes\ket{1_{L-1}0_L}_c\bra{1_{L-1}1_L}+1\otimes\ket{1_{L-1}1_L}_c\bra{1_{L-1}1_L}
\end{aligned}
\end{align}
Now we consider the representation by $\{a_i,a^\dagger_i\}$. It is easy to see that 
\begin{equation}
    H_\text{clock}\Leftrightarrow \sum_{\tau=1}^{L-1}(1-n_{\tau})n_{\tau+1}
\end{equation}
is 1 if and only if it acts on $\ket{0_\tau 1_{\tau+1}}_c$. And we find  
\begin{equation}
    H_{\text{clock init}}\Leftrightarrow n_1
\end{equation}
returns 1 if and only if it acts on $\ket{1_1}_c$. Moreover the input Hamiltonian corresponds to
\begin{equation}
    H_\text{input}\Leftrightarrow \sum_{i=1}^Nn_{i}\otimes (1-n_1),
\end{equation}
where $n_i$ acts on logical qubits and $1-n_1$ acts on the clock state. As mentioned previously, any unitary operator $U_i$ can be reconstructed by CNOT and a single qubit operators. So each gate operation is, for example, 
\begin{align}
\begin{aligned}
    H_\tau\Leftrightarrow &1\otimes n_{\tau-1}(1-n_\tau)(1-n_{\tau+1})-U_\tau\otimes n_{\tau-1}a^\dagger_\tau(1-n_{\tau+1})\\
    &-U^\dagger_\tau\otimes n_{\tau-1}a_\tau (1-n_{\tau+1})+1\otimes n_{\tau-1}n_\tau(1-n_{\tau+1}). 
\end{aligned}    
\end{align}
In summary, we find that it is possible to do universal computation by $\{a_i,a^\dagger_i\}$. To implement the universal adiabatic quantum computation, we want to seek for the simplest Hamiltonian. It is known that the 5-local Hamiltonian above can be well approximated by the following QMA-complete 2-local Hamiltonian \cite{PhysRevA.78.012352} 
\begin{align}
    H&=\sum_{i}{h_z}_{i} Z_i+\sum_{i}{h_x}_{i}X_i+\sum_{ij}J_{ij}Z_iZ_j+\sum_{ij}\Gamma_{ij}X_iX_j\\
    &=\sum_{i}{h_z}_{i} [a_i,a^\dagger_i]+\sum_{i}{h_x}_{i}(a_i+a^\dagger_i)+\sum_{ij}J_{ij}[a_i,a^\dagger_i][a_j,a^\dagger_j]+\sum_{ij}\Gamma_{ij}(a_i+a^\dagger_i)(a_j+a^\dagger_j). 
\end{align}
Instead of using eigenvalues $s_i\in \{-1 ,+1\}$ of the spin $Z_i$ operator, we may use variables in $\{0,1\}$, that are the eigenvalues of $\frac{1- Z_i}{2}=n_i$ operators. Then the corresponding representation of the Hamiltonian is 
\begin{equation}
    H=\sum_{i}{h_z}_{i} n_i+\sum_{i}{h_x}_{i}X_i+\sum_{ij}J_{ij}n_in_j+\sum_{ij}\Gamma_{ij}X_iX_j. 
\end{equation}

\begin{rem}
Another important way of quantum field theoretical universal computation is called matchgate \cite{2008RSPSA.464.3089J}. Free fermions play a crucial role. A standard representation of the fermionic creation $b^\dagger_j$ and annihilation $b_j$ operators is given by the Jordan-Wigner representation \cite{Jordan1928}
\begin{equation}
    b_j=Z_1Z_2\cdots Z_{j-1}a_j~~~~b^\dagger_j =Z_1Z_2\cdots Z_{j-1}a^\dagger_j,
\end{equation}
They obey $\{b_i,b^\dagger_j\}=\delta_{ij}$ and $\{b_i,b_j\}=0$. And Majorana fermions are represented by 
\begin{equation}\label{eq:MJ}
    c_{2k-1}=b_k+b^\dagger_k~~~~c_{2k}=-i(b_k-b^\dagger_k), 
\end{equation}
which satisfy $\{c_\mu,c_\nu\}=2\delta_{\mu,\nu}$. 

\end{rem}

\subsection{Example}
Combinatorial optimization is all about finding an optimal object from a finite set of objects. A lot of combinatorial optimization problems are NP-complete or NP-hard, and are widely studied from a perspective of computational complexity theory. One of the most famous examples of combinatorial optimization problems is the traveling salesman problem (TSP), which is NP-hard. As an example we formulate the TSP with our model. Let a salesman visit cities $i=1,\cdots N$ step by step only one time. Let $D_{ij}$ be the distance between $i,j$, hence it is symmetric $D_{ij}=D_{ji}$. Then the $H_0$ term which respects those constraints is given by  
\begin{equation}
    H_0=\sum_{t=1}^{N-1}\sum_{ij}D_{ij}n_{i,t+1}n_{j,t}+\sum_{t=1}^N\lambda_1(t)\left(\sum_{i}^Nn_{i,t}-1\right)+\sum_{i=1}^N\lambda_2(i)\left(\sum_{t}^Nn_{i,t}-1\right),
\end{equation}
where $\lambda_1(t)$ and $\lambda_2(i)$ are Lagrange multipliers. We redefine the operators by  $n_{i,t}=n_{i+N(t-1)}$ and label the problem by $N^2$ indexes. An eigenstate $\ket{\psi}$ satisfying the equality constraints of $H_0$ looks like 
\begin{equation}
\ket{\psi}=\prod_{t=1}^Na^\dagger_{i_t+N(t-1)}\ket{\emptyset},~~\{i_t\}_t\subset\{1,2,\cdots,N\}  
\end{equation}
and has the corresponding eigenvalue 
\begin{equation}
    \sum_{t=1}^{N-1}D_{i_{t+1}i_t}. 
\end{equation}
 Therefore the ground state of $H_0$ is the solution of the TSP. Many of generic classical combinatorial optimization problems are solvable with some $H_0=H_0(n)$. It is straightforward to translate Ising formulation of NP problems \cite{10.3389/fphy.2014.00005} into ours.

\if{
\subsection{Dynamics of Electron System}
Another example is a study of quantum physics. The quantum annealing is commonly recognized as a solver of combinatorial optimization problems, and indeed we showed they can be solved in our way before. In addition to that, we show our method is also fairly useful to simulate fermionic systems. This is an advantage of our method. Although it would not be impossible to formulate ferimonic systems with $Z_i$ basis, coding can be messy in general. Here we aim at demonstrating a Bloch electron system under uniform magnetic flux $\phi$ perpendicular to the system. This system is one of the most important ones in condensed matter physics. Especially the following Hamiltonian \eqref{eq:ferm} is widely used for studying the two dimensional integer quantum Hall effect, which exhibits the most fundamental topological property, therefore plays a crucial role to study not only general topological matter physics but also high energy physics and mathematical physics \cite{Ikeda:2018tlz,Ikeda:2017uce}. We define the coupling by the associated $U(1)$ gauge field $\theta$. Then the Hamiltonian is 
\begin{equation}\label{eq:ferm}
H_0=-\left(\sum_{m,n}e^{2\pi i \theta^x_{m,n}}c^\dagger_{m+1,n}c_{m,n}+e^{2\pi i \theta^y_{m,n}}c^\dagger_{m,n+1}c_{m,n}+h.c\right),     
\end{equation}
where $c_i,c^\dagger_i$ are fermionic annihilation and creation operators $\{c_i,c^\dagger_j\}=\delta_{ij}, \{c_i,c_j\}=\{c^\dagger_i,c^\dagger_j\}=0$. Now let us reproduce physics of this model with our method. To this end, we work on a tight-binding Hamiltonian on a two dimensional square lattice:
\begin{equation}\label{eq:tight}
    \widetilde{H_0}=-\left(\sum_{\langle ij\rangle}\gamma_{ij}a^\dagger_i a_j+\gamma^*_{ji}a^\dagger_ja_i \right),
\end{equation}
where the summation is taken over the nearest neighbor pairs $i=(i_m, i_n)$. Then the tight-binding Hamiltonian can be written as 
\begin{equation}\label{eq:qhe}
    \widetilde{H_0}=-\left(\sum_{m,n}e^{2\pi i \theta^x_{m,n}}a^\dagger_{m+1,n}a_{m,n}+e^{2\pi i \theta^y_{m,n}}a^\dagger_{m,n+1}a_{m,n}+h.c\right), 
\end{equation}
where $(i_m,i_n)$ in \eqref{eq:tight} is abbreviated by $(m,n)$. For a single particle state
\begin{equation}
\ket{\phi}=\sum_{m,n}\phi_{m,n}a^\dagger_{m,n}\ket{\emptyset},    
\end{equation}
the hopping energy from one site $a^\dagger_{m,n}\ket{\emptyset}=\ket{m,n}$ to another $\ket{m+1,n}$ is
\begin{equation}
    \bra{m+1,n}\widetilde{H_0}\ket{m,n}=-e^{2\pi i\theta^x_{m,n}},
\end{equation}
which corresponds to a matrix element of $H_0$. Fig.\ref{fig:butt} shows flux dependence of energy spectrum for the open boundary case. The fractral structure in the figure is realized by the interplay of Bragg's reflection and Landau's quantization of Bloch electrons on a lattice \cite{PhysRevB.14.2239}. It attracts the interest of many authors from viewpoints of condensed matter physics, high energy physics \cite{Hatsuda:2016mdw} and mathematical physics \cite{doi:10.1063/1.4998635}. In our case with boundary, the butterflies accommodate energy spectra of edge states. 
\begin{figure}[H]
    \centering
    \includegraphics[width=7cm,bb=0 0 432 288]{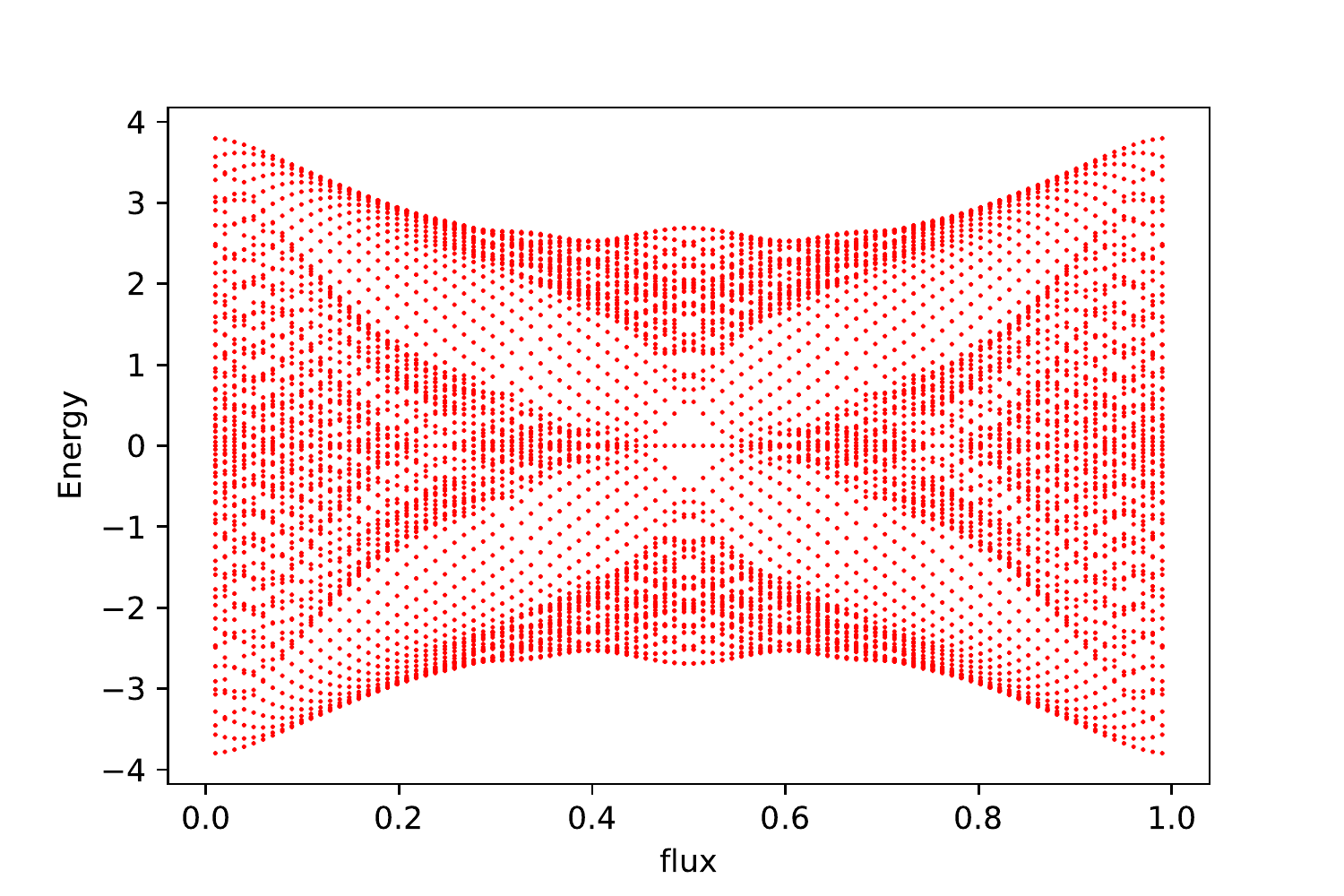}
    \caption{Energy spectra of a Bloch electron system on a square lattice with fractional magnetic flux $\phi$ perpendicular to the system. $\phi$ runs over $1/101,2/101,\cdots, 100/101$. The Landau gauge $(\theta^x_{m,n}, \theta^y_{m,n})=(0,m\phi)$ was used for computation.}
    \label{fig:butt}
\end{figure}
\begin{figure}[H]
    \centering
    \includegraphics[width=7cm]{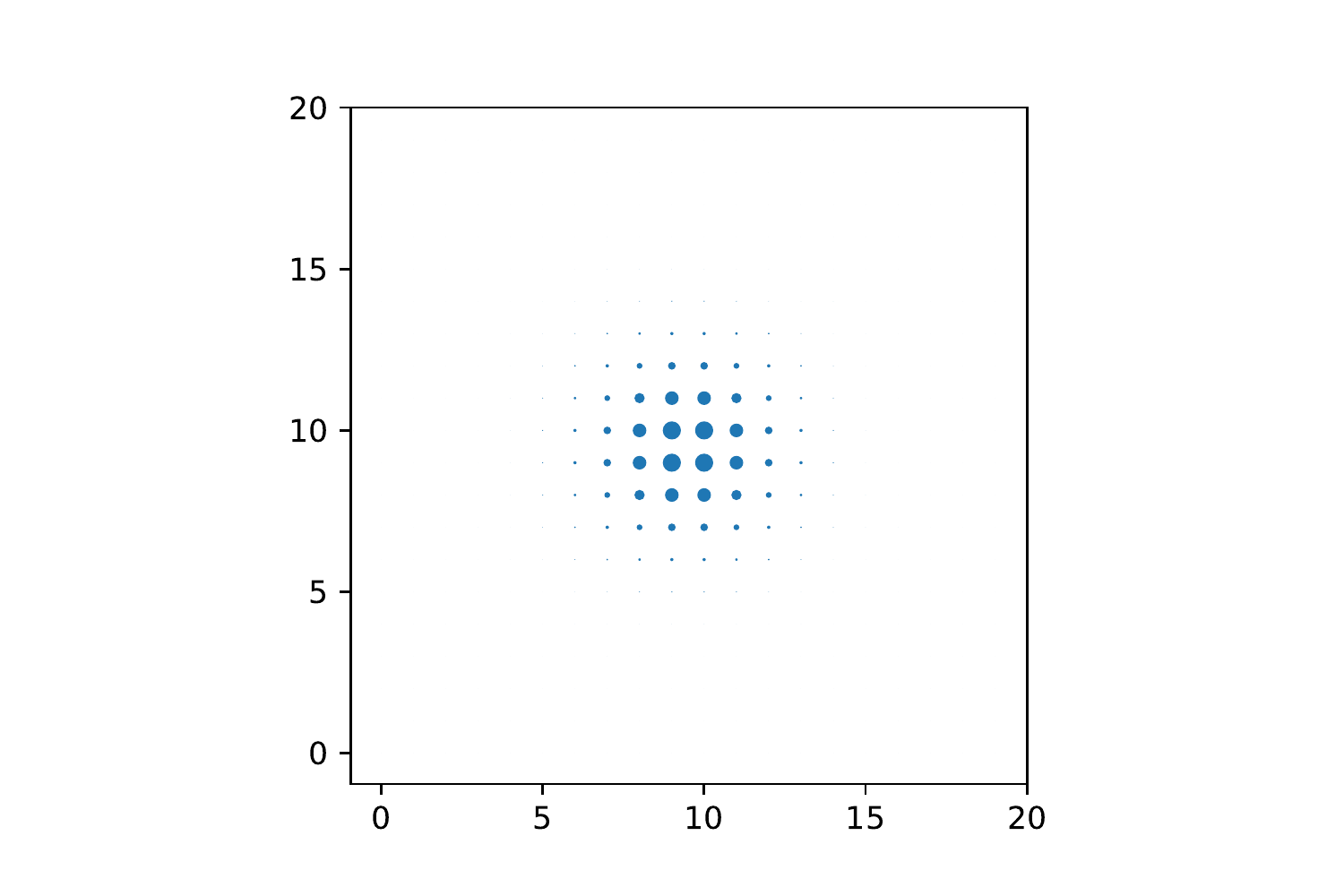}
    \caption{The density distribution of the ground state of Bloch electrons on a 20 by 20 square lattice under uniform magnetic flux $\phi=1/11$.}
    \label{fig:dos_g}
\end{figure}
Fig. \ref{fig:dos_g} shows the density distribution of the ground state of Bloch electrons on a 20 by 20 square lattice under uniform magnetic flux $\phi=1/11$. The distribution is computed with the standard fermionic tight-binding Hamiltonian \eqref{eq:ferm}. Fig. \ref{fig:dos} shows time dependence of the density distribution of the ground state of $H(t)=(1-\Gamma(t))H_0+\Gamma(t)H_1$ defined over the same lattice. At the initial time $t=0$, the ground sate of the ferromagnetic $H_1=-XX$ interaction is uniformly distributed to all regions of the lattice. As time pass by, states gather at center of the bulk. Comparing Fig. \ref{fig:dos_g} and the last figure in Fig. \ref{fig:dos}, we find that the model approximates the density distribution accurately at some large $t$. 

\begin{figure}[H]
\begin{minipage}{0.329\hsize}
\centering
    \includegraphics[width=\hsize]{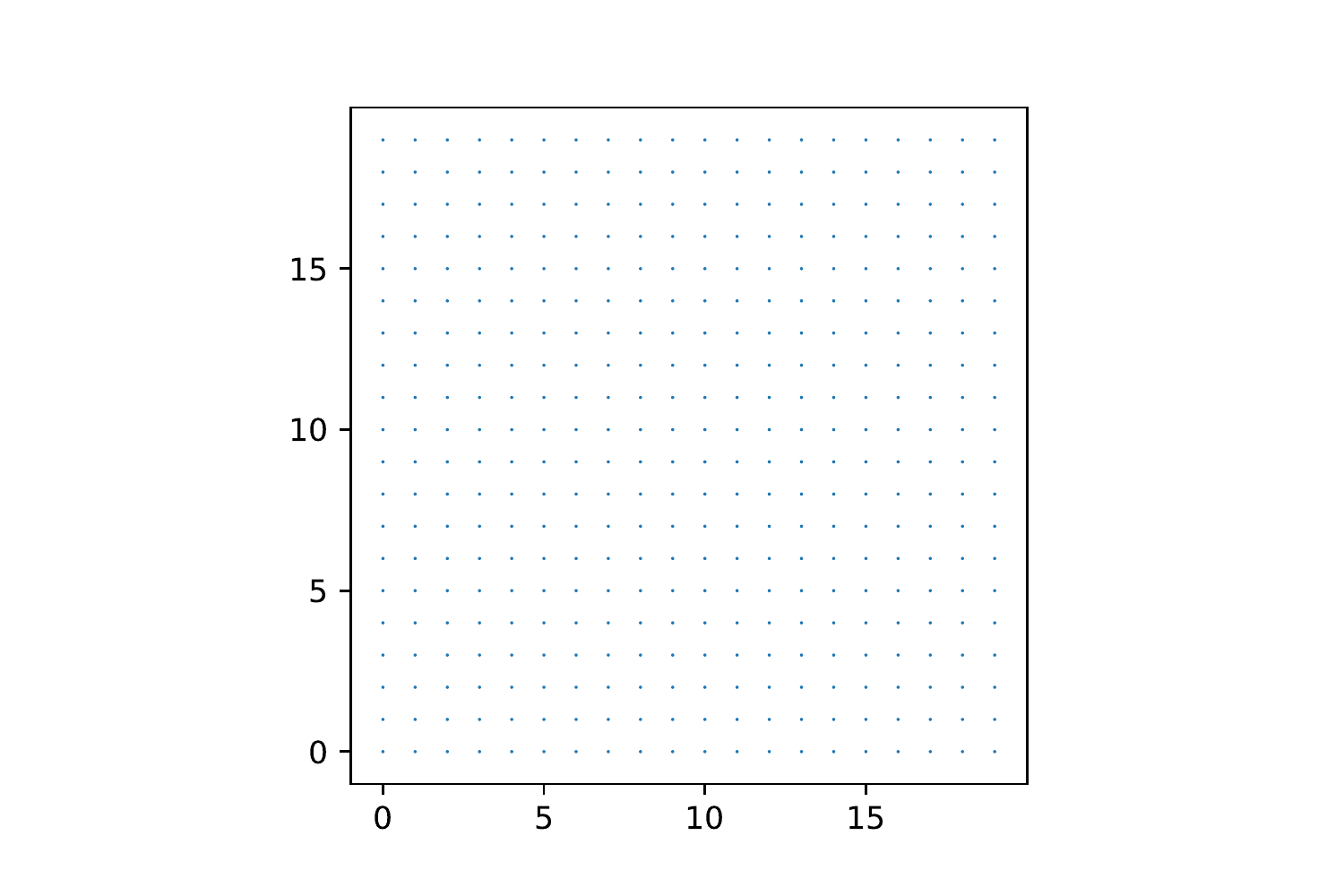}
    $t=0$
\end{minipage}
\begin{minipage}{0.329\hsize}
\centering
    \includegraphics[width=\hsize]{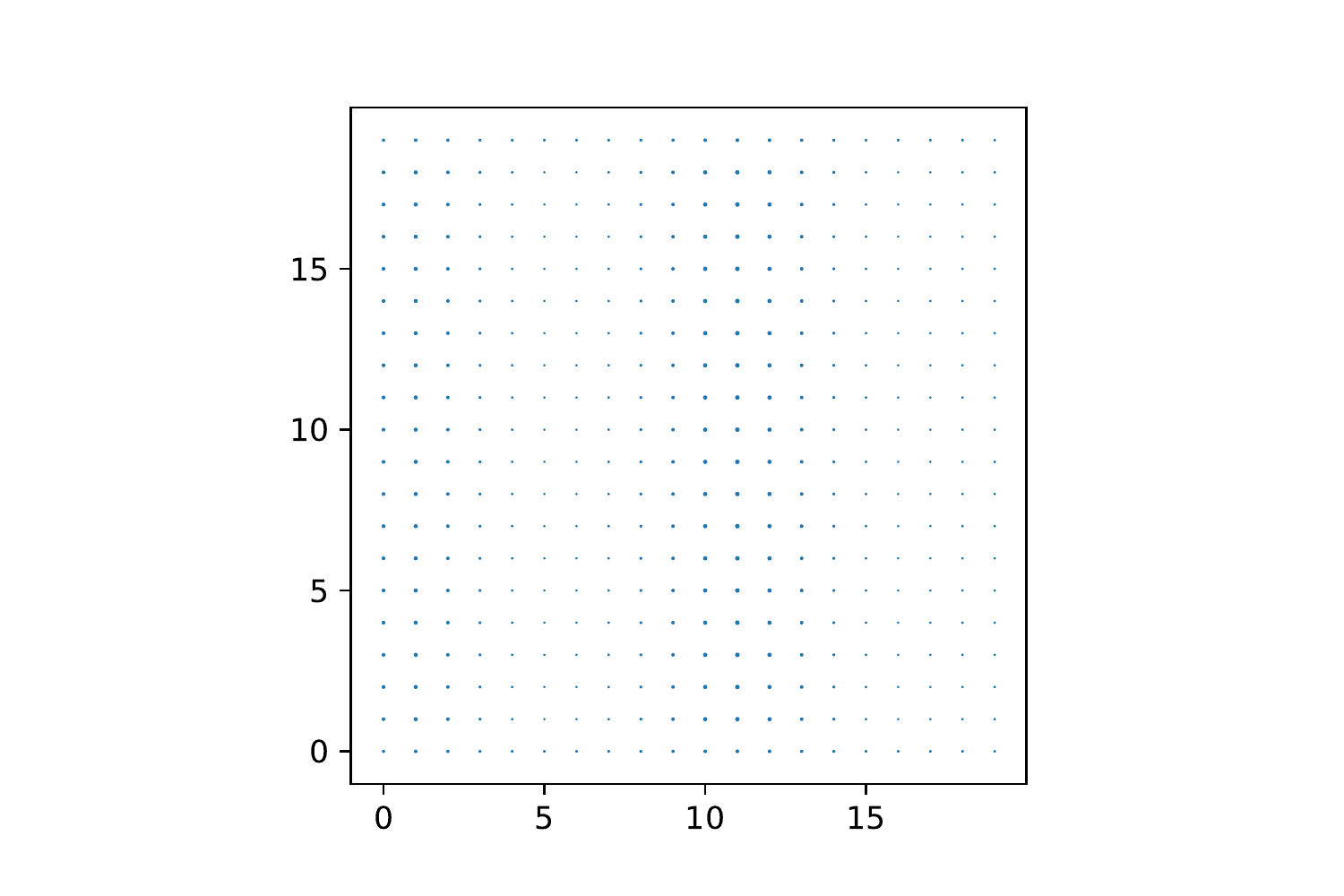}
\end{minipage}
\begin{minipage}{0.329\hsize}
\centering
    \includegraphics[width=\hsize]{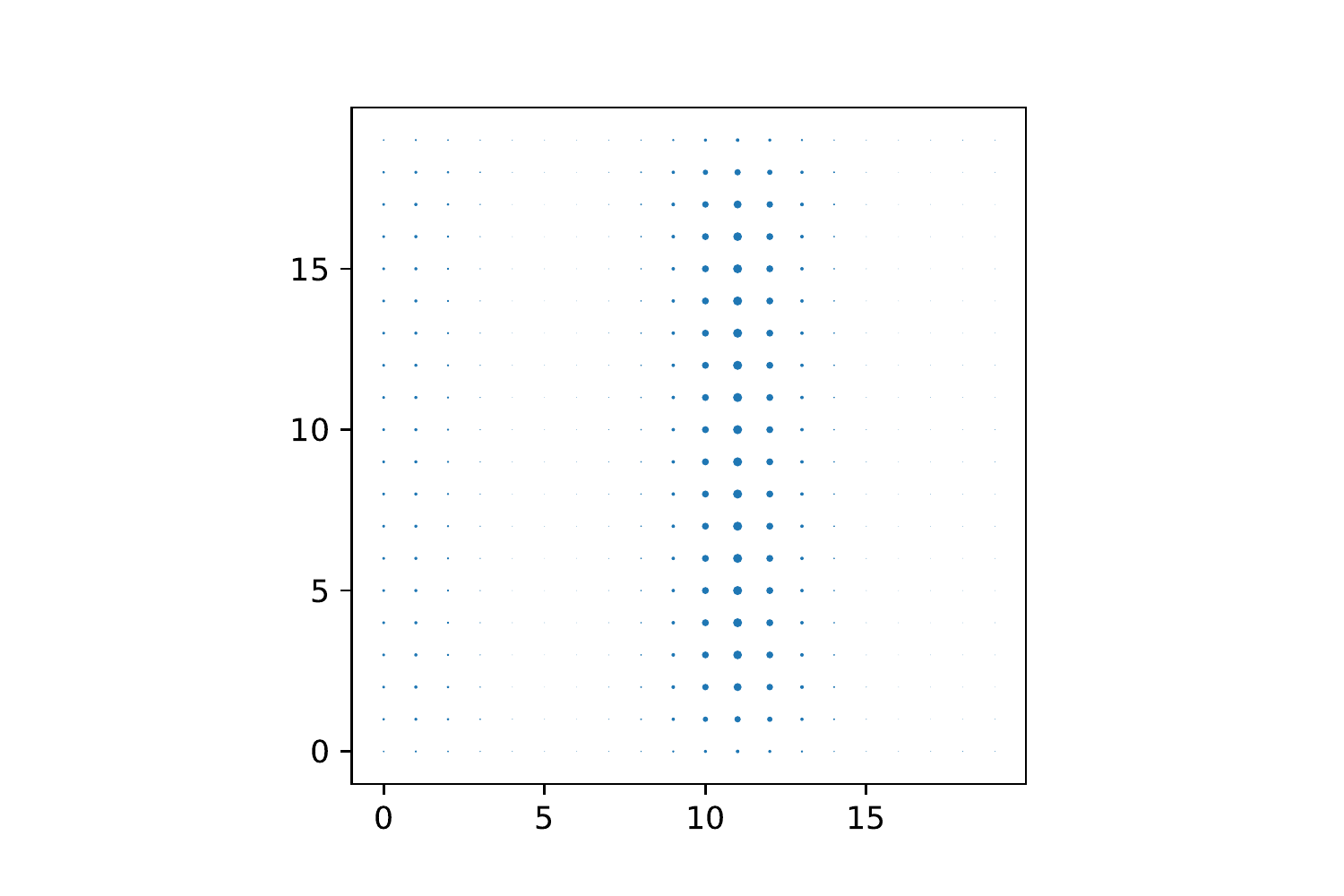}
\end{minipage}
\begin{minipage}{0.329\hsize}
\centering
    \includegraphics[width=\hsize]{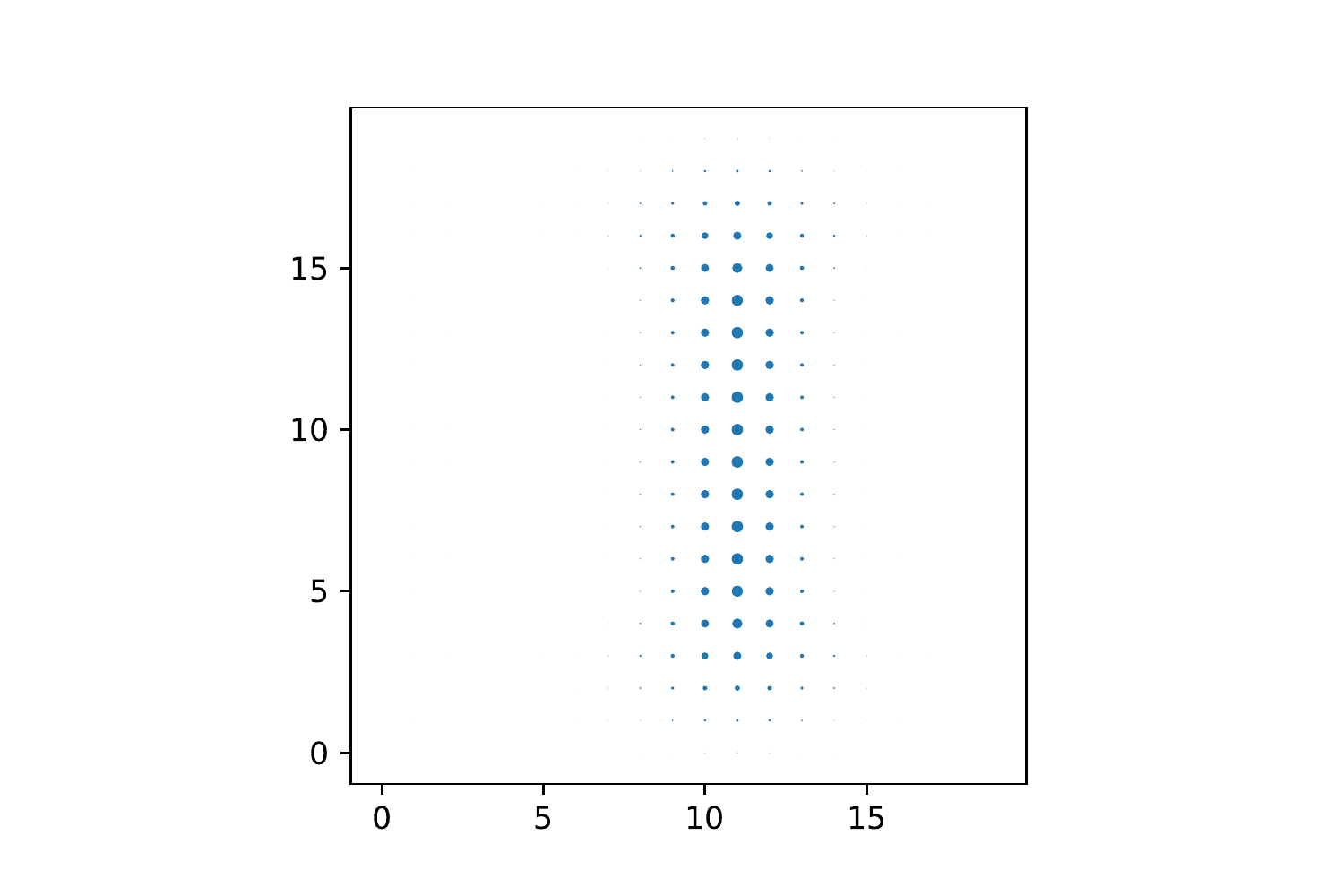}
\end{minipage}
\begin{minipage}{0.329\hsize}
\centering
    \includegraphics[width=\hsize]{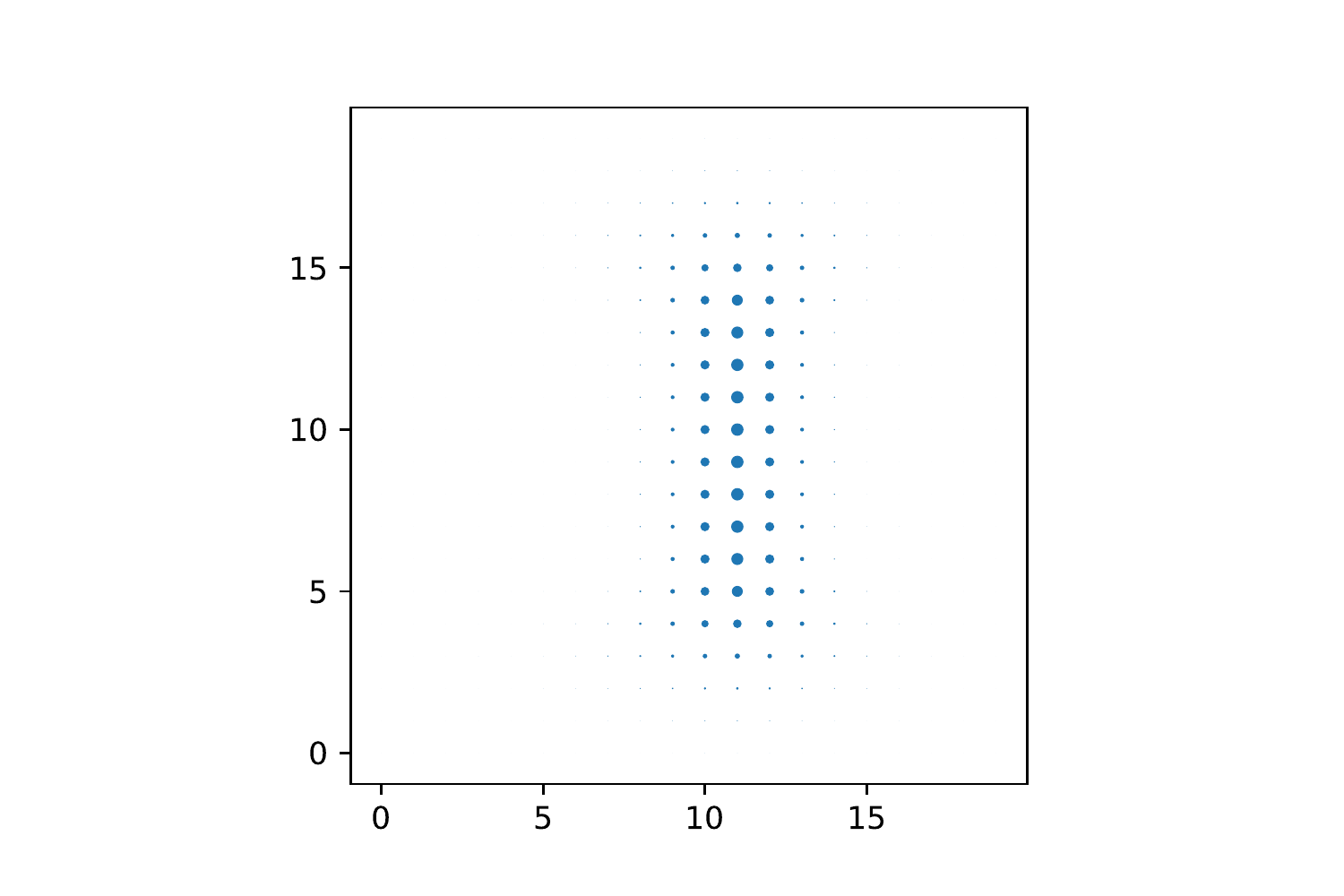}
\end{minipage}
\begin{minipage}{0.329\hsize}
\centering
    \includegraphics[width=\hsize]{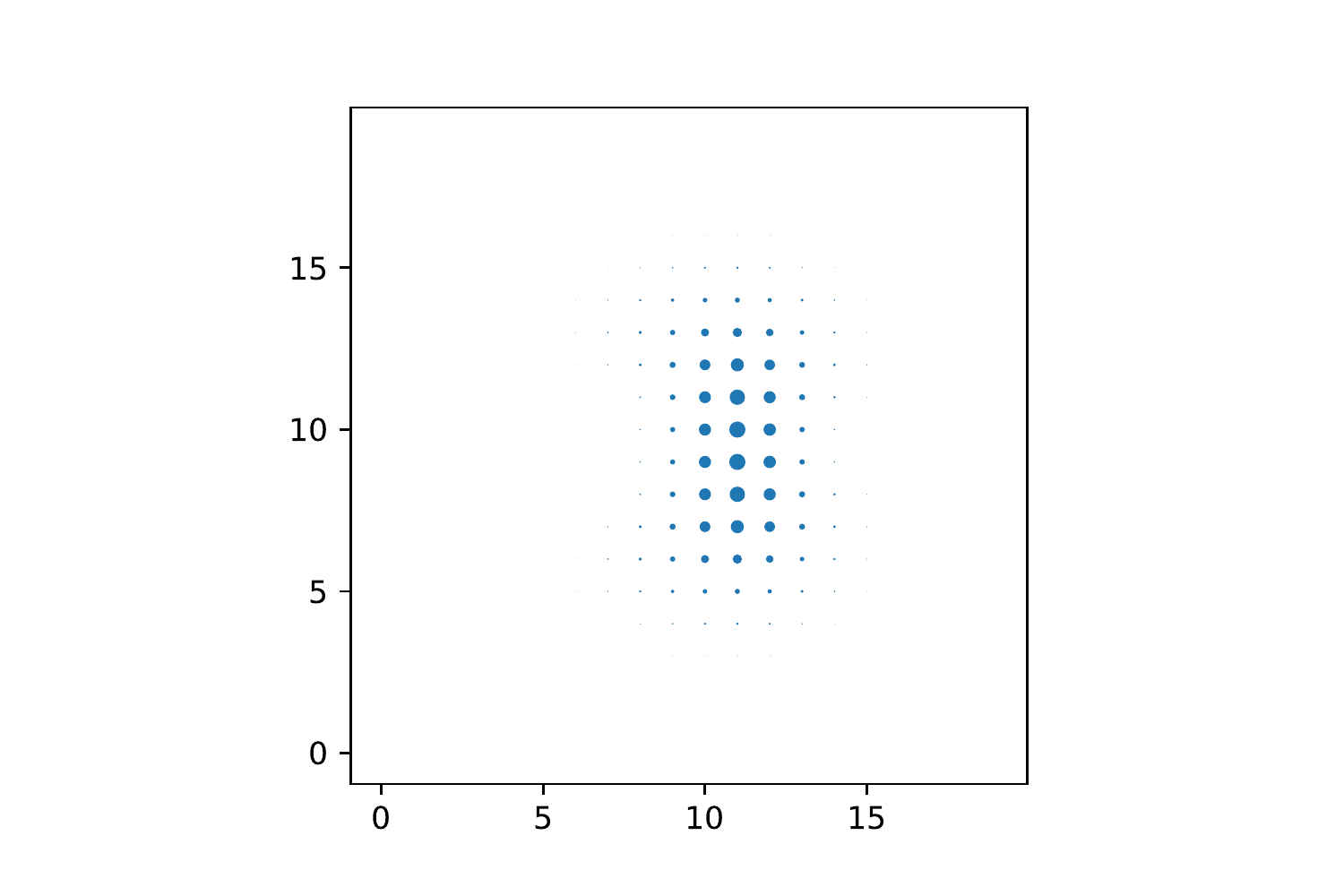}
\end{minipage}
\begin{minipage}{0.329\hsize}
\centering
    \includegraphics[width=\hsize]{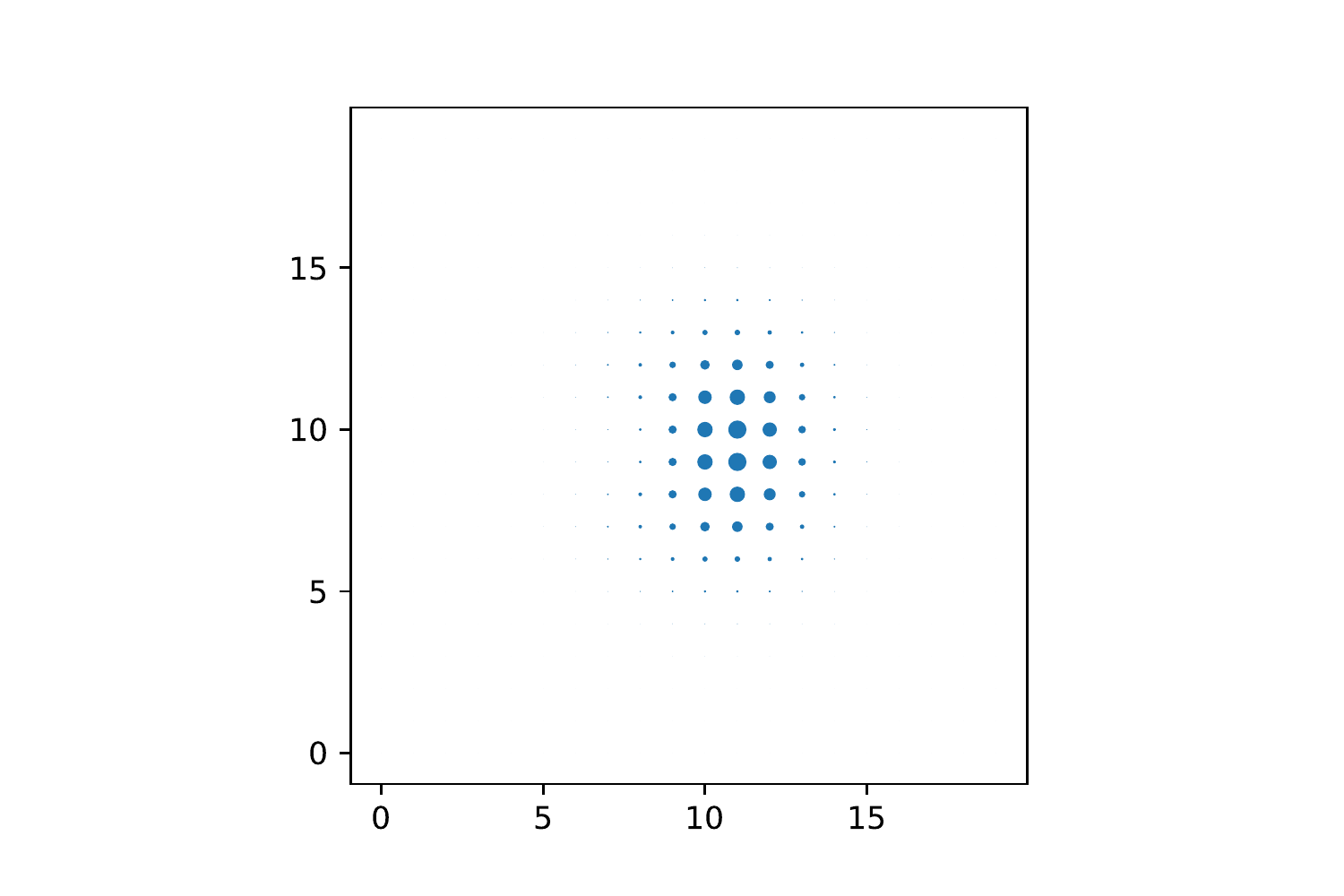}
\end{minipage}
\begin{minipage}{0.329\hsize}
\centering
    \includegraphics[width=\hsize]{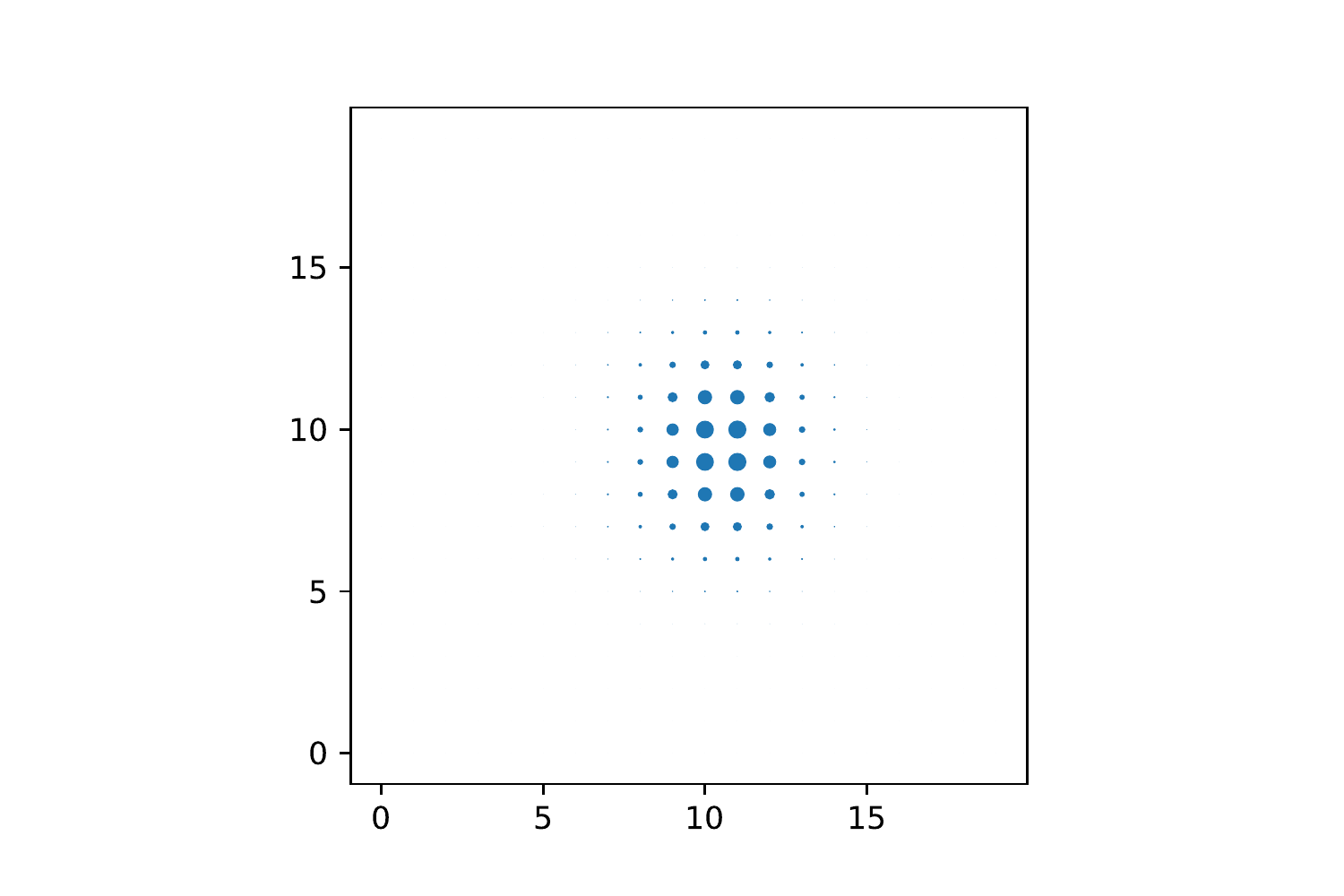}
\end{minipage}
\begin{minipage}{0.329\hsize}
\centering
    \includegraphics[width=\hsize]{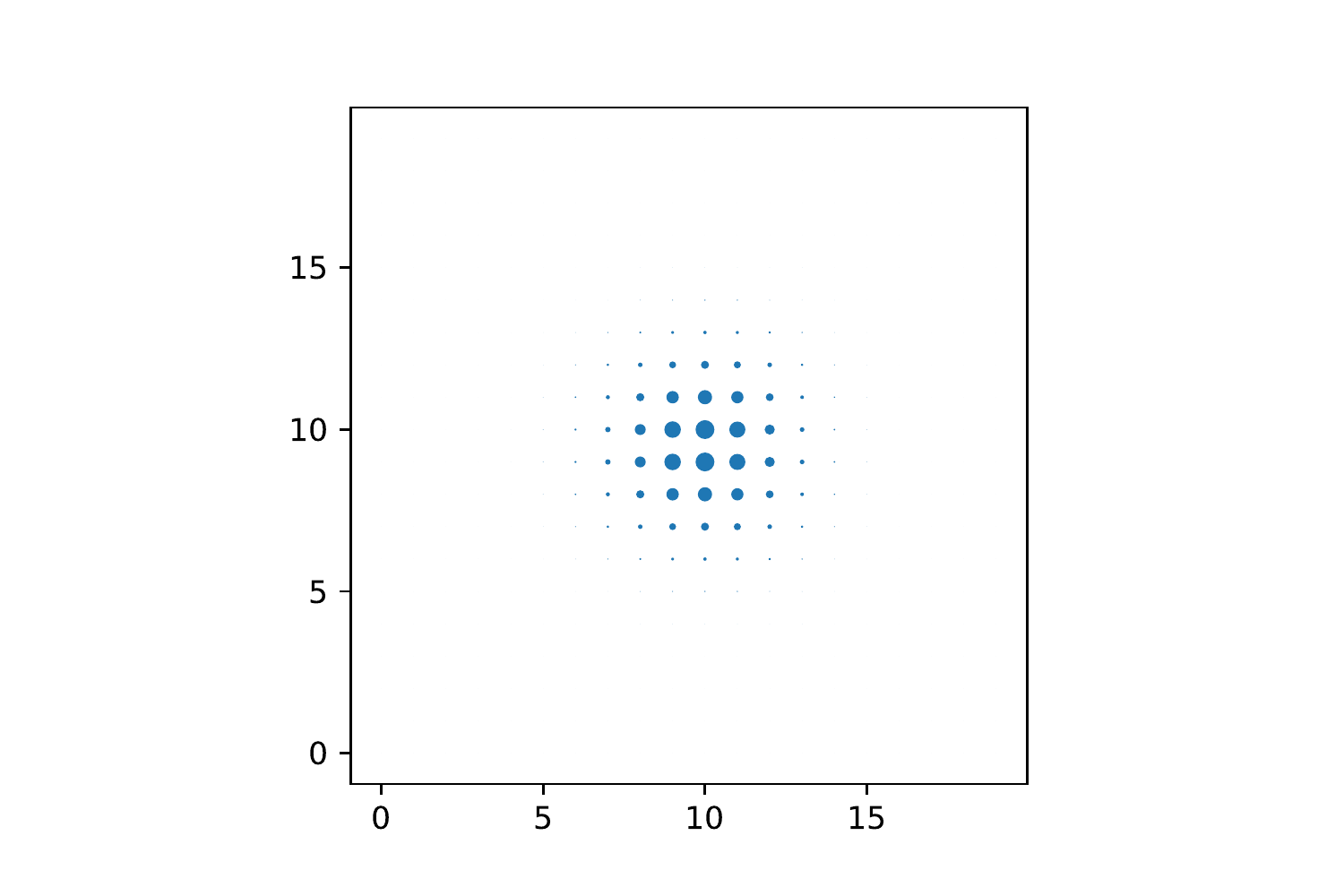}
    $t=20$
\end{minipage}
\caption{Density distribution of the ground states of the Hamiltonian $H(t)$ with $\phi=1/11$ on a 20 by 20 square lattice ($\Gamma(t)=e^{-t}, t\in [0,20]$). Size of a disk represents the density.}
    \label{fig:dos}
\end{figure}

\begin{figure}[H]
    \centering
    \includegraphics[width=7cm]{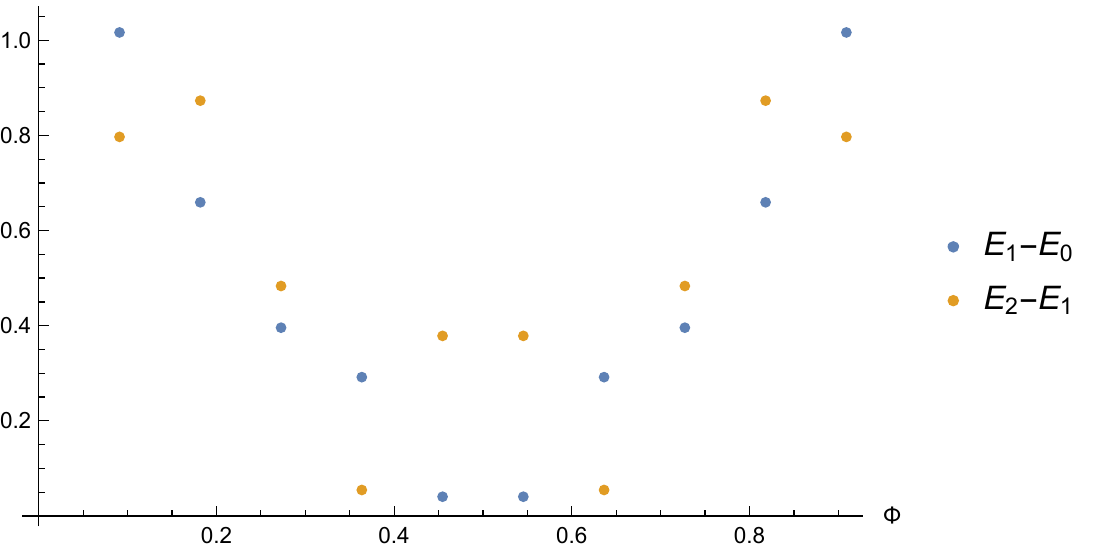}
    \caption{Energy spacing between subsequent levels: between the 1st excited levels and the ground levels $E_1-E_0$, and between the 2nd and the 1st excited levels $E_2-E_1$. }
    \label{fig:es}
\end{figure}

 Fig.\ref{fig:probs} and \ref{fig:probs2} show time dependence of finding some eigenstates of the same Hamiltonian $H(t)$. The behavior can be explained as follows. Computation starts with the ground state of $H_1$, whose energy is sufficiently separated from other energy levels, and as the $H_0$ is introduced other energy levels may become closer to the ground state. Intuitively, the closer they get, the higher the probability that an excited state is obtained. However introducing $H_1$ generally breaks the symmetry of the vacuum, hence a ground state may not be obtained with high probability. The figures of the probability of finding 3rd excited states imply that ground states, 1st excited states, 2nd excited stats are dominant solutions of this method with those schedules $\Gamma(t)$. One important feature of AQC is that the longer the computational time is the higher the probability of finding the ground state becomes. This is in fact exhibited in both Fig.\ref{fig:probs} and \ref{fig:probs2}. As expected, if $\Gamma(t)=e^{-t}$ is used, the probability converges much faster than when $\Gamma(t)=1/\text{log}(t+e)$ is used. In addition, Fig. \ref{fig:es} shows energy spacing between consecutive energy levels $\{E_{i+1}(\phi)-E_{i}(\phi)\}_{i=0,1}$. The energy spectra are symmetric $E_i(\phi+1/2)=E_i(-\phi+1/2)$ and $E_1(\phi)-E_0(\phi)$ gets close to 0 with $\phi\to 1/2$. So it is not surprising that the second excited state is more easily obtained as $\phi$ comes close to $1/2$. Moreover, by the same reason, the probability of obtaining the second excited at $\phi=4/11$ is as high as that of the first excited state. A problem of this method is that the performance heavily depends on the choice of schedules $\Gamma(t)$ as well as of the kinetic terms $H_1$. Moreover, in general, a wide flat (local) minimum solution is preferably chosen by AQC, hence solutions are not obtained with equal probability, even if they belong to the same energy level. This is called unfair sampling. In fact, the ground states of the $\phi=1/2$ case are degenerated, but one of them cannot be obtained equally. The wide-flatness, however, is sometimes useful. In some sense, model-independent features can be found in the wide-flat regime. Machine learning is a good application of AQC. For example, a narrow global minimum should be interpreted as an effect of over-training in machine learning.  
 \begin{figure}[H]
\begin{minipage}{0.50\hsize}
\centering
    \includegraphics[width=\hsize]{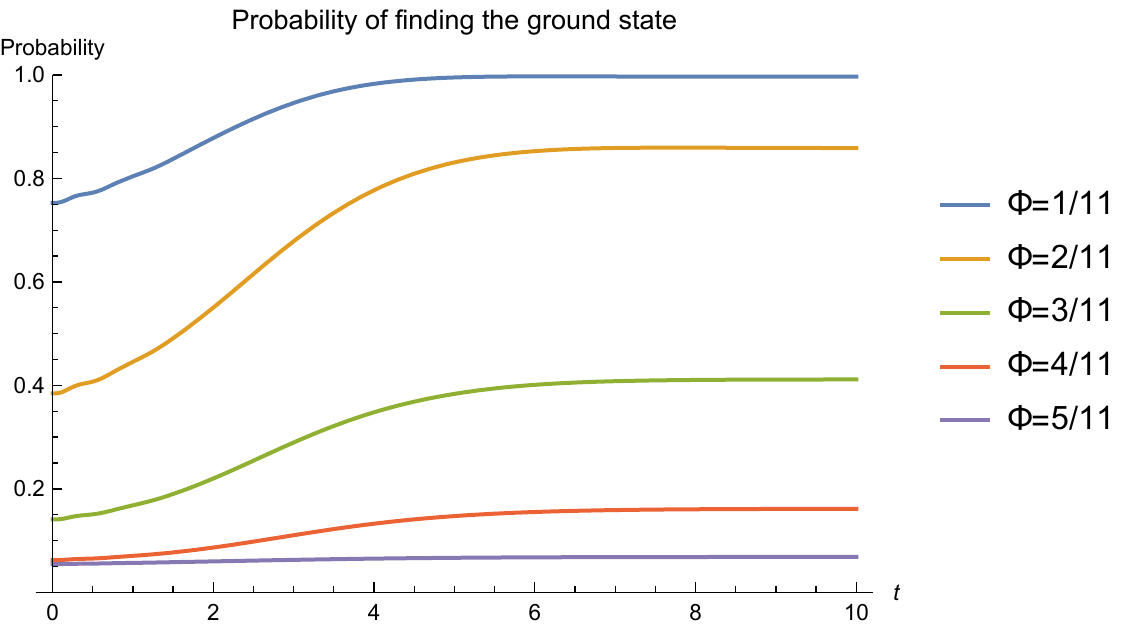}
\end{minipage}
\begin{minipage}{0.50\hsize}
\centering
    \includegraphics[width=\hsize]{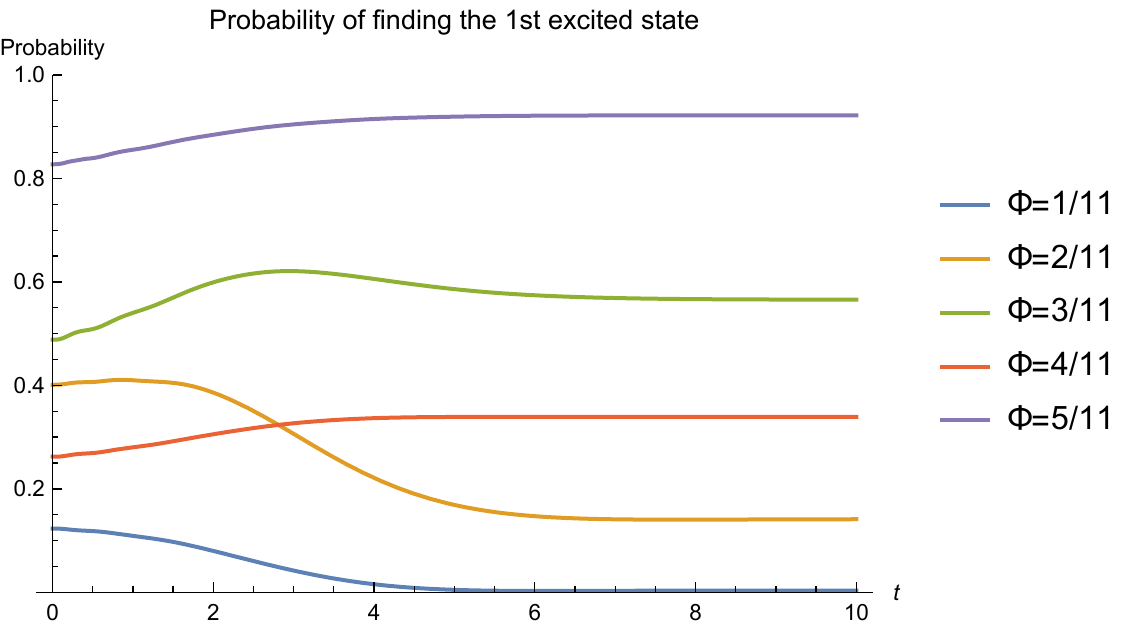}
\end{minipage}
\begin{minipage}{0.50\hsize}
\centering
    \includegraphics[width=\hsize]{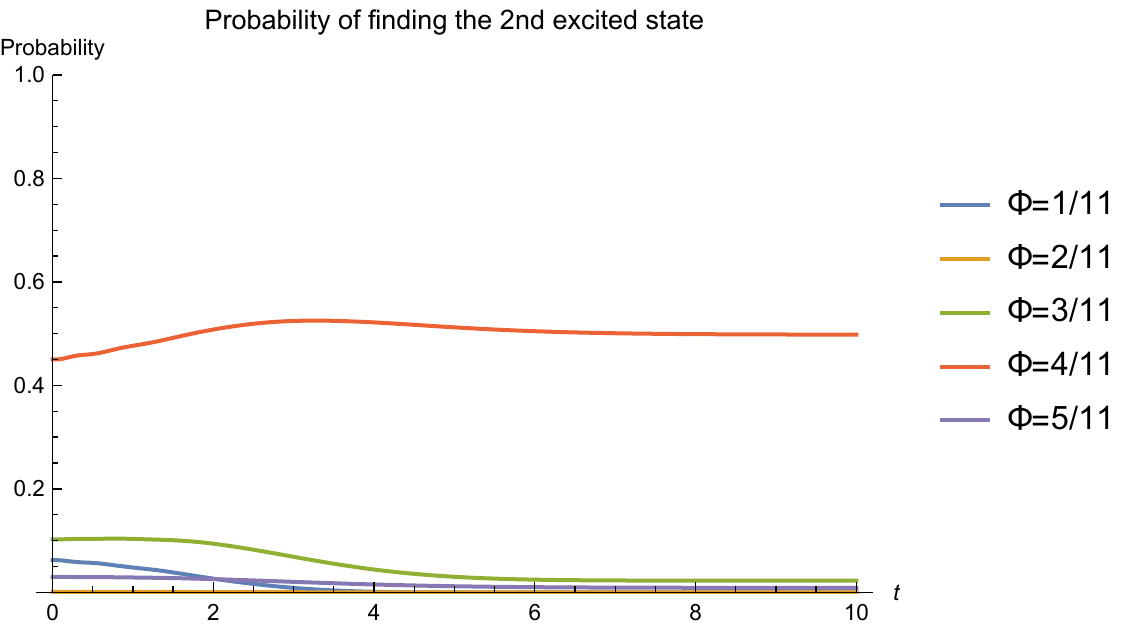}
\end{minipage}
\begin{minipage}{0.50\hsize}
\centering
    \includegraphics[width=\hsize]{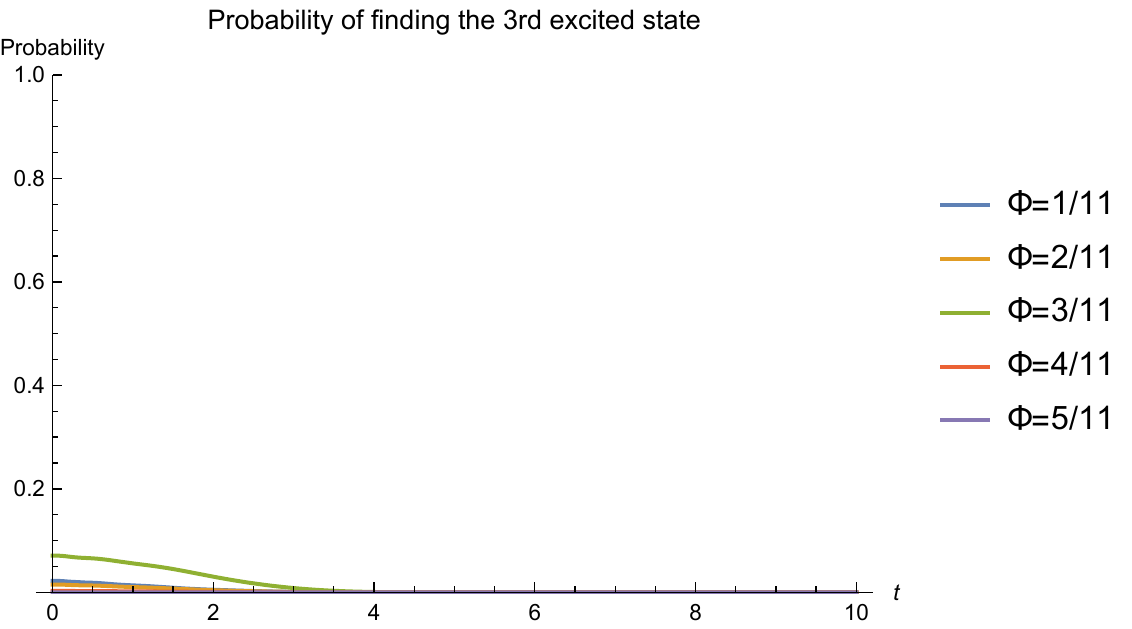}
\end{minipage}
\caption{Numerical results of probability of obtaining states on a square lattice under uniform magnetic flux $\phi$. The system obeys the Hamiltonian $H(t)=(1-\Gamma(t))H_0+\Gamma(t)H_1$ whose time dependence is described by  $\Gamma(t)=e^{-t}$.}
    \label{fig:probs}
\end{figure}

\begin{figure}[H]
\begin{minipage}{0.5\hsize}
\centering
    \includegraphics[width=\hsize]{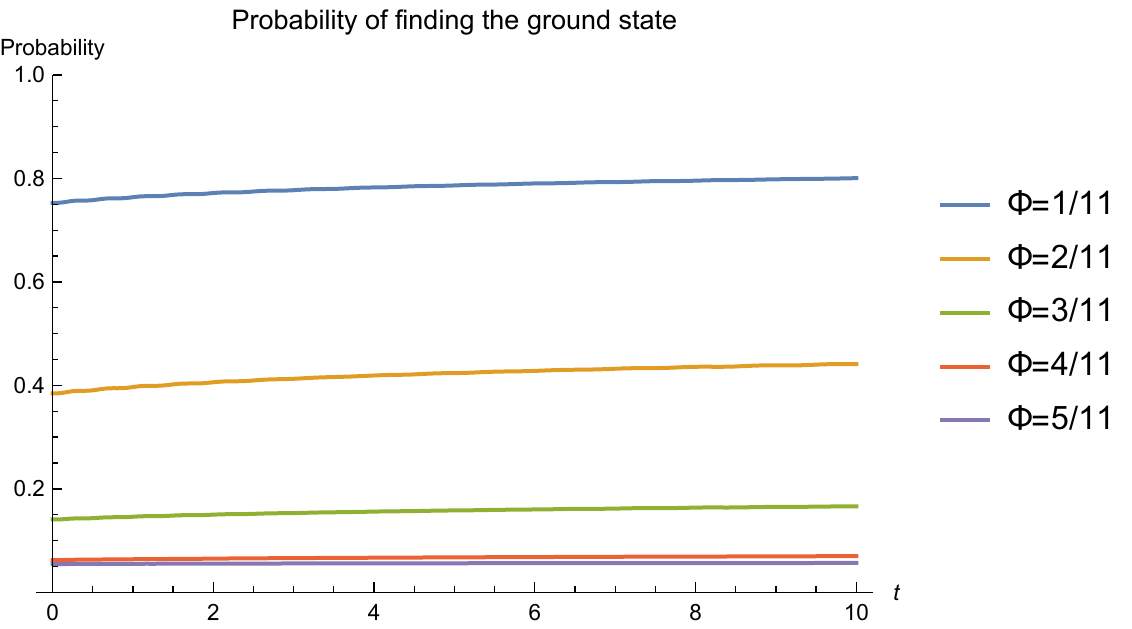}
\end{minipage}
\begin{minipage}{0.5\hsize}
\centering
    \includegraphics[width=\hsize]{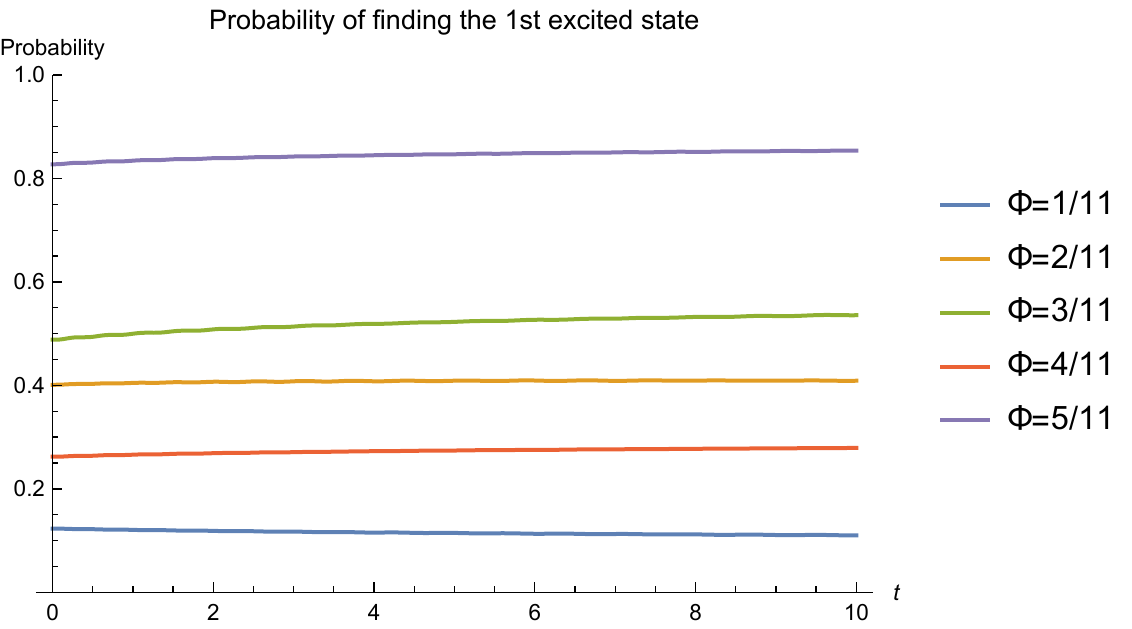}
\end{minipage}
\begin{minipage}{0.5\hsize}
\centering
    \includegraphics[width=\hsize]{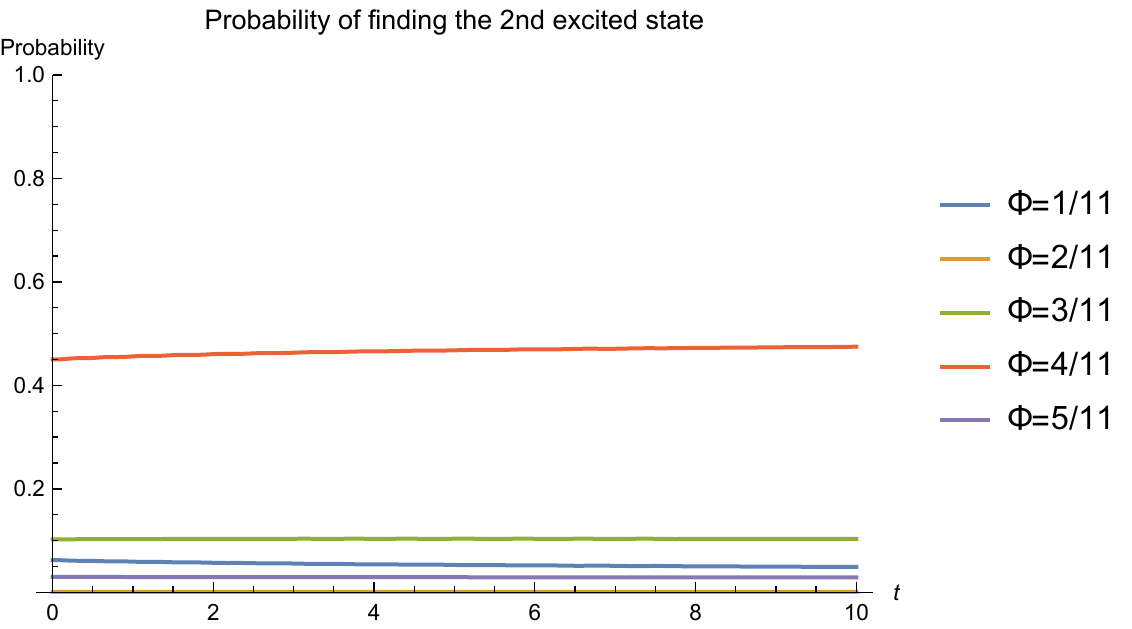}
\end{minipage}
\begin{minipage}{0.5\hsize}
\centering
    \includegraphics[width=\hsize]{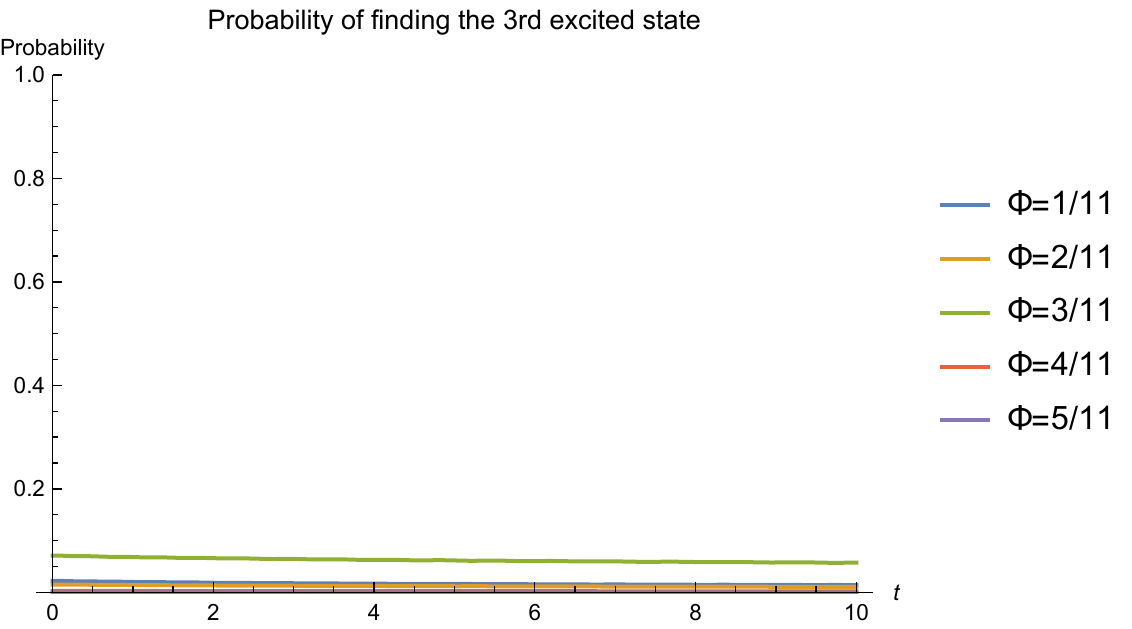}
\end{minipage}
\caption{Numerical results of probability of obtaining states on a square lattice under uniform magnetic flux $\phi$. The system obeys the Hamiltonian $H(t)=(1-\Gamma(t))H_0+\Gamma(t)H_1$ whose time dependence is described by  $\Gamma(t)=1/\log(t+e)$.}
    \label{fig:probs2}
\end{figure}
In general, performance of AQC depends on schedules. So in what follows we try several other choices and explore more on the schedule dependence of the probability. Let $\{\mathcal{E}_i(t)\}$ be a one-parameter family of energy eigenvalues $(i=0,\cdots, \dim H(t))$ of $H(t)$. We define the energy gap between the ground state and the first excited state by 
 \begin{equation}
     \Delta=\inf_{t}(\mathcal{E}_1(t)-\mathcal{E}_0(t)). 
 \end{equation}
 According to the adiabatic theorem, the ground state of the target Hamiltonian $H_0$ should be found with probability arbitrarily close to 1, after sufficiently long time $T\gg O(1/\Delta^2)$. However, looking at the figures \ref{fig:probs} and \ref{fig:probs2}, one may guess it is by no means possible to find the ground states frequently even after sufficiently long time. To solve this puzzle, we consider the schedule dependence of the probability. We address two cases: a finite schedule ($\Gamma(\tau)=0$ at some $\tau<\infty$) and an infinite schedule ($\Gamma(t)>0$ for any $t$ and $\lim_{t\to\infty}\Gamma(t)=0$). For the infinite schedule, we use $\Gamma(t)=\exp(-at)$ and control the speed by tuning $a>0$. Fig. \ref{fig:probs4} exhibits the numerical results of finding the ground states. As the finite case \eqref{eq:f}, the probability successfully increases as the computation speed decreases. In both of two cases in Fig. \ref{fig:probs4}, computation stops with the same value of $\Gamma(t)>0$. 

\begin{figure}[H]
\begin{minipage}{0.5\hsize}
\centering
    \includegraphics[width=\hsize]{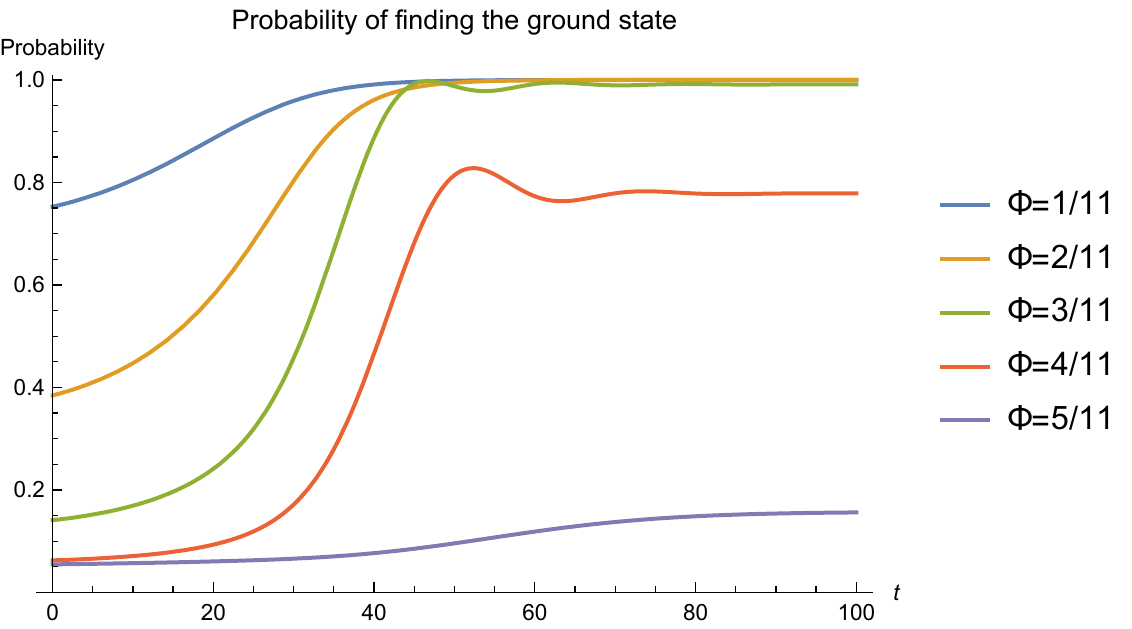}
\end{minipage}
\begin{minipage}{0.5\hsize}
\centering
    \includegraphics[width=\hsize]{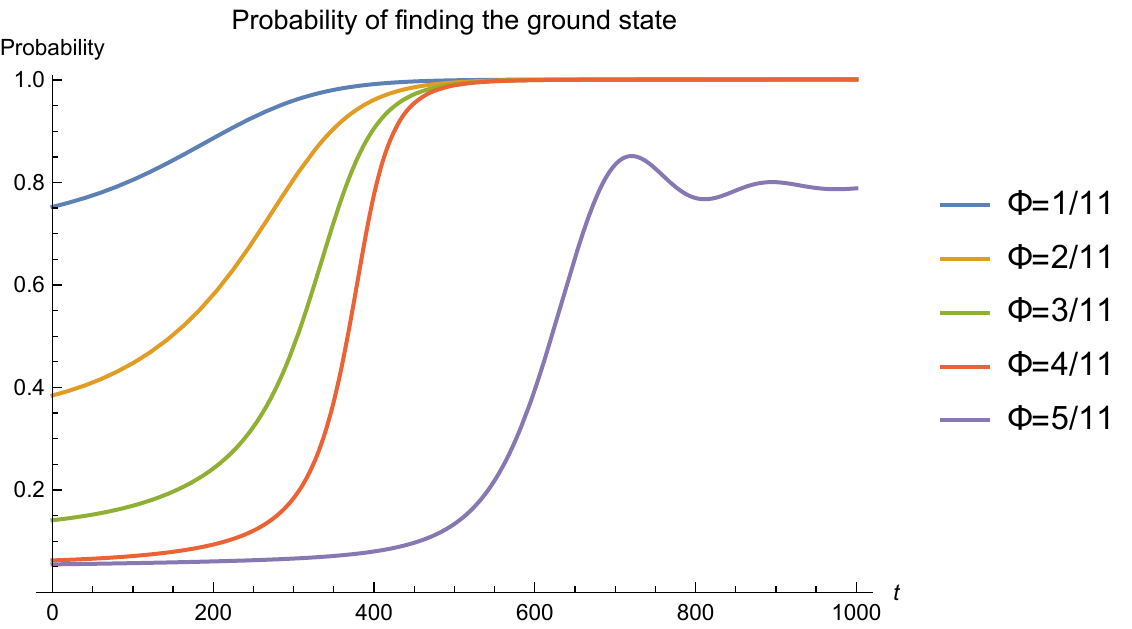}
\end{minipage}
\caption{Numerical results of probability of obtaining the ground states on a 3 by 3 square lattice under uniform magnetic flux $\phi$. The system obey the Hamiltonian $H(t)=(1-\Gamma(t))H_0+\Gamma(t)H_1$ whose time dependence is described by $\Gamma(t)=\exp(-at)$. [Left] $a=0.1$ [Right] $a=0.01$}
    \label{fig:probs4}
\end{figure}

For the finite schedule, we introduce the following function \begin{equation}\label{eq:f}
    \Gamma(t)=1-\frac{1}{2}\left(\frac{\arctan{(t-\tau/2)}}{|\arctan{(-\tau/2)}|}+1\right), 
\end{equation}
where $\tau(<\infty)$ is finite computational time. This monotonic function slowly begins to decrease and gradually banishes at $t=\tau$. A difference from previous functions is that it satisfy $\Gamma''(t)<0$ for $0<t<\tau/2$ and $\Gamma''(t)>0$ for $\tau/2<t< \tau$ (see Fig. \ref{fig:f}). 

\begin{figure}[H]
    \centering
    \includegraphics[width=8cm]{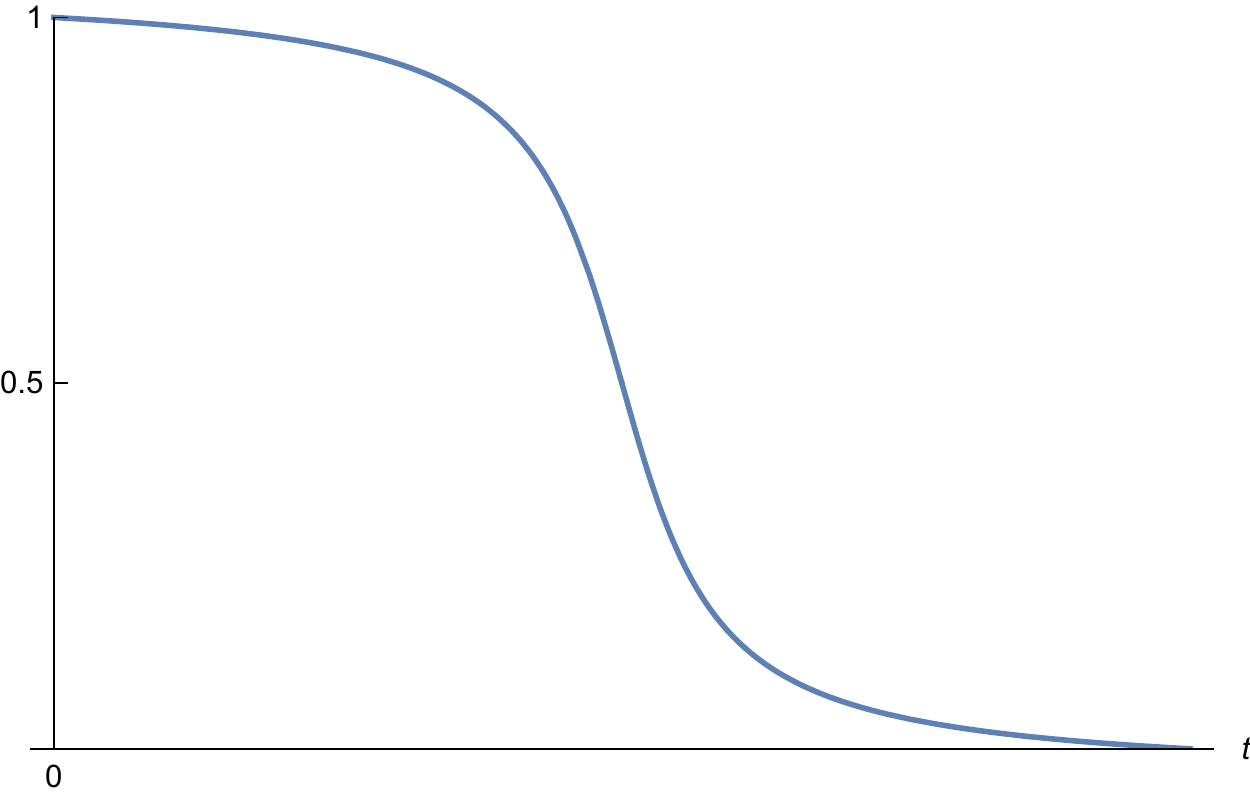}
    \caption{Behavior of \eqref{eq:f}}
    \label{fig:f}
\end{figure}

The probability of finding ground states is shown in Fig. \ref{fig:probs3}. As expected, taking a longer runtime $\tau$ helps one find the ground states accurately for all $\phi$, thereby the other states are unlikely obtained. It would be typical to this schedule $\Gamma(t)$ \eqref{eq:f} that the probability rapidly increases around $t$ when $\Gamma''(t)=0$.

\begin{figure}[H]
\begin{minipage}{0.329\hsize}
\centering
    \includegraphics[width=\hsize]{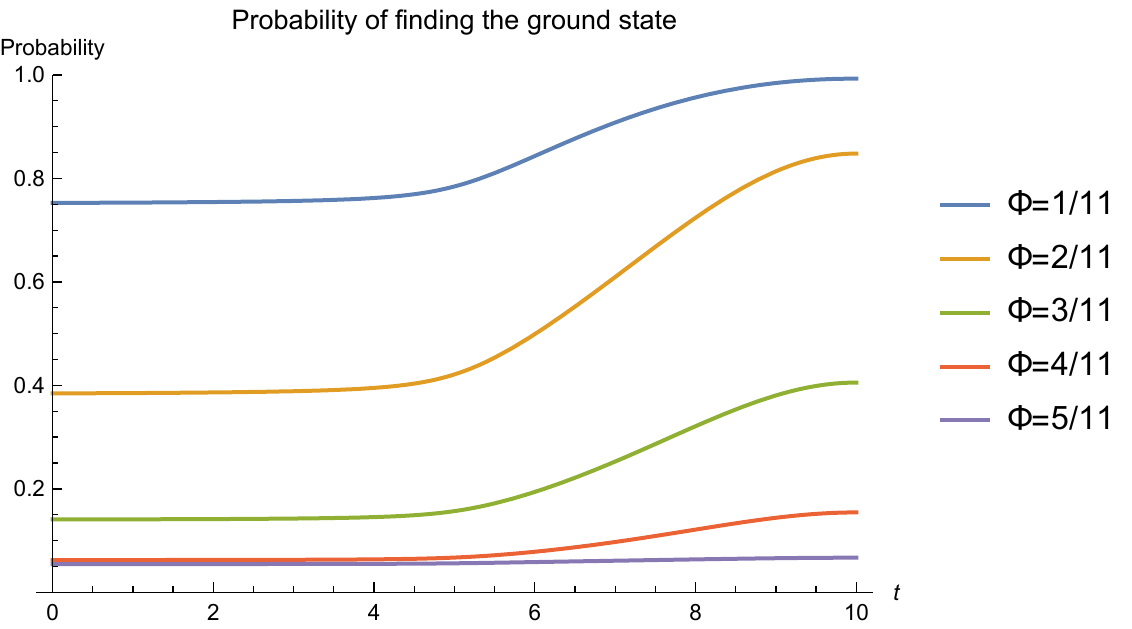}
\end{minipage}
\begin{minipage}{0.329\hsize}
\centering
    \includegraphics[width=\hsize]{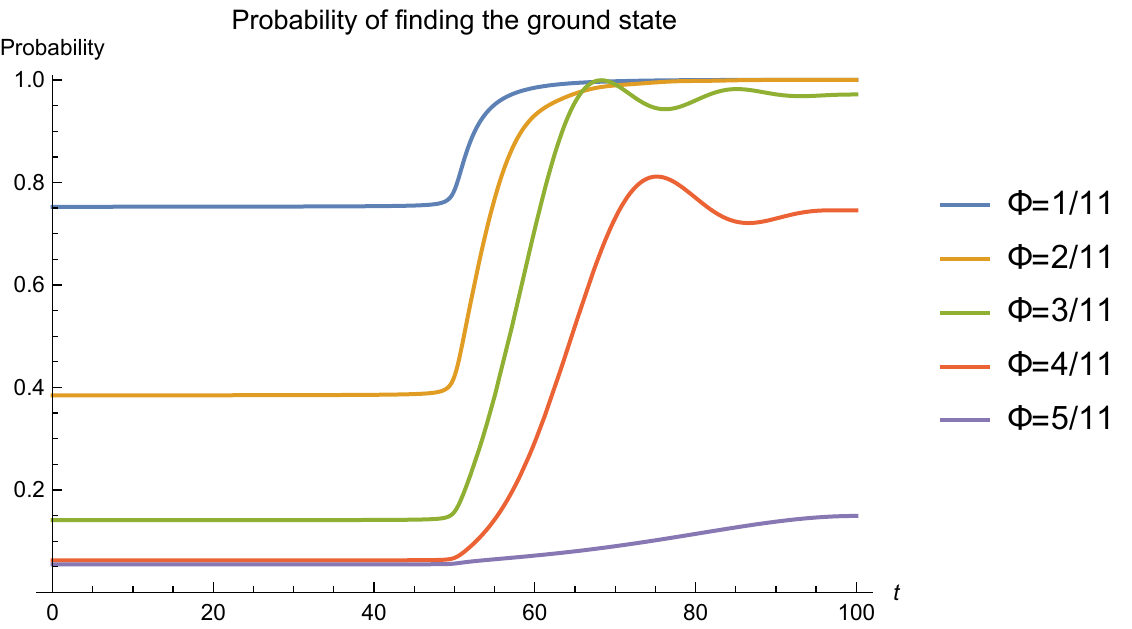}
\end{minipage}
\begin{minipage}{0.329\hsize}
\centering
    \includegraphics[width=\hsize]{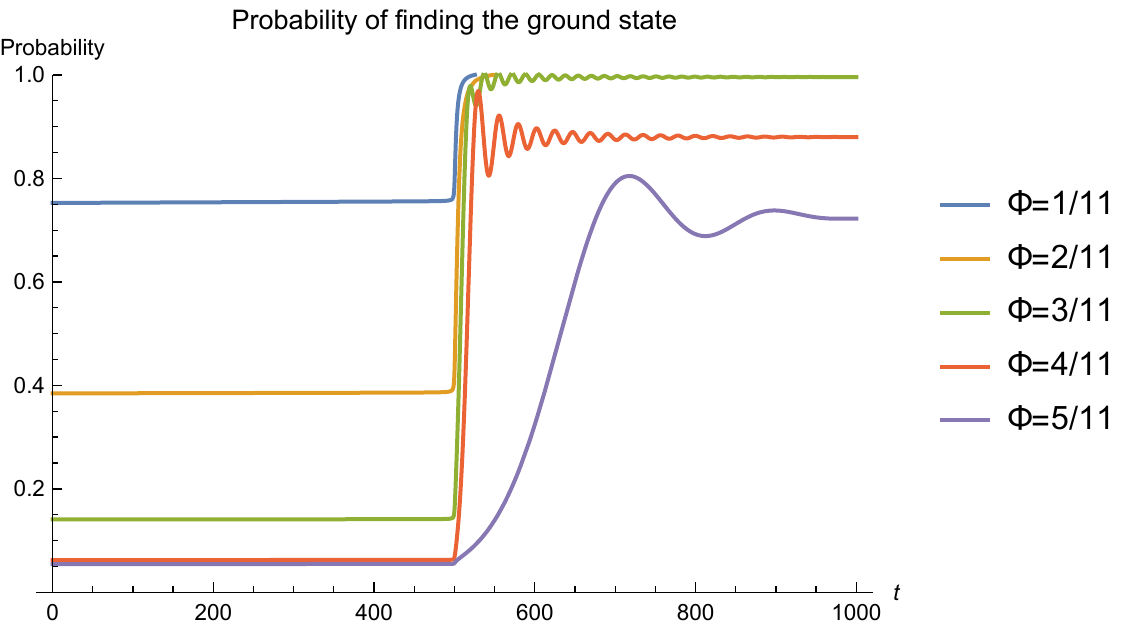}
\end{minipage}
\caption{Numerical results of probability of obtaining the ground states on a 3 by 3 square lattice under uniform magnetic flux $\phi$. The system obey the Hamiltonian $H(t)=(1-\Gamma(t))H_0+\Gamma(t)H_1$ whose time dependence is described by \eqref{eq:f}. [Left] $\tau=10$ [Middle] $\tau=100$ [Right] $\tau=1000$}
    \label{fig:probs3}
\end{figure}
}\fi

\section{Dynamics of the Butterfly}\label{sec:ex}
Another example of AQC is a study of quantum physics. The quantum annealing is commonly recognized as a solver of combinatorial optimization problems, and indeed we showed they can be solved in our way before. In addition to that, we show our method is also fairly useful to simulate fermionic systems. Although it would not be impossible to formulate ferimonic systems with $Z_i$ basis, coding can be messy in general. Here we aim at demonstrating a Bloch electron system under uniform magnetic flux $\phi$ perpendicular to the system. This system is one of the most important ones in condensed matter physics. Especially the following Hamiltonian \eqref{eq:ferm} is widely used for studying the two dimensional integer quantum Hall effect, which exhibits the most fundamental topological property, therefore plays a crucial role to study not only general topological matter physics but also high energy physics and mathematical physics \cite{doi:10.1063/1.4998635,Ikeda:2017uce,Ikeda:2018tlz}. We define the coupling by the associated $U(1)$ gauge field $\theta$. Then the Hamiltonian is 
\begin{equation}\label{eq:ferm}
H_0=-\left(\sum_{m,n}e^{2\pi i \theta^x_{m,n}}c^\dagger_{m+1,n}c_{m,n}+e^{2\pi i \theta^y_{m,n}}c^\dagger_{m,n+1}c_{m,n}+h.c\right),     
\end{equation}
where $c_i,c^\dagger_i$ are fermionic annihilation and creation operators $\{c_i,c^\dagger_j\}=\delta_{ij}, \{c_i,c_j\}=\{c^\dagger_i,c^\dagger_j\}=0$. Now let us reproduce physics of this model with our method. To this end, we work on a tight-binding Hamiltonian on a two dimensional square lattice:
\begin{equation}\label{eq:tight}
    \widetilde{H_0}=-\left(\sum_{\langle ij\rangle}\gamma_{ij}a^\dagger_i a_j+\gamma^*_{ji}a^\dagger_ja_i \right),
\end{equation}
where the summation is taken over the nearest neighbor pairs $i=(i_m, i_n)$. Then the tight-binding Hamiltonian can be written as 
\begin{equation}\label{eq:qhe}
    \widetilde{H_0}=-\left(\sum_{m,n}e^{2\pi i \theta^x_{m,n}}a^\dagger_{m+1,n}a_{m,n}+e^{2\pi i \theta^y_{m,n}}a^\dagger_{m,n+1}a_{m,n}+h.c\right), 
\end{equation}
where $(i_m,i_n)$ in \eqref{eq:tight} is abbreviated by $(m,n)$. For a single particle state
\begin{equation}
\ket{\phi}=\sum_{m,n}\phi_{m,n}a^\dagger_{m,n}\ket{\emptyset},    
\end{equation}
the hopping energy from one site $a^\dagger_{m,n}\ket{\emptyset}=\ket{m,n}$ to another $\ket{m+1,n}$ is
\begin{equation}
    \bra{m+1,n}\widetilde{H_0}\ket{m,n}=-e^{2\pi i\theta^x_{m,n}},
\end{equation}
which corresponds to a matrix element of $H_0$.

In our case with boundary, the butterflies accommodate energy spectra of edge states. 
\begin{figure}[H]
    \centering
    \includegraphics[width=7cm]{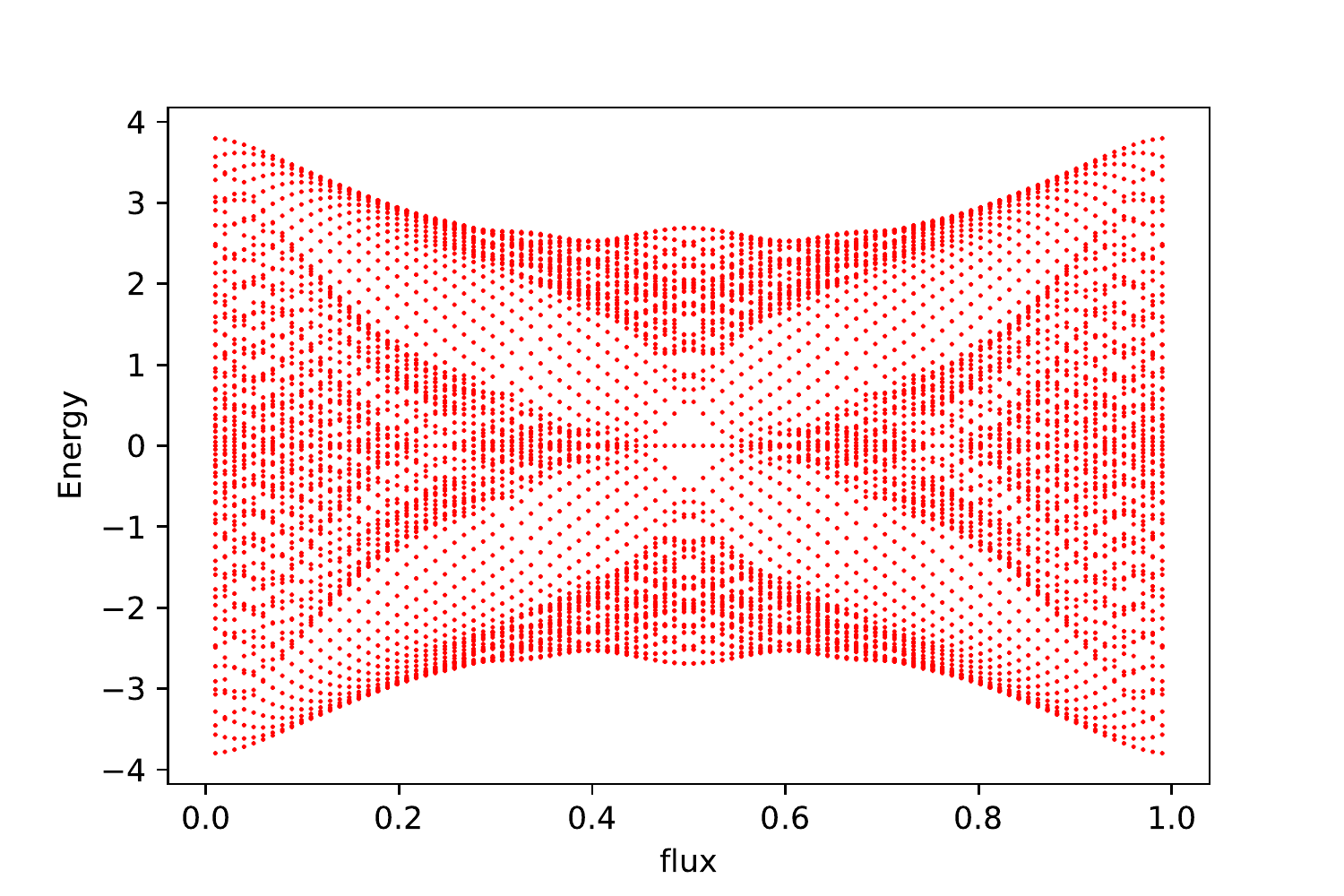}
    \caption{Energy spectra of a Bloch electron system on a square lattice with fractional magnetic flux $\phi$ perpendicular to the system. $\phi$ runs over $1/101,2/101,\cdots, 100/101$. The Landau gauge $(\theta^x_{m,n}, \theta^y_{m,n})=(0,m\phi)$ was used for computation.}
    \label{fig:butt}
\end{figure}
\begin{figure}[H]
    \centering
    \includegraphics[width=7cm]{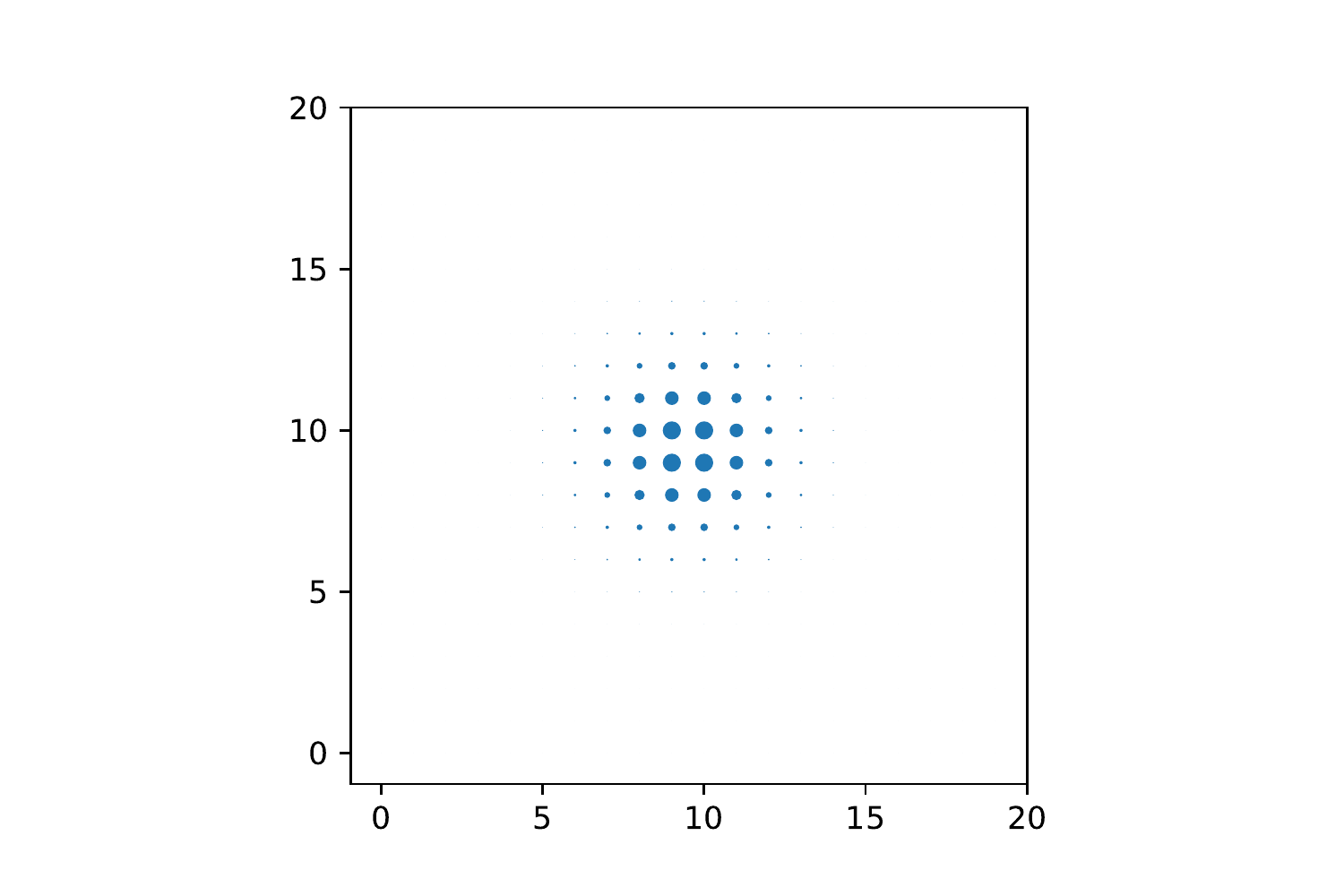}
    \caption{The density distribution of the ground state of Bloch electrons on a 20 by 20 square lattice under uniform magnetic flux $\phi=1/11$.}
    \label{fig:dos_g}
\end{figure}
Fig. \ref{fig:dos_g} shows the density distribution of the ground state of Bloch electrons on a 20 by 20 square lattice under uniform magnetic flux $\phi=1/11$. The distribution is computed with the standard fermionic tight-binding Hamiltonian \eqref{eq:ferm}. Fig. \ref{fig:dos} shows time dependence of the density distribution of the ground state of $H(t)=(1-\Gamma(t))H_0+\Gamma(t)H_1$ defined over the same lattice. At the initial time $t=0$, the ground sate of the ferromagnetic $H_1=-XX$ interaction is uniformly distributed to all regions of the lattice. As time pass by, states gather at center of the bulk. Comparing Fig. \ref{fig:dos_g} and the last figure in Fig. \ref{fig:dos}, we find that the model approximates the density distribution accurately at some large $t$.

The fractral structure in the figure is realized by the interplay of Bragg's reflection and Landau's quantization of Bloch electrons on a lattice \cite{PhysRevB.14.2239}. It attracts the interest of many authors from viewpoints of condensed matter physics, high energy physics \cite{Hatsuda:2016mdw} and mathematical physics \cite{doi:10.1063/1.4998635}. In our case with boundary, the butterflies accommodate energy spectra of edge states. The distribution is computed with the standard fermionic tight-binding Hamiltonian \eqref{eq:ferm}. Fig. \ref{fig:dos} shows time dependence of the density distribution of the ground state of $H(t)=(1-\Gamma(t))H_0+\Gamma(t)H_1$ defined over the same lattice. At the initial time $t=0$, the ground sate of the ferromagnetic $H_1=-XX$ interaction is uniformly distributed to all regions of the lattice. As time pass by, states gather at center of the bulk. We find that the model approximates the density distribution accurately at some large $t$. 

\begin{figure}[H]
\begin{minipage}{0.329\hsize}
\centering
    \includegraphics[width=\hsize]{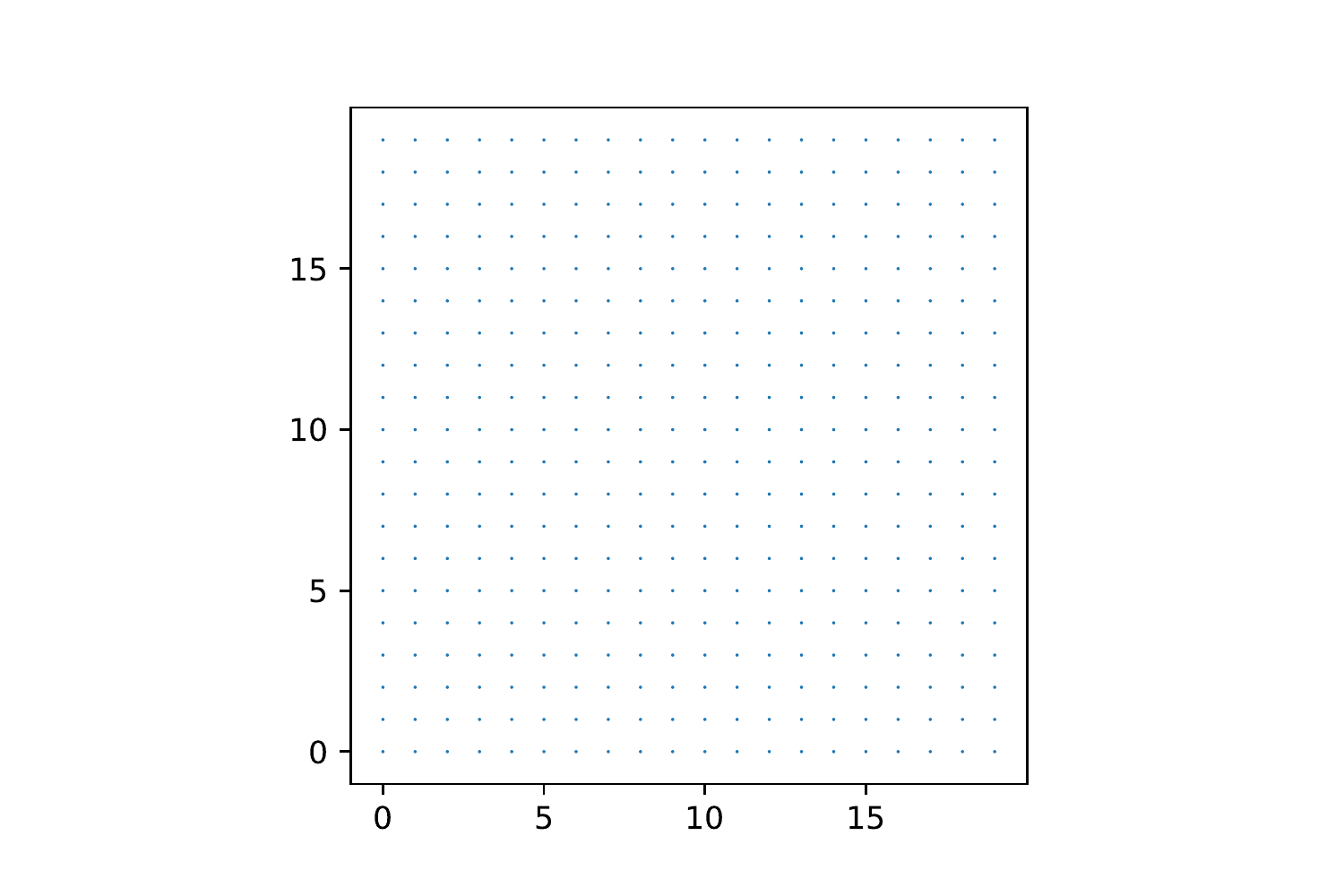}
    $t=0$
\end{minipage}
\begin{minipage}{0.329\hsize}
\centering
    \includegraphics[width=\hsize]{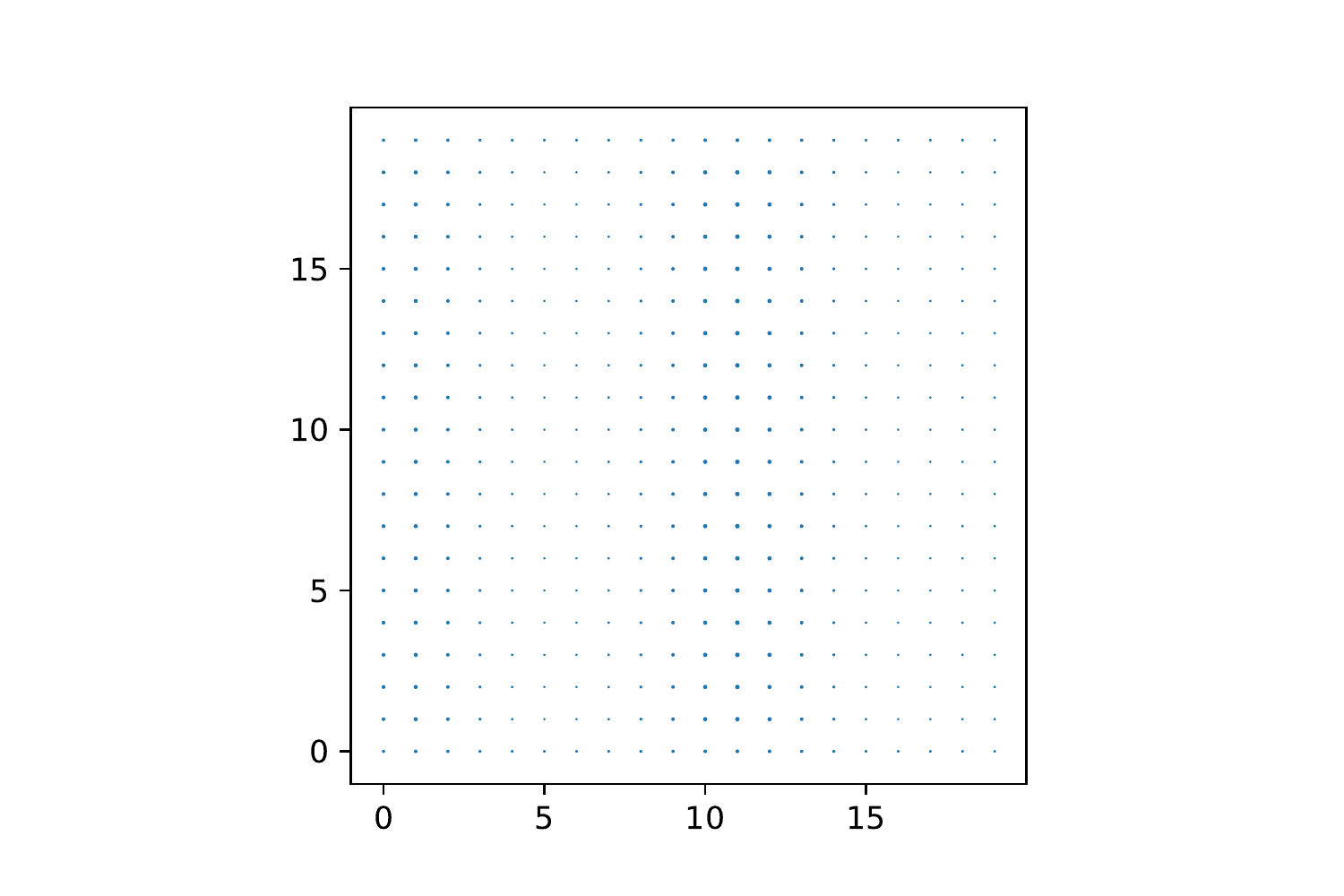}
\end{minipage}
\begin{minipage}{0.329\hsize}
\centering
    \includegraphics[width=\hsize]{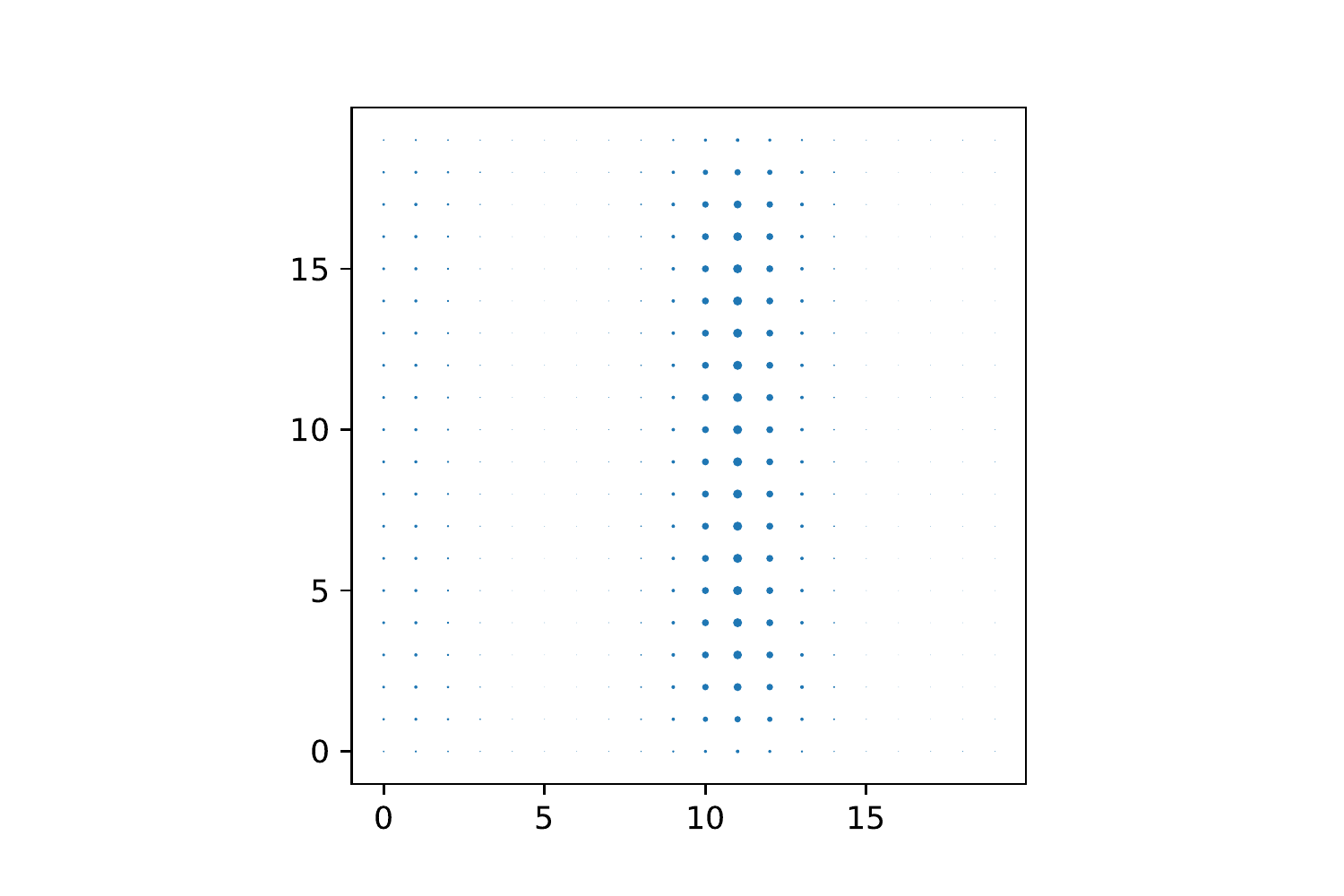}
\end{minipage}
\begin{minipage}{0.329\hsize}
\centering
    \includegraphics[width=\hsize]{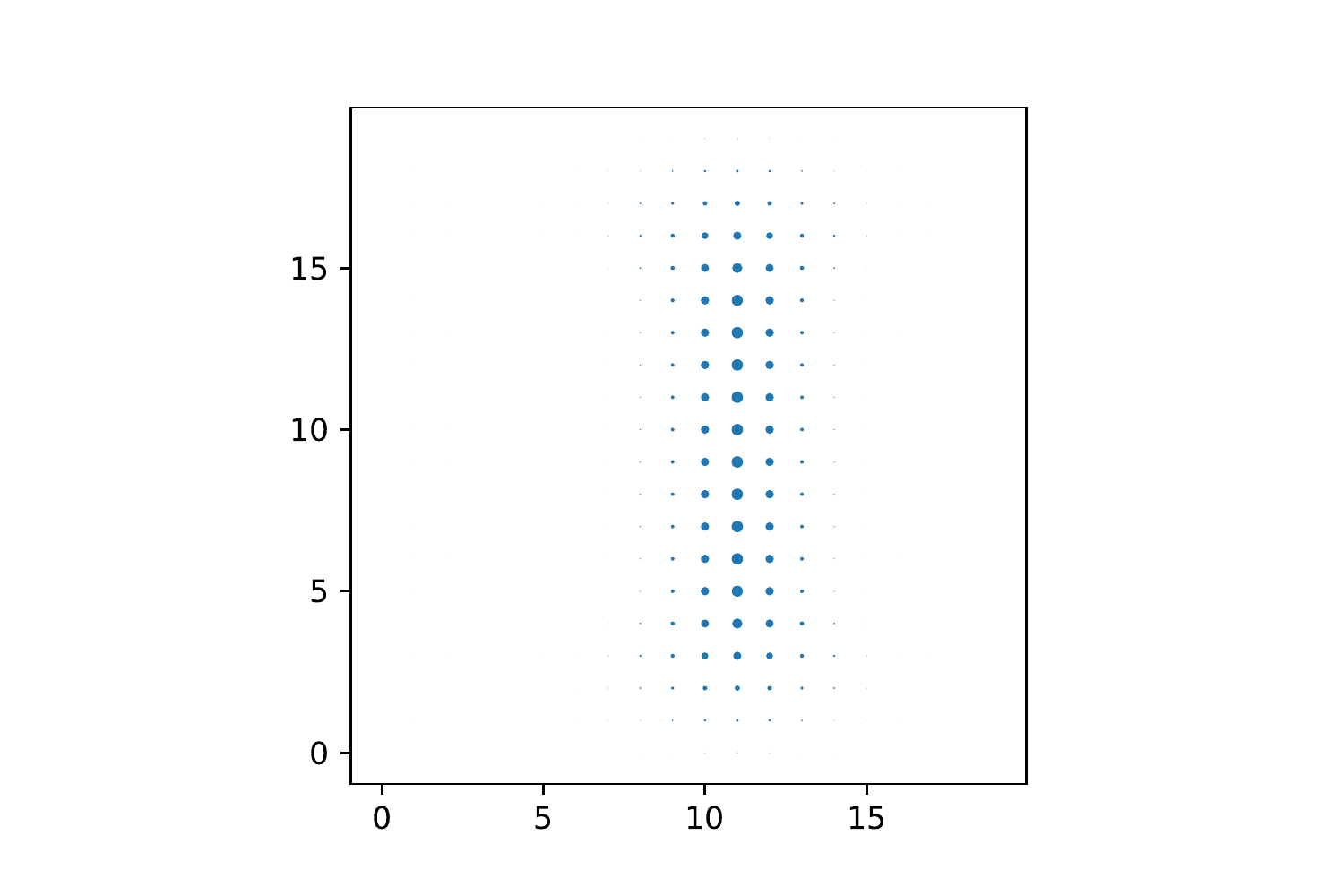}
\end{minipage}
\begin{minipage}{0.329\hsize}
\centering
    \includegraphics[width=\hsize]{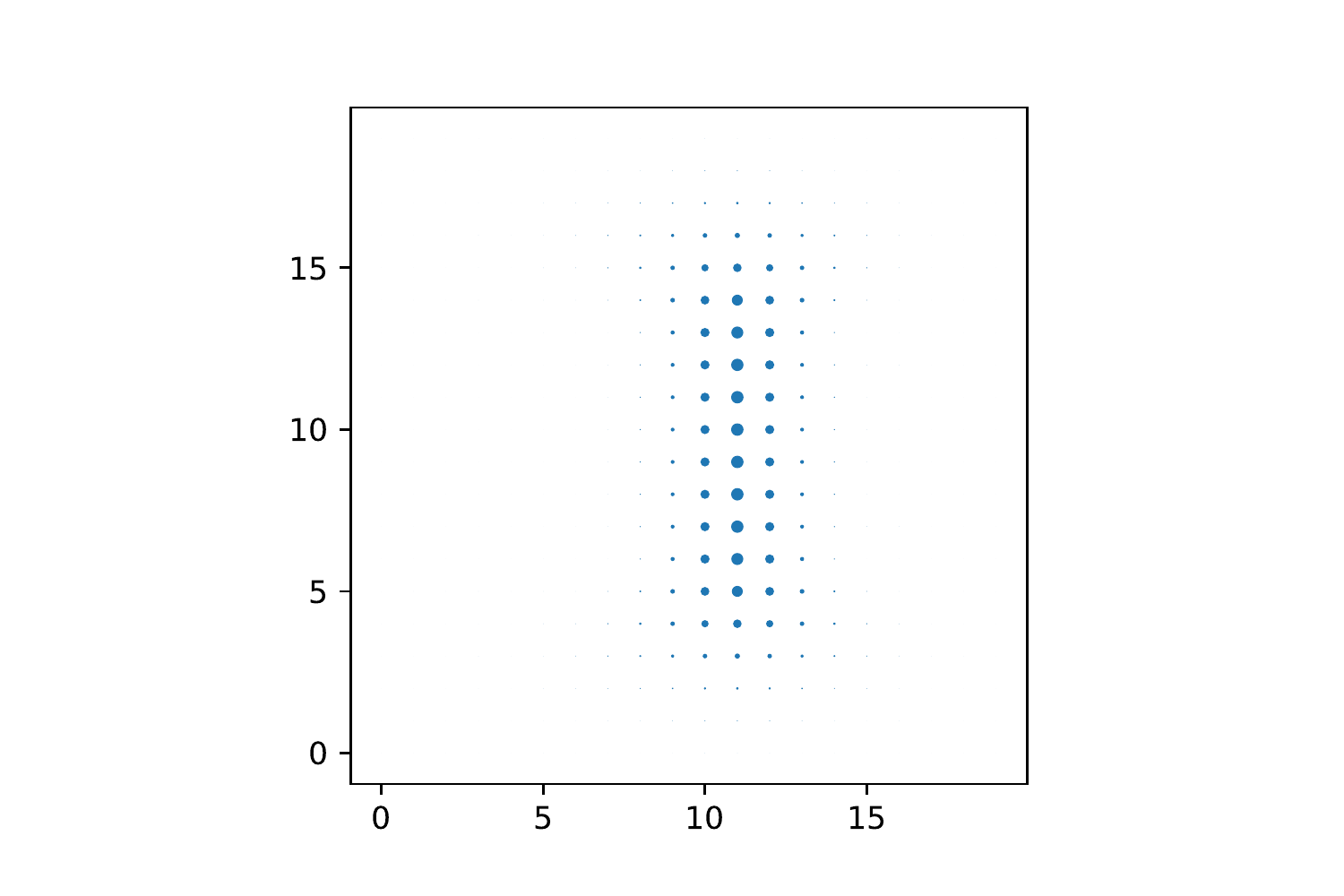}
\end{minipage}
\begin{minipage}{0.329\hsize}
\centering
    \includegraphics[width=\hsize]{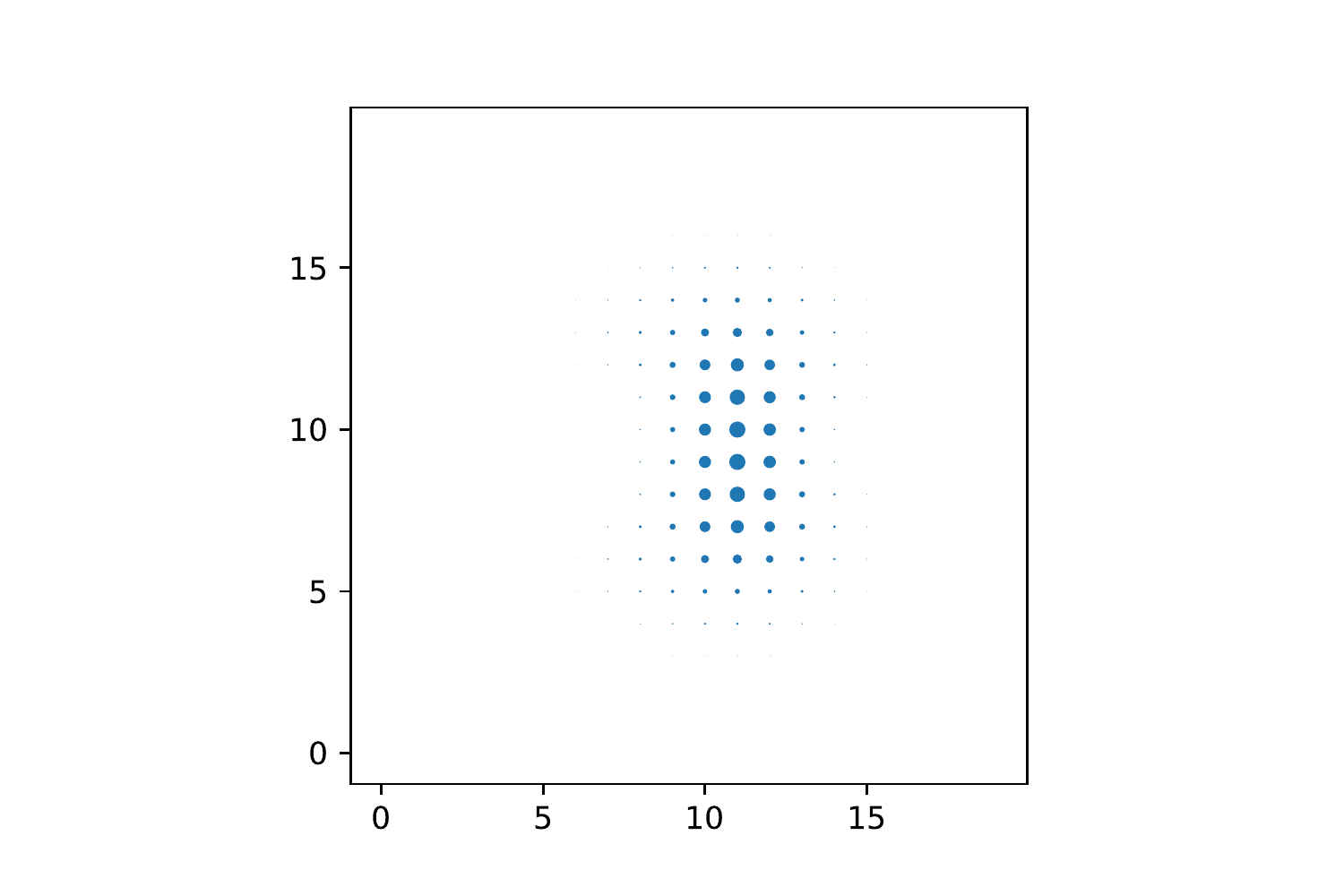}
\end{minipage}
\begin{minipage}{0.329\hsize}
\centering
    \includegraphics[width=\hsize]{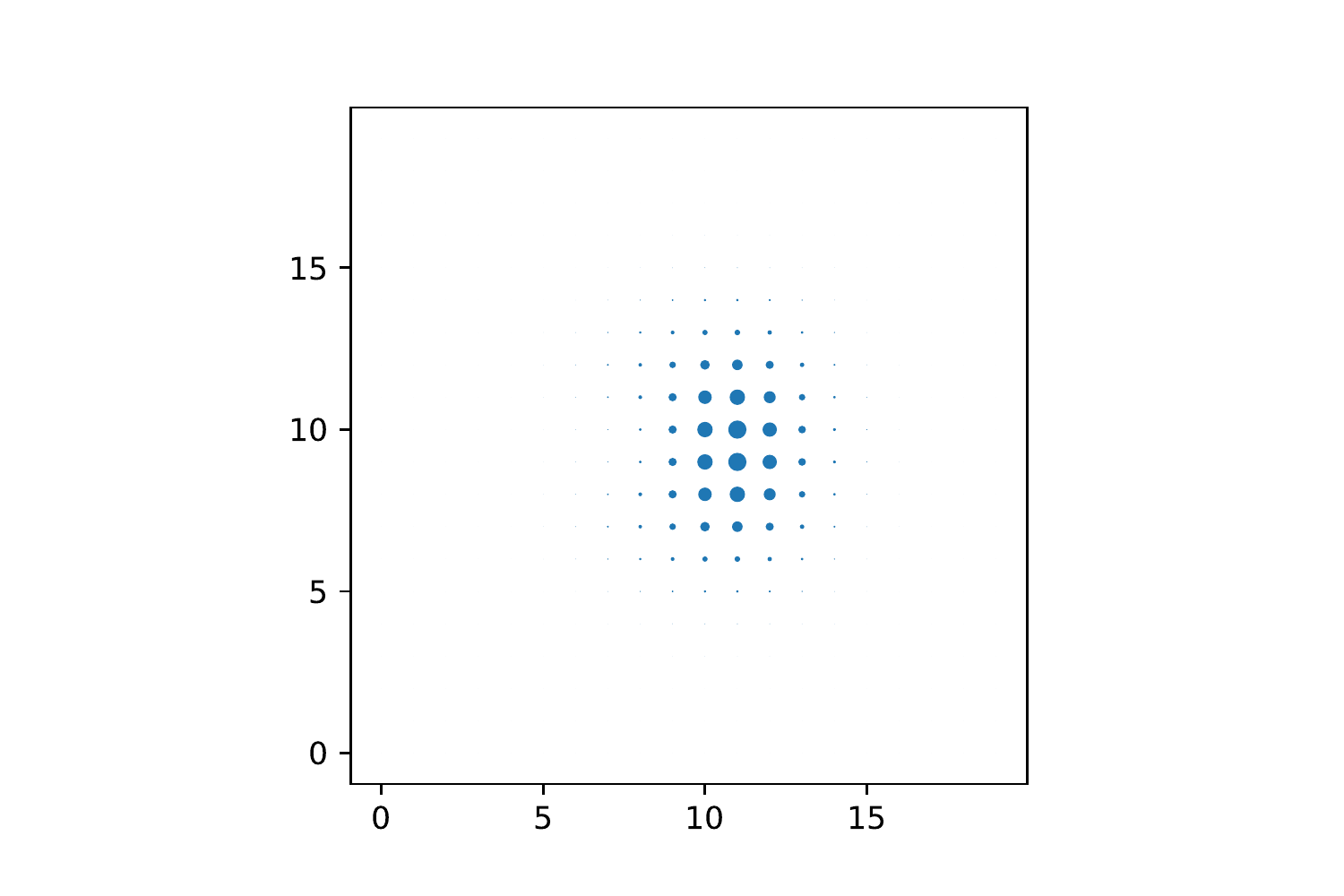}
\end{minipage}
\begin{minipage}{0.329\hsize}
\centering
    \includegraphics[width=\hsize]{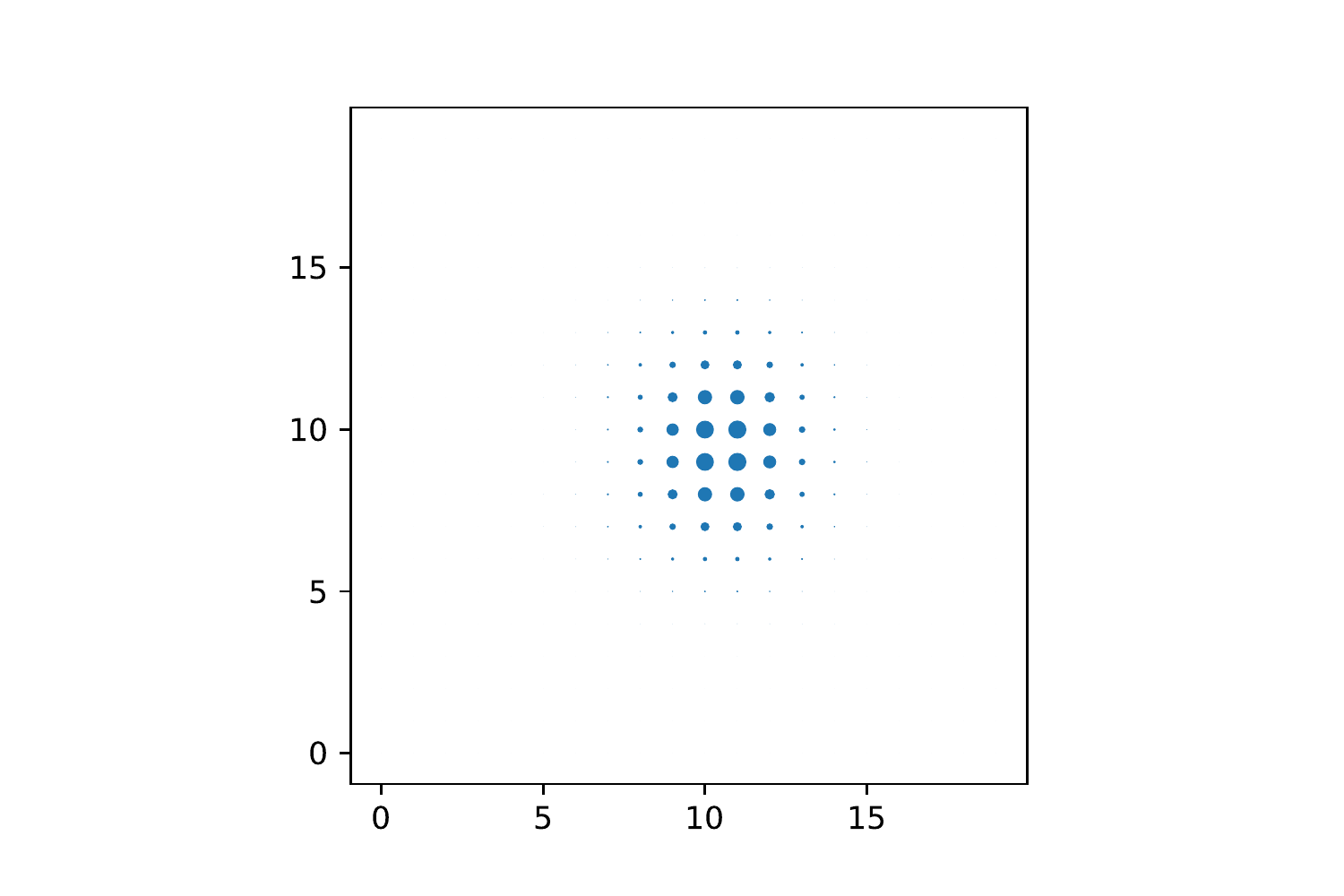}
\end{minipage}
\begin{minipage}{0.329\hsize}
\centering
    \includegraphics[width=\hsize]{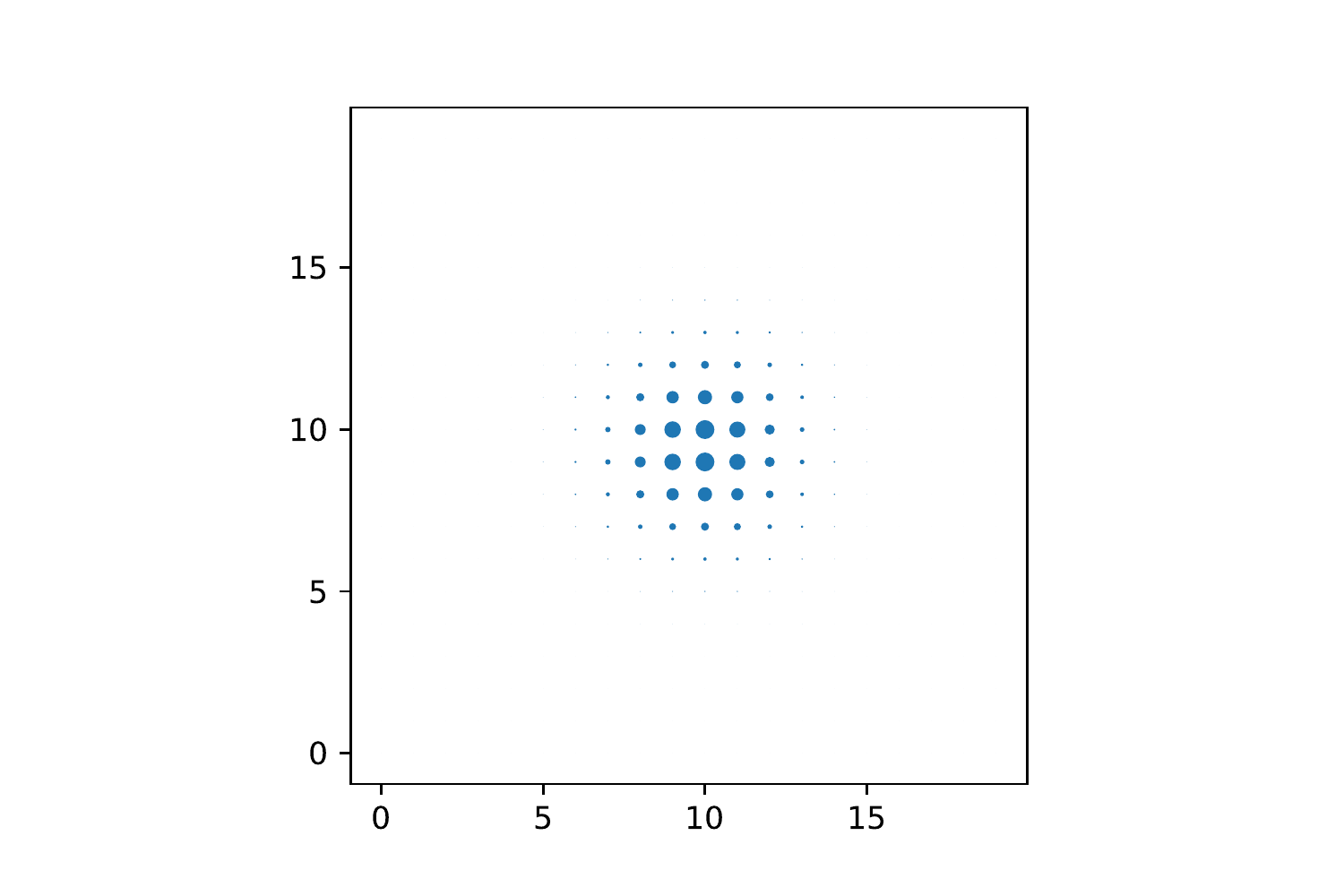}
    $t=20$
\end{minipage}
\caption{Density distribution of the ground states of the Hamiltonian $H(t)$ with $\phi=1/11$ on a 20 by 20 square lattice ($\Gamma(t)=e^{-t}, t\in [0,20]$). Size of a disk represents the density.}
    \label{fig:dos}
\end{figure}

\begin{figure}[H]
\begin{minipage}{0.245\hsize}
        \centering
    \includegraphics[width=\hsize]{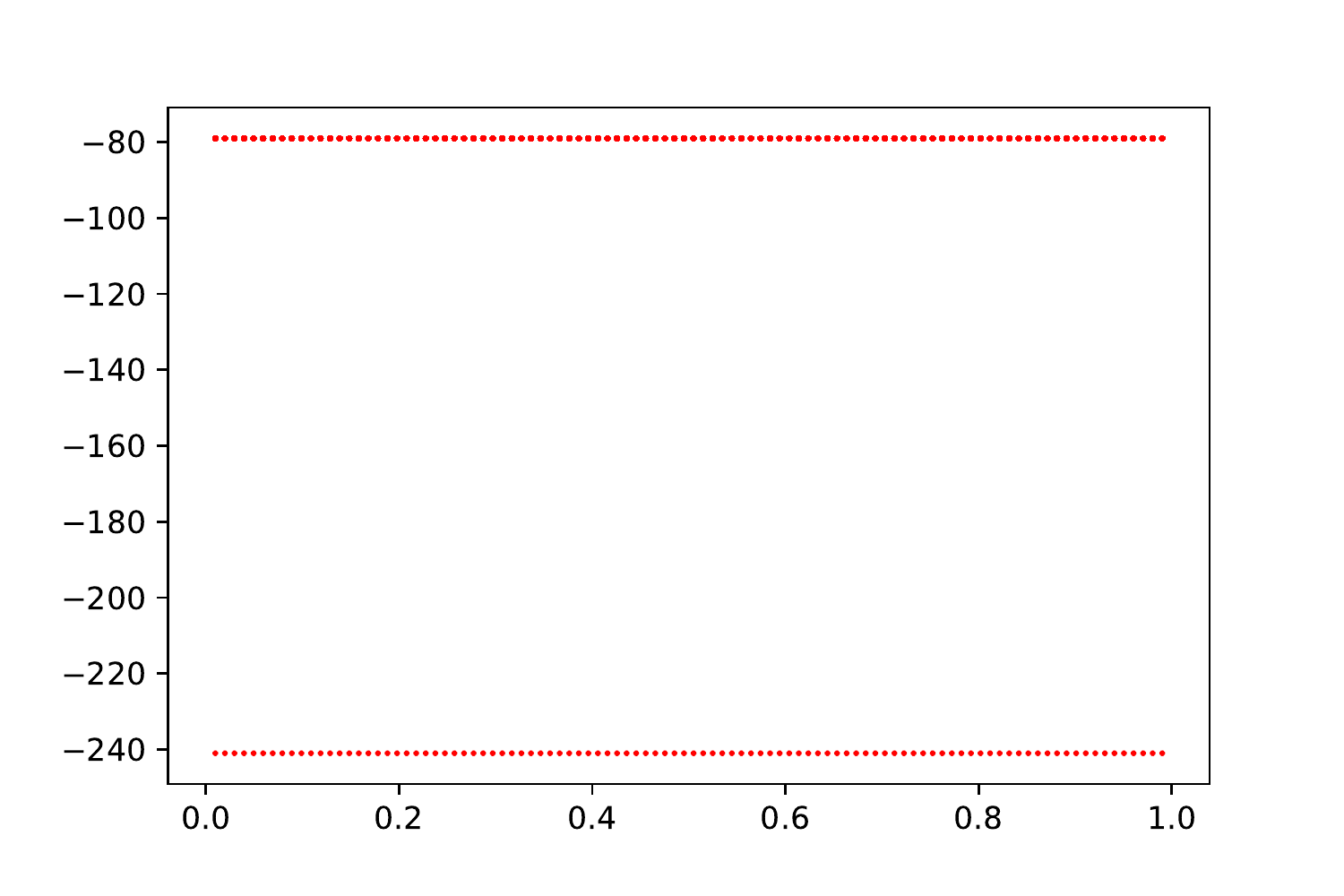}
\end{minipage}
\begin{minipage}{0.245\hsize}
        \centering
    \includegraphics[width=\hsize]{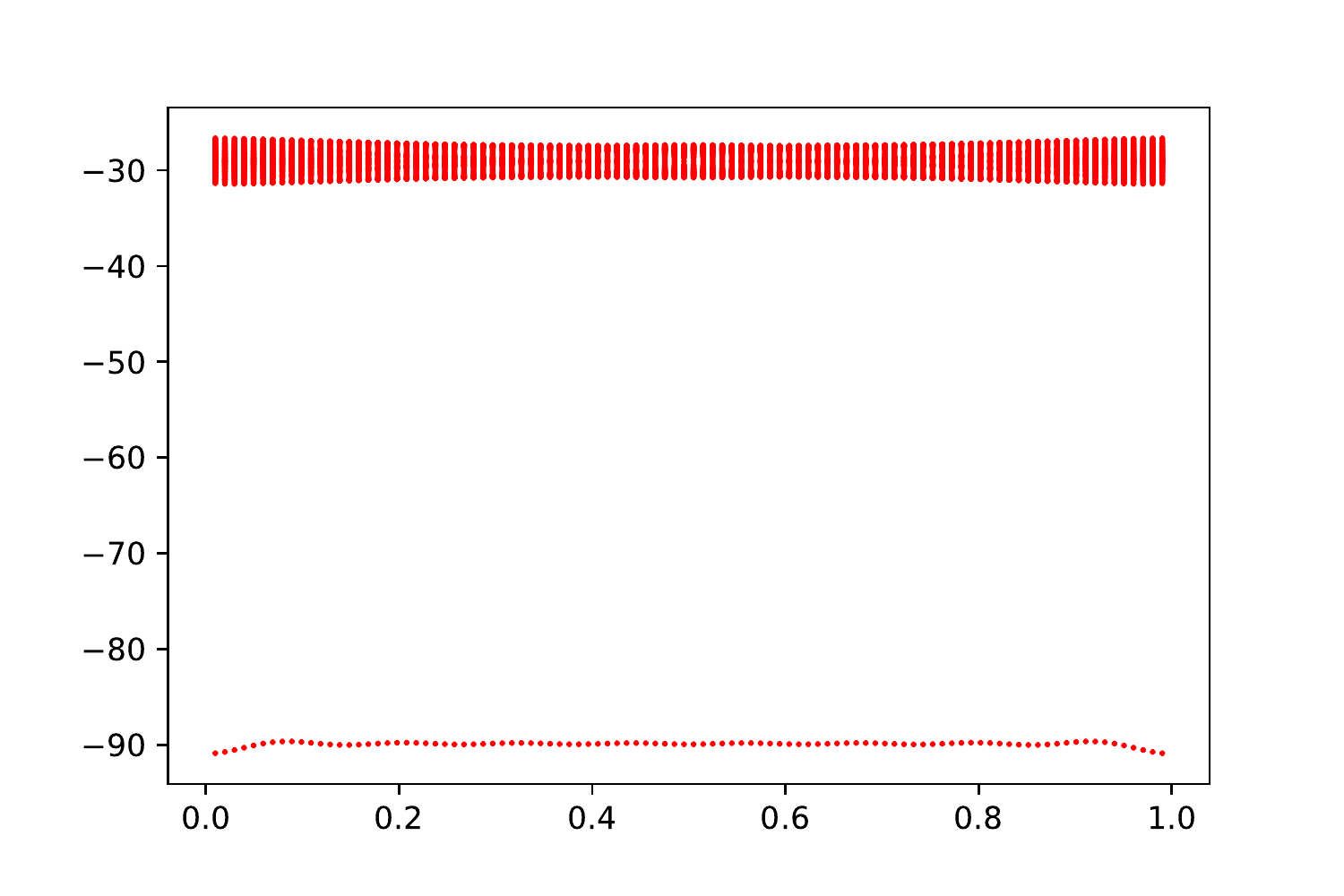}
\end{minipage}
\begin{minipage}{0.245\hsize}
        \centering
    \includegraphics[width=\hsize]{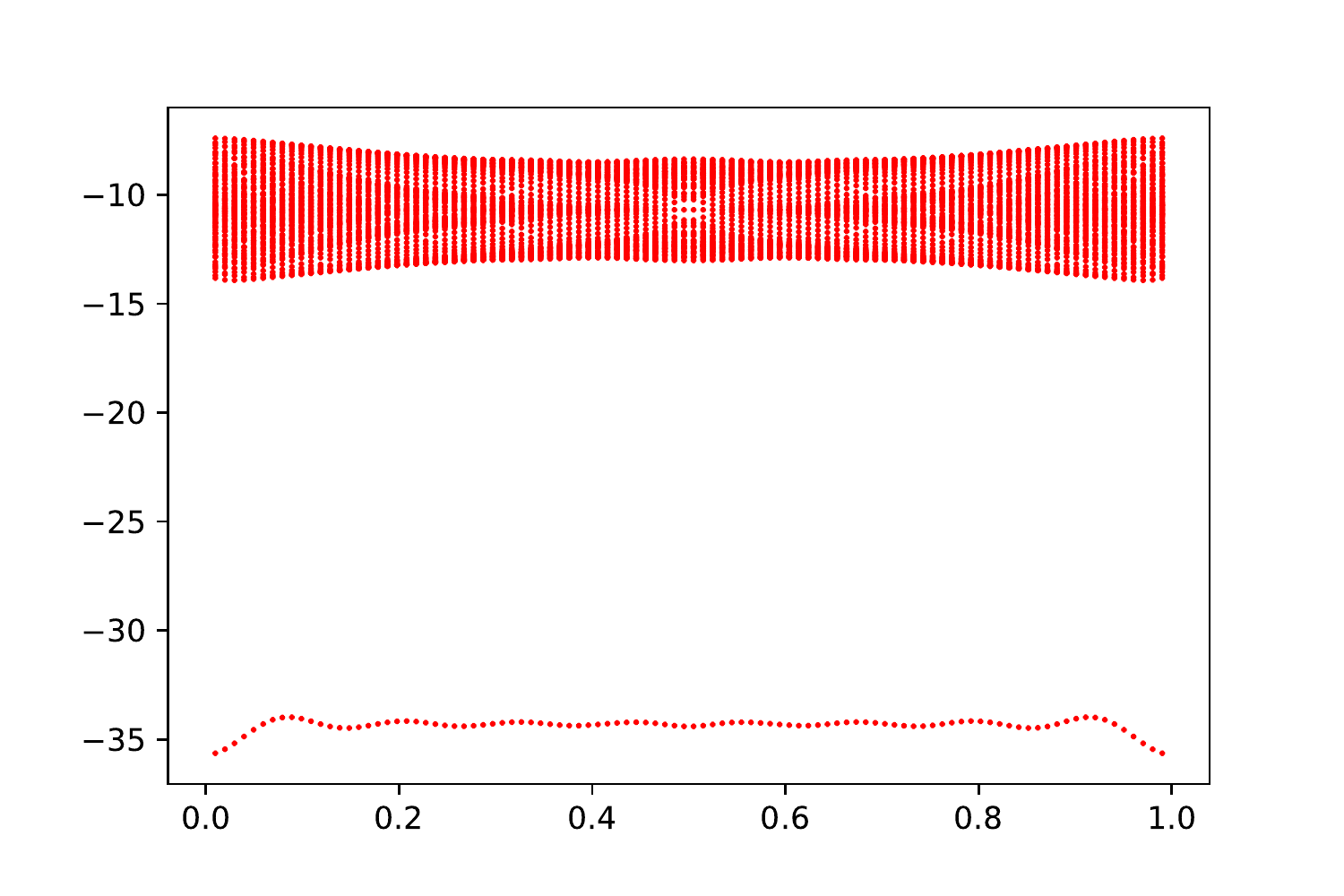}
\end{minipage}
\begin{minipage}{0.245\hsize}
        \centering
    \includegraphics[width=\hsize]{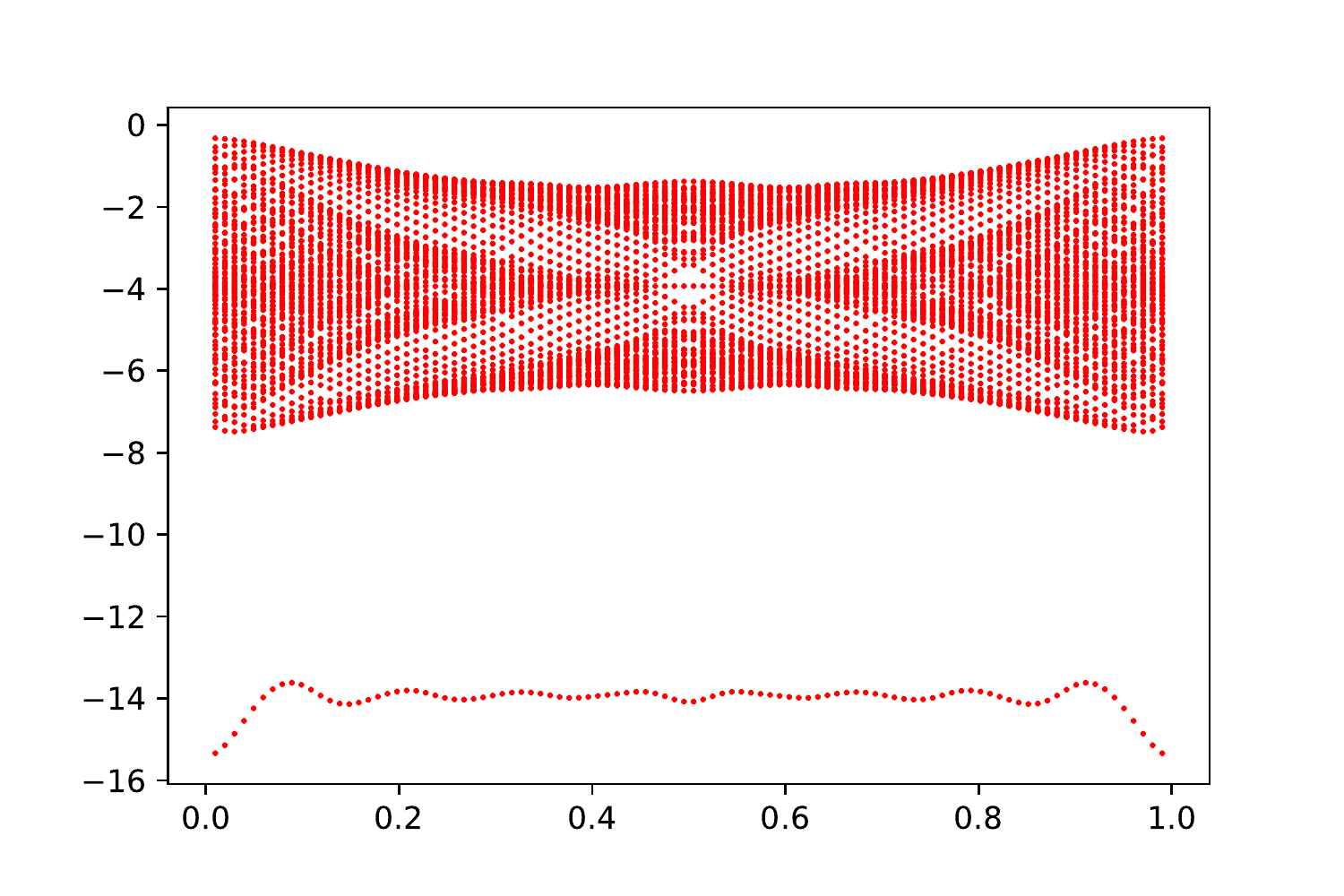}
\end{minipage}
\begin{minipage}{0.245\hsize}
        \centering
    \includegraphics[width=\hsize]{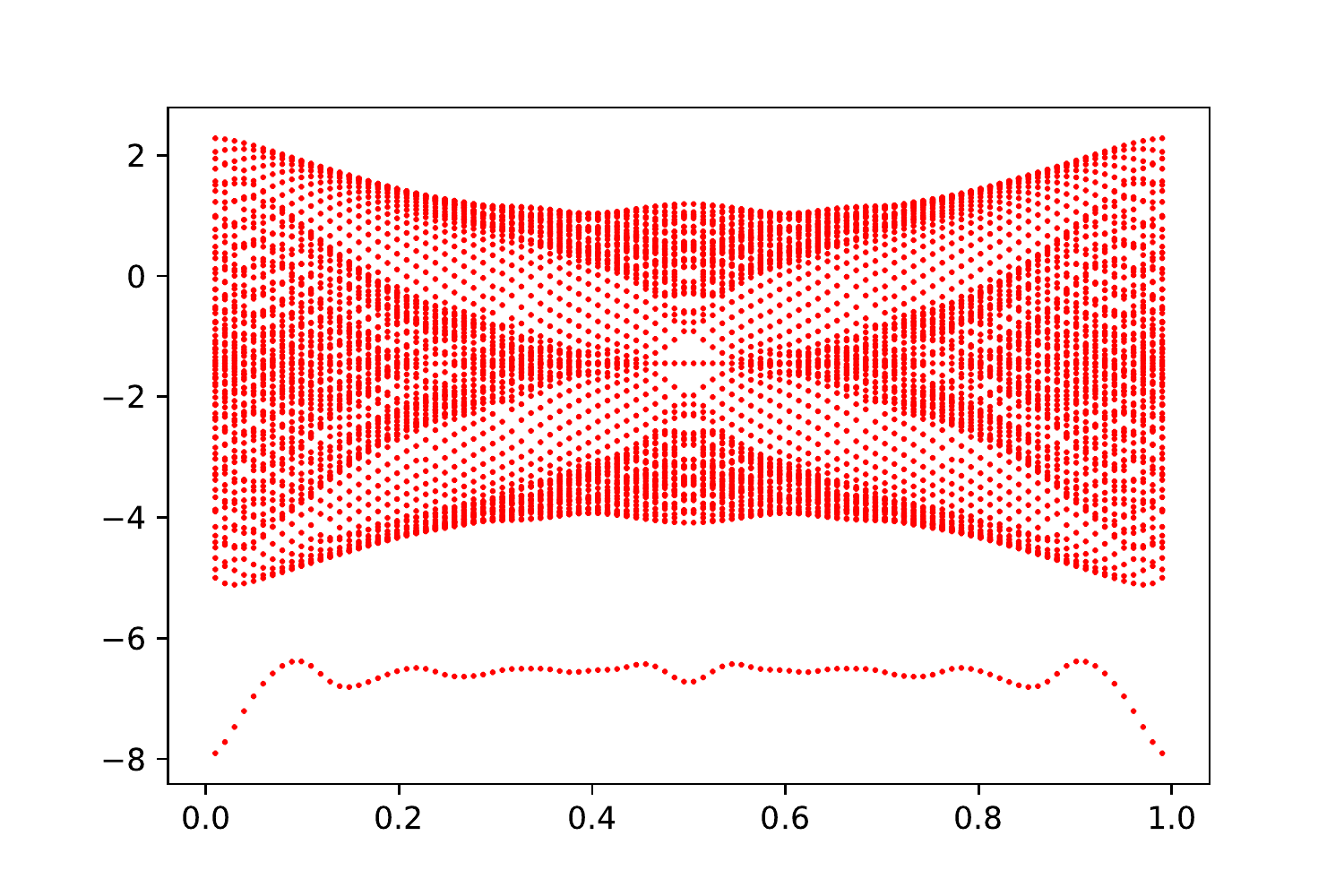}
\end{minipage}
    \begin{minipage}{0.245\hsize}
        \centering
    \includegraphics[width=\hsize]{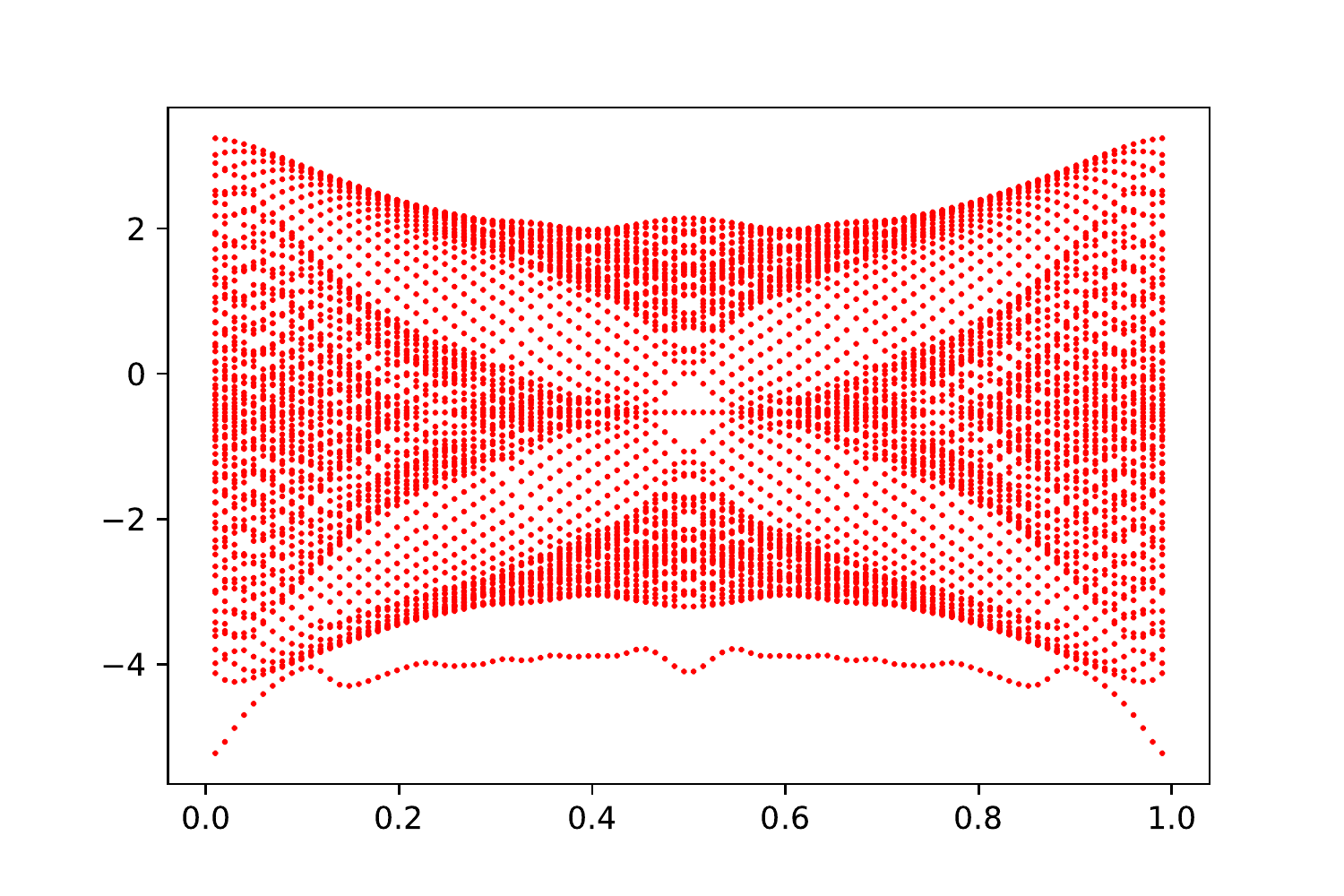}
\end{minipage}
\begin{minipage}{0.245\hsize}
        \centering
    \includegraphics[width=\hsize]{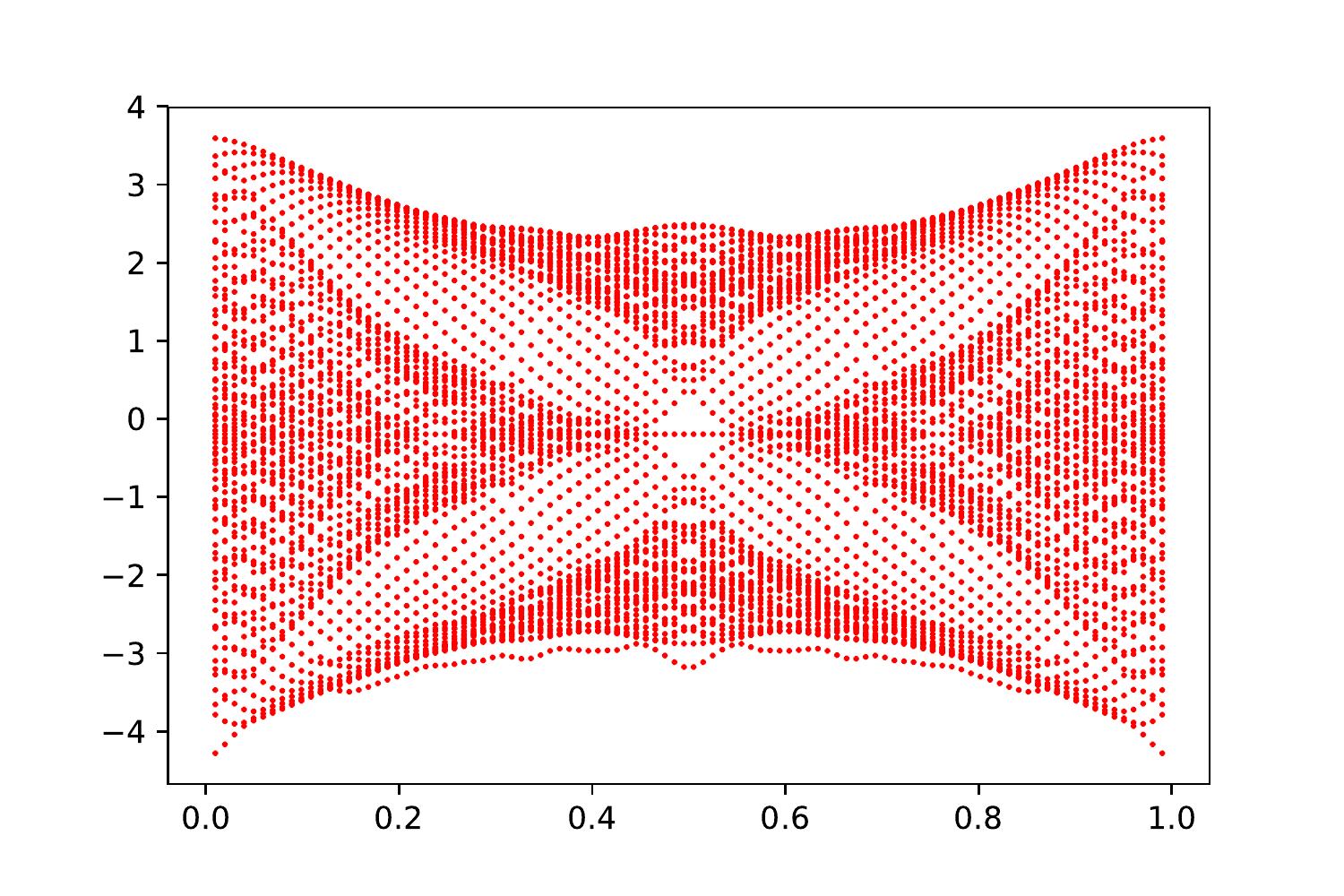}
\end{minipage}
\begin{minipage}{0.245\hsize}
        \centering
    \includegraphics[width=\hsize]{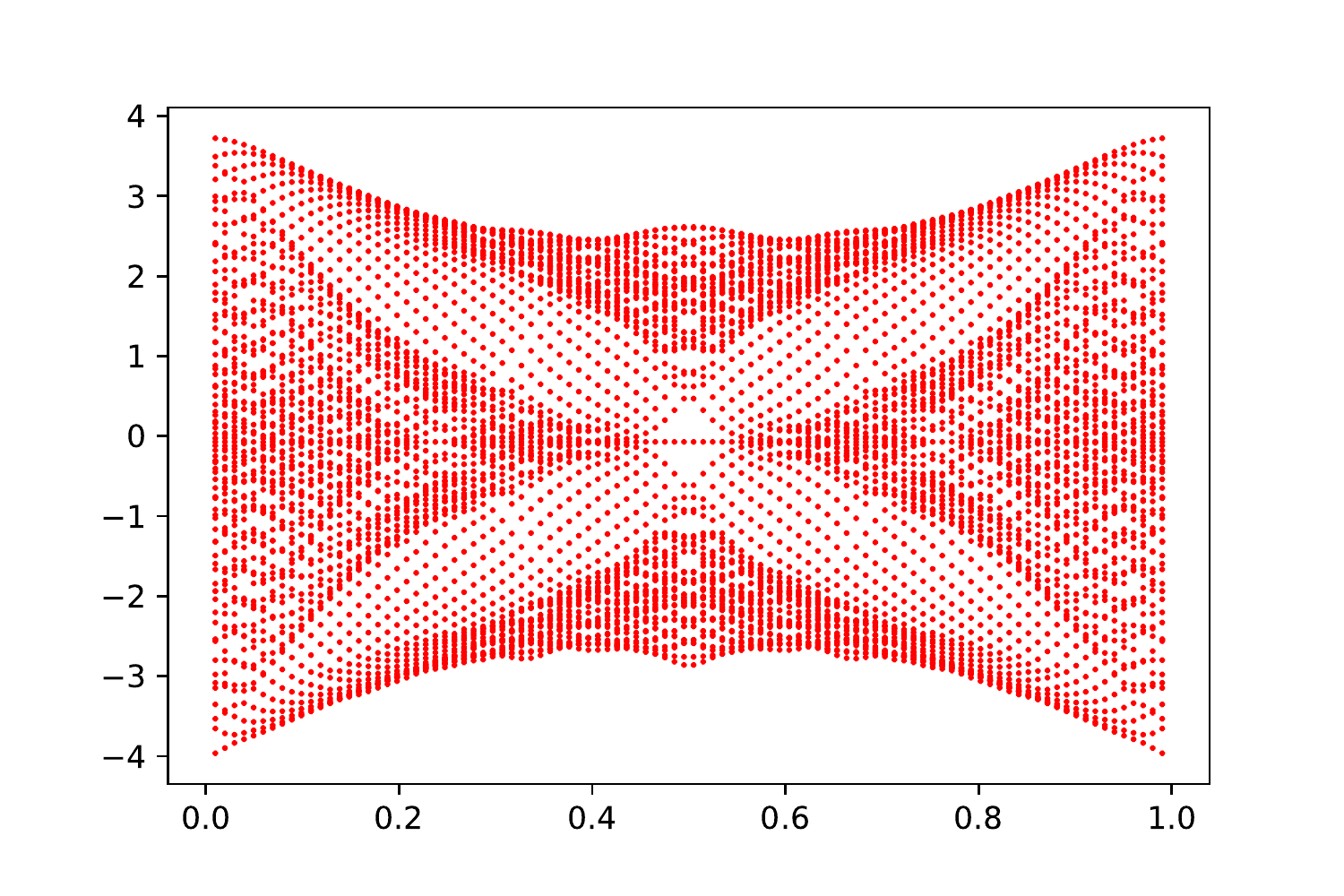}
\end{minipage}
\caption{$\Gamma(t)=e^{-t}, t\in[0,7]$}
    \label{fig:my_label}
\end{figure}

\if{
\begin{figure}[H]
    \centering
    \includegraphics[width=7cm]{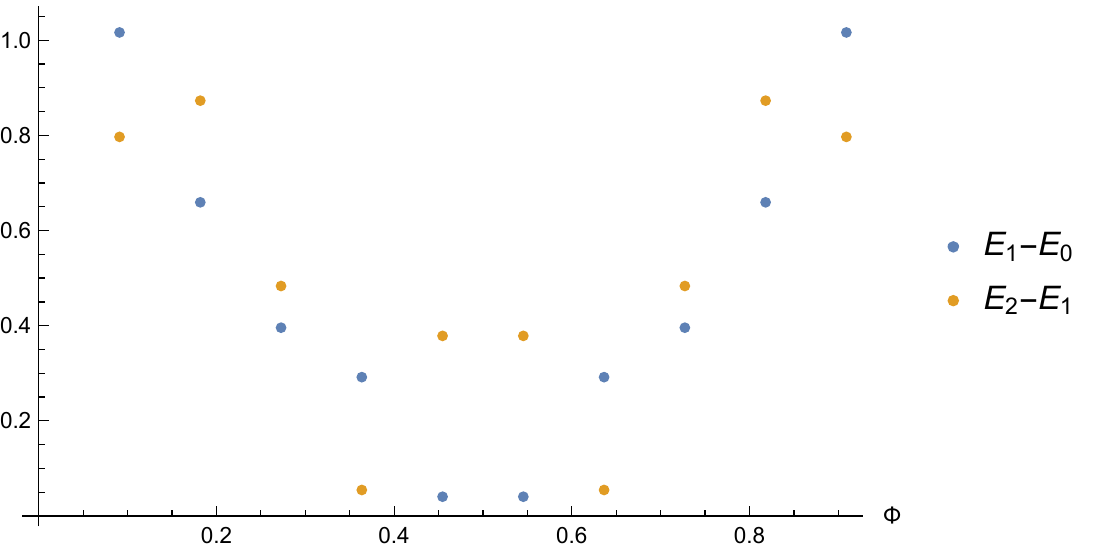}
    \caption{Energy spacing between subsequent levels: between the 1st excited levels and the ground levels $E_1-E_0$, and between the 2nd and the 1st excited levels $E_2-E_1$. }
    \label{fig:es}
\end{figure}

 Fig.\ref{fig:probs} and \ref{fig:probs2} show time dependence of finding some eigenstates of the same Hamiltonian $H(t)$. The behavior can be explained as follows. Computation starts with the ground state of $H_1$, whose energy is sufficiently separated from other energy levels, and as the $H_0$ is introduced other energy levels may become closer to the ground state. Intuitively, the closer they get, the higher the probability that an excited state is obtained. However introducing $H_1$ generally breaks the symmetry of the vacuum, hence a ground state may not be obtained with high probability. The figures of the probability of finding 3rd excited states imply that ground states, 1st excited states, 2nd excited stats are dominant solutions of this method with those schedules $\Gamma(t)$. One important feature of AQC is that the longer the computational time is the higher the probability of finding the ground state becomes. This is in fact exhibited in both Fig.\ref{fig:probs} and \ref{fig:probs2}. As expected, if $\Gamma(t)=e^{-t}$ is used, the probability converges much faster than when $\Gamma(t)=1/\text{log}(t+e)$ is used. In addition, Fig. \ref{fig:es} shows energy spacing between consecutive energy levels $\{E_{i+1}(\phi)-E_{i}(\phi)\}_{i=0,1}$. The energy spectra are symmetric $E_i(\phi+1/2)=E_i(-\phi+1/2)$ and $E_1(\phi)-E_0(\phi)$ gets close to 0 with $\phi\to 1/2$. So it is not surprising that the second excited state is more easily obtained as $\phi$ comes close to $1/2$. Moreover, by the same reason, the probability of obtaining the second excited at $\phi=4/11$ is as high as that of the first excited state. A problem of this method is that the performance heavily depends on the choice of schedules $\Gamma(t)$ as well as of the kinetic terms $H_1$. Moreover, in general, a wide flat (local) minimum solution is preferably chosen by AQC, hence solutions are not obtained with equal probability, even if they belong to the same energy level. This is called unfair sampling. In fact, the ground states of the $\phi=1/2$ case are degenerated, but one of them cannot be obtained equally. The wide-flatness, however, is sometimes useful. In some sense, model-independent features can be found in the wide-flat regime. Machine learning is a good application of AQC. For example, a narrow global minimum should be interpreted as an effect of over-training in machine learning.  
 \begin{figure}[H]
\begin{minipage}{0.50\hsize}
\centering
    \includegraphics[width=\hsize]{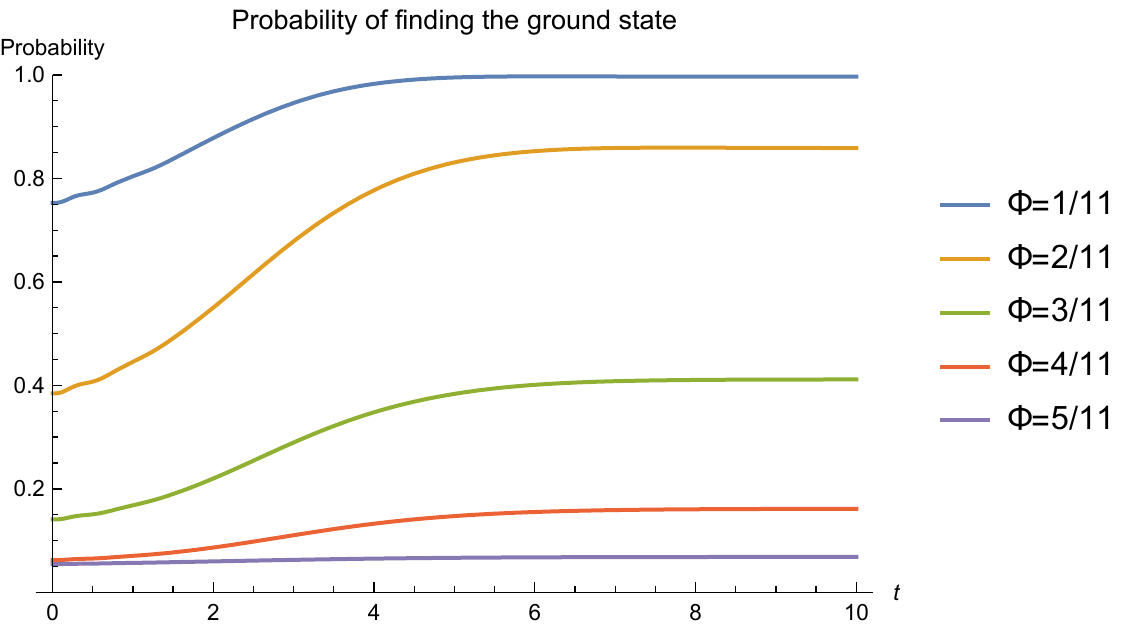}
\end{minipage}
\begin{minipage}{0.50\hsize}
\centering
    \includegraphics[width=\hsize]{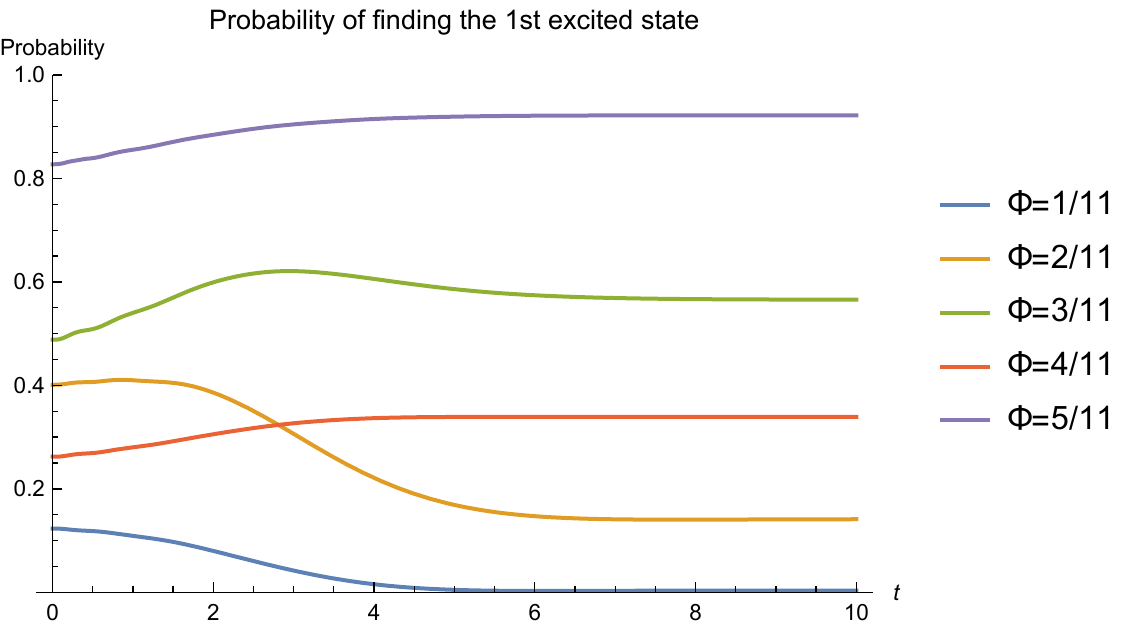}
\end{minipage}
\begin{minipage}{0.50\hsize}
\centering
    \includegraphics[width=\hsize]{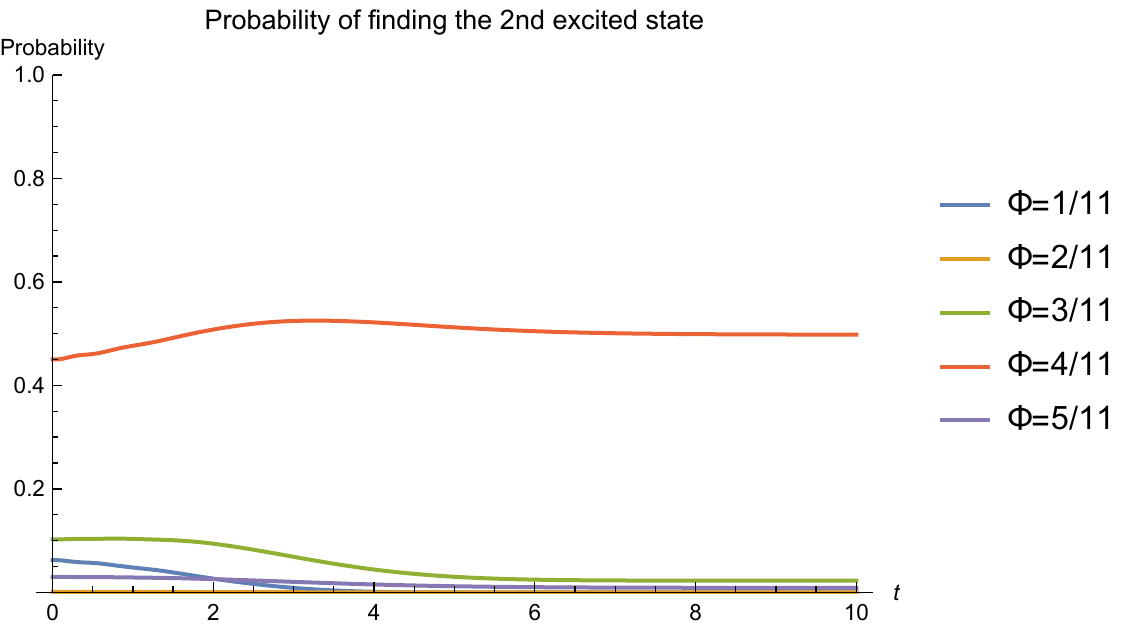}
\end{minipage}
\begin{minipage}{0.50\hsize}
\centering
    \includegraphics[width=\hsize]{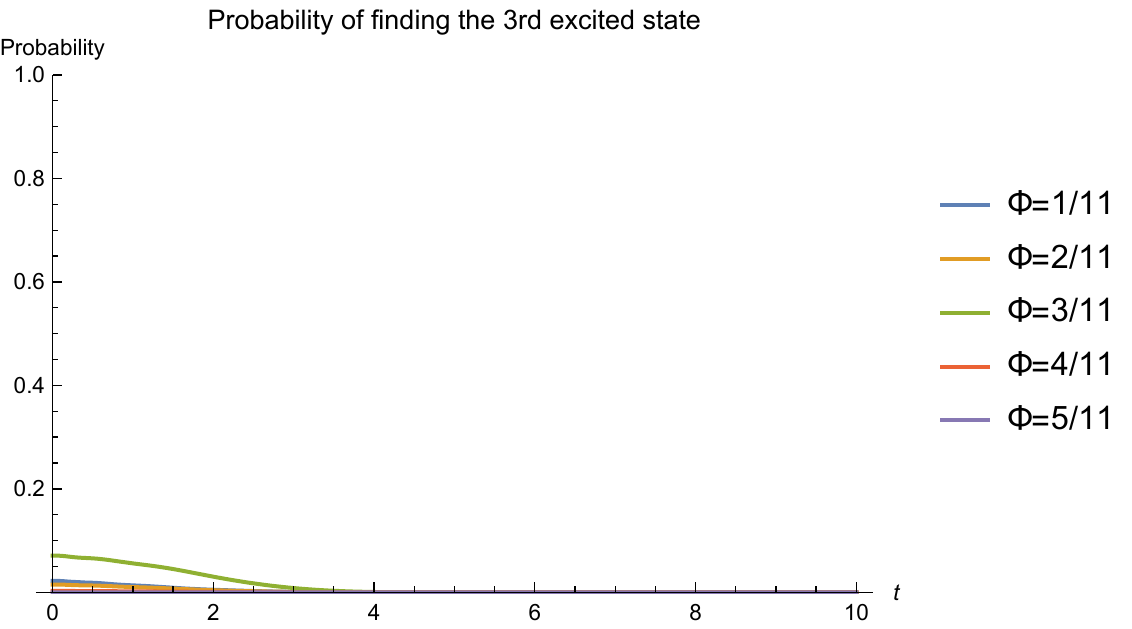}
\end{minipage}
\caption{Numerical results of probability of obtaining states on a square lattice under uniform magnetic flux $\phi$. The system obeys the Hamiltonian $H(t)=(1-\Gamma(t))H_0+\Gamma(t)H_1$ whose time dependence is described by  $\Gamma(t)=e^{-t}$.}
    \label{fig:probs}
\end{figure}

\begin{figure}[H]
\begin{minipage}{0.5\hsize}
\centering
    \includegraphics[width=\hsize]{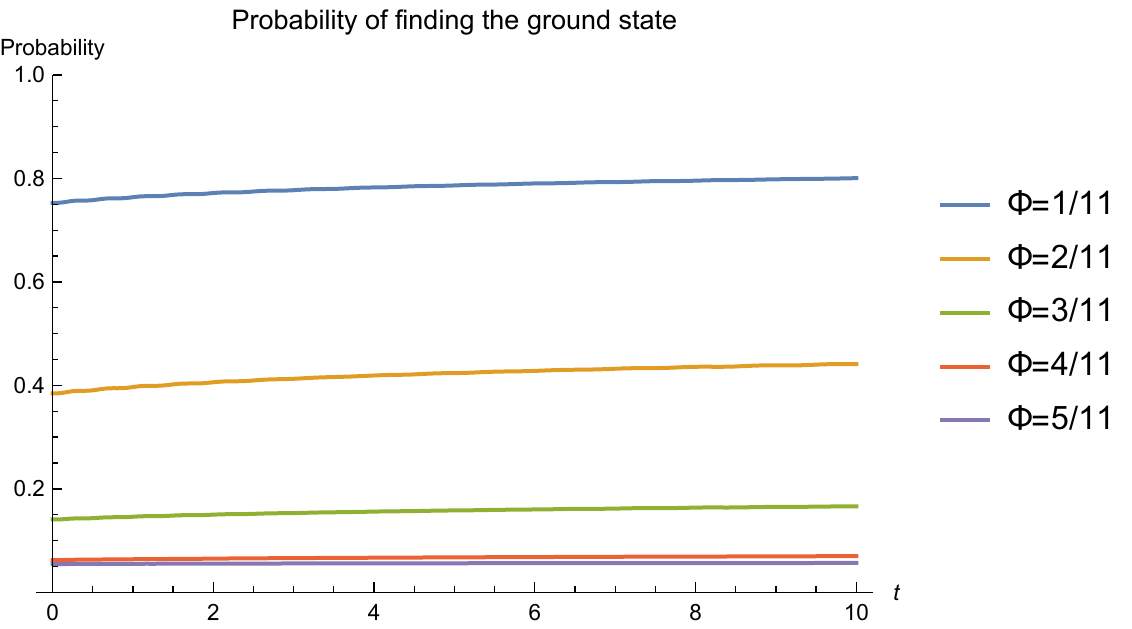}
\end{minipage}
\begin{minipage}{0.5\hsize}
\centering
    \includegraphics[width=\hsize]{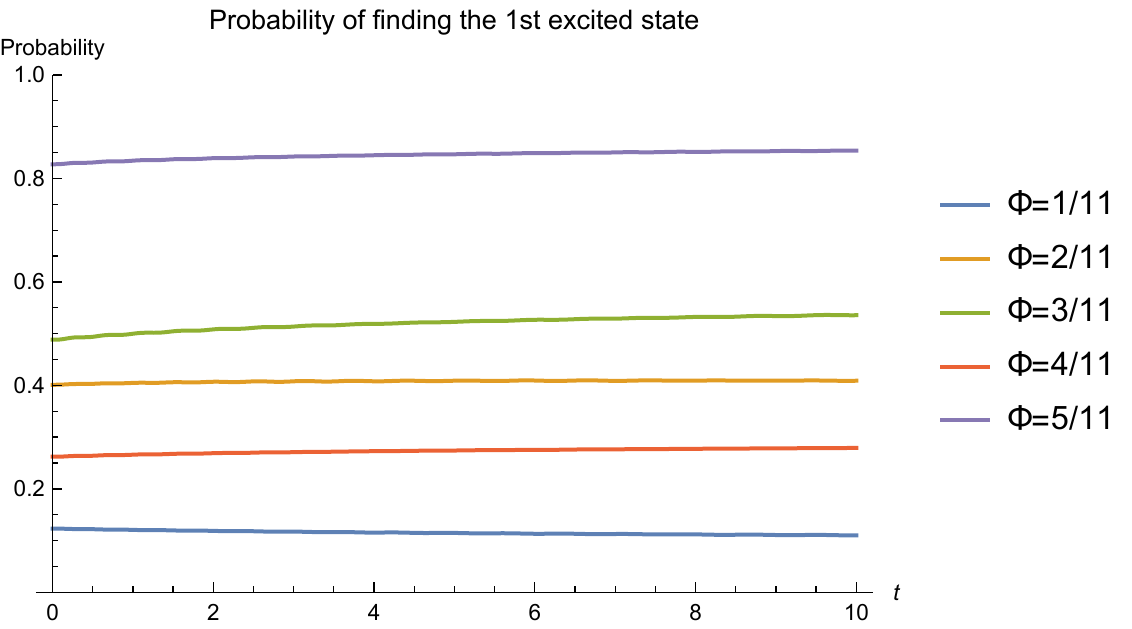}
\end{minipage}
\begin{minipage}{0.5\hsize}
\centering
    \includegraphics[width=\hsize]{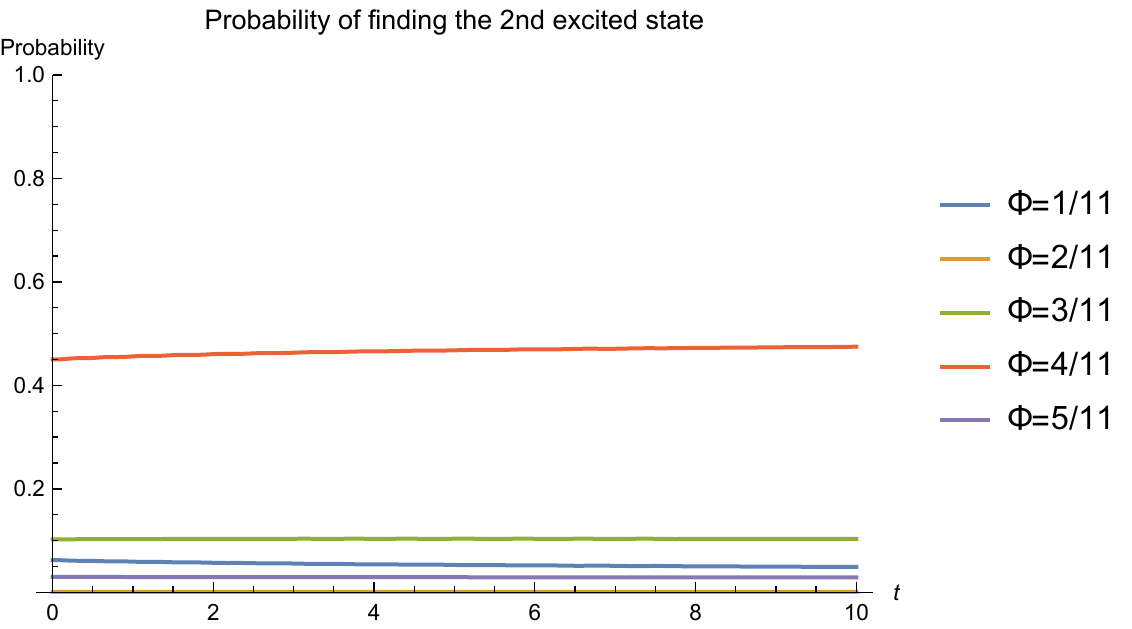}
\end{minipage}
\begin{minipage}{0.5\hsize}
\centering
    \includegraphics[width=\hsize]{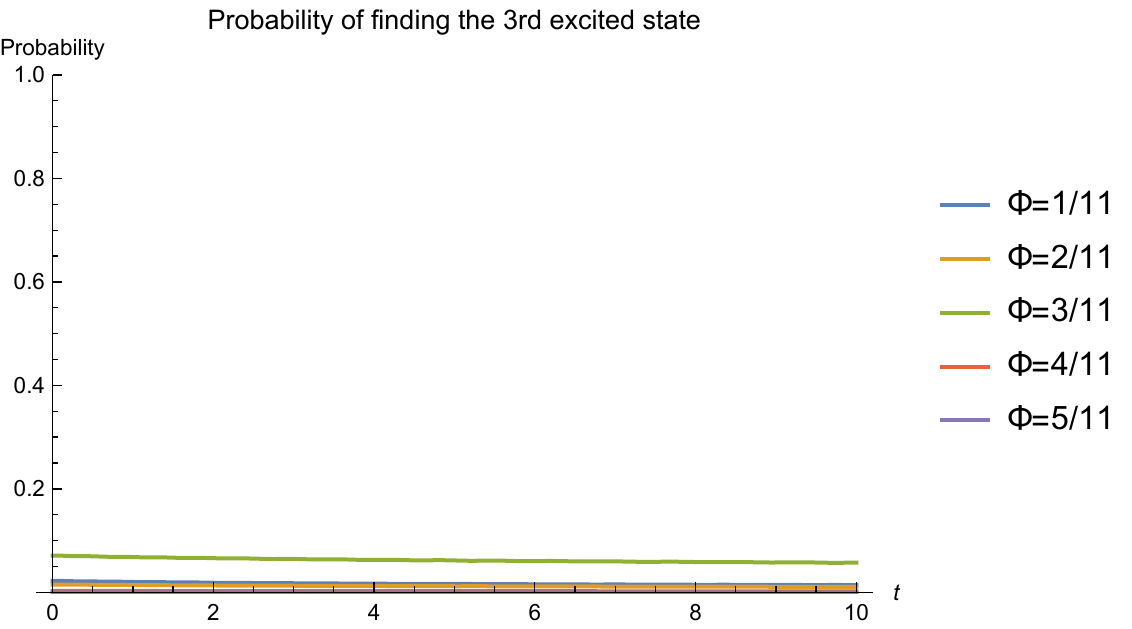}
\end{minipage}
\caption{Numerical results of probability of obtaining states on a square lattice under uniform magnetic flux $\phi$. The system obeys the Hamiltonian $H(t)=(1-\Gamma(t))H_0+\Gamma(t)H_1$ whose time dependence is described by  $\Gamma(t)=1/\log(t+e)$.}
    \label{fig:probs2}
\end{figure}
}\fi

In general, performance of AQC depends on schedules. So in what follows we try several other choices and explore more on the schedule dependence of the probability. Let $\{\mathcal{E}_i(t)\}$ be a one-parameter family of energy eigenvalues $(i=0,\cdots, \dim H(t))$ of $H(t)$. We define the energy gap between the ground state and the first excited state by 
 \begin{equation}
     \Delta=\inf_{t}(\mathcal{E}_1(t)-\mathcal{E}_0(t)). 
 \end{equation}
 According to the adiabatic theorem, the ground state of the target Hamiltonian $H_0$ should be found with probability arbitrarily close to 1, after sufficiently long time $T\gg O(1/\Delta^2)$. We address two cases: a finite schedule ($\Gamma(\tau)=0$ at some $\tau<\infty$) and an infinite schedule ($\Gamma(t)>0$ for any $t$ and $\lim_{t\to\infty}\Gamma(t)=0$). For the infinite schedule, we use $\Gamma(t)=\exp(-at)$ and control the speed by tuning $a>0$. Fig. \ref{fig:probs4} exhibits the numerical results of finding the ground states. As the finite case \eqref{eq:f}, the probability successfully increases as the computation speed decreases. In both of two cases in Fig. \ref{fig:probs4}, computation stops with the same value of $\Gamma(t)>0$. 

\begin{figure}[H]
\begin{minipage}{0.5\hsize}
\centering
    \includegraphics[width=\hsize]{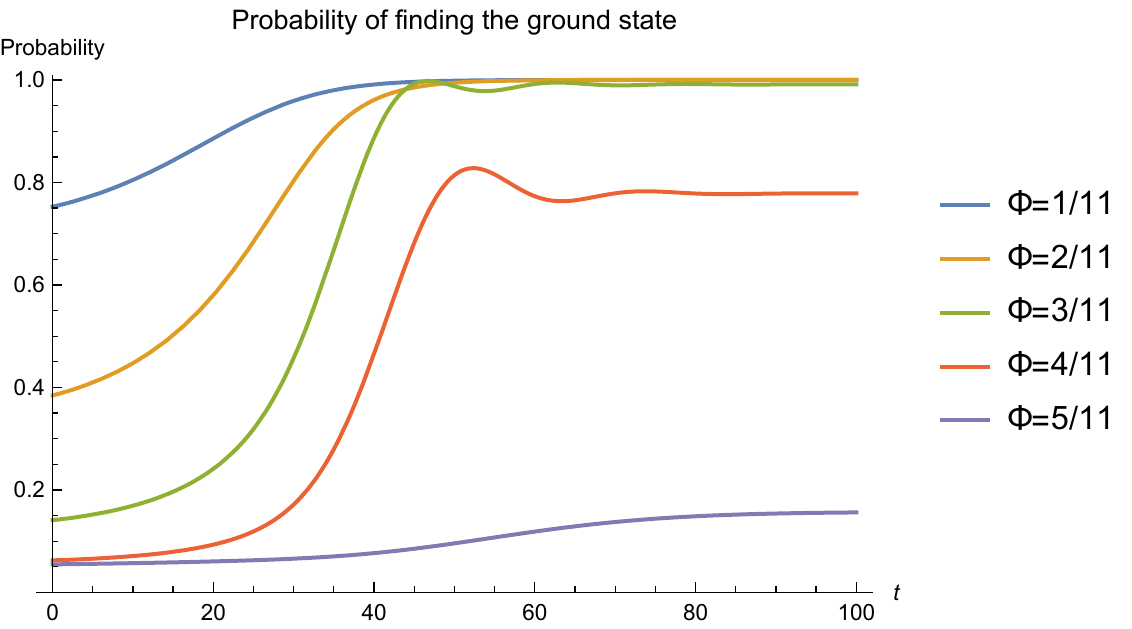}
\end{minipage}
\begin{minipage}{0.5\hsize}
\centering
    \includegraphics[width=\hsize]{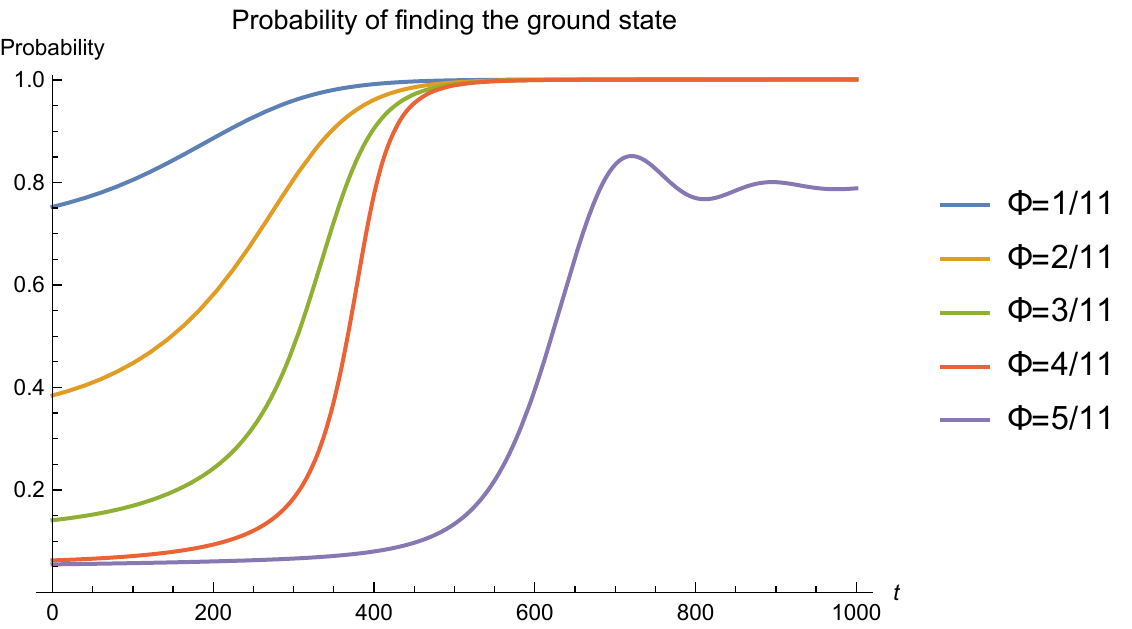}
\end{minipage}
\caption{Numerical results of probability of obtaining the ground states on a 3 by 3 square lattice under uniform magnetic flux $\phi$. The system obey the Hamiltonian $H(t)=(1-\Gamma(t))H_0+\Gamma(t)H_1$ whose time dependence is described by $\Gamma(t)=\exp(-at)$. [Left] $a=0.1$ [Right] $a=0.01$}
    \label{fig:probs4}
\end{figure}

For the finite schedule, we introduce the following function \begin{equation}\label{eq:f}
    \Gamma(t)=1-\frac{1}{2}\left(\frac{\arctan{(t-\tau/2)}}{|\arctan{(-\tau/2)}|}+1\right), 
\end{equation}
where $\tau(<\infty)$ is finite computational time. This monotonic function slowly begins to decrease and gradually banishes at $t=\tau$. A difference from previous functions is that it satisfy $\Gamma''(t)<0$ for $0<t<\tau/2$ and $\Gamma''(t)>0$ for $\tau/2<t< \tau$ (see Fig. \ref{fig:f}). 

\begin{figure}[H]
    \centering
    \includegraphics[width=8cm]{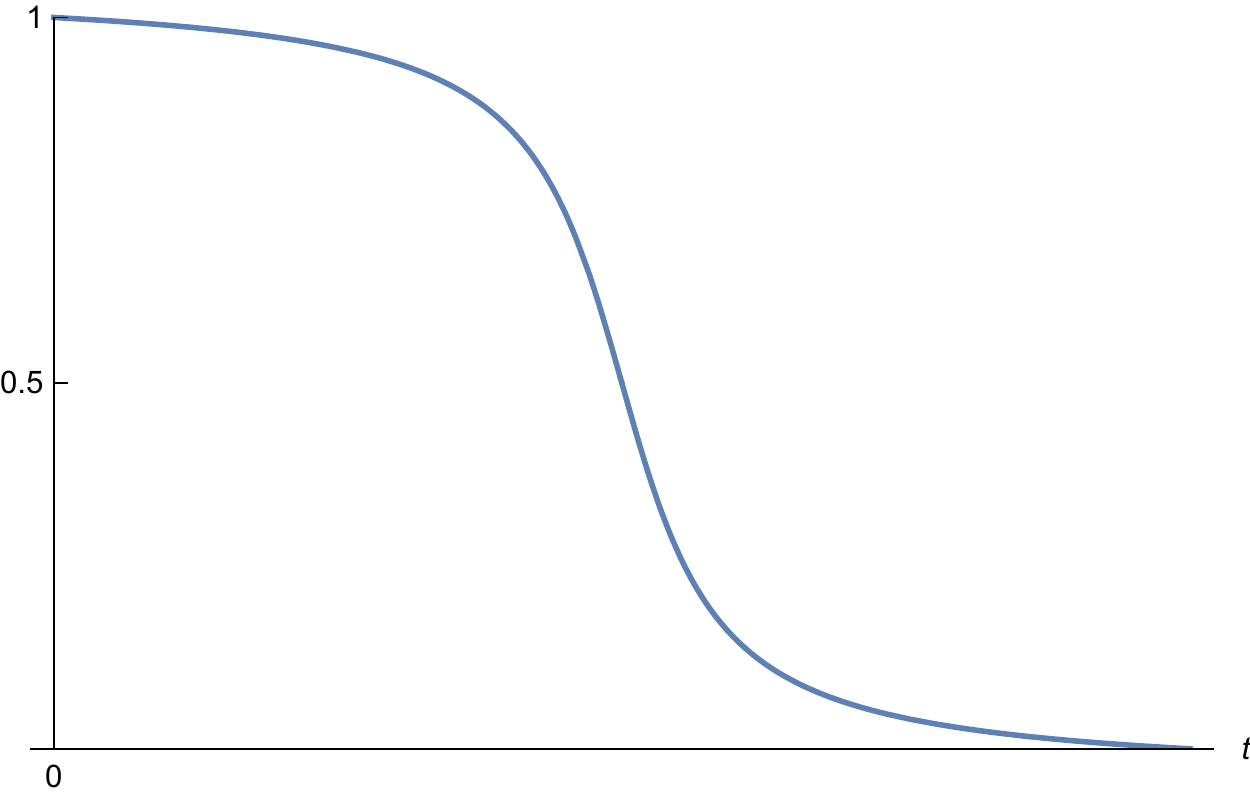}
    \caption{Behavior of \eqref{eq:f}}
    \label{fig:f}
\end{figure}

The probability of finding ground states is shown in Fig. \ref{fig:probs3}. As expected, taking a longer runtime $\tau$ helps one find the ground states accurately for all $\phi$, thereby the other states are unlikely obtained. It would be typical to this schedule $\Gamma(t)$ \eqref{eq:f} that the probability rapidly increases around $t$ when $\Gamma''(t)=0$.

\begin{figure}[H]
\begin{minipage}{0.329\hsize}
\centering
    \includegraphics[width=\hsize]{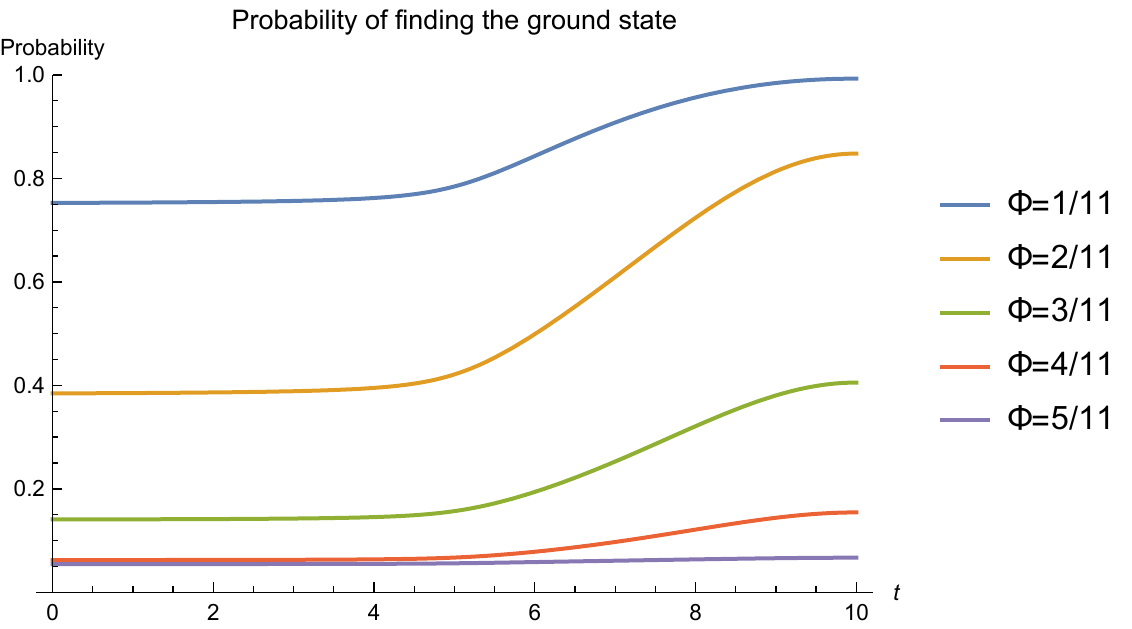}
\end{minipage}
\begin{minipage}{0.329\hsize}
\centering
    \includegraphics[width=\hsize]{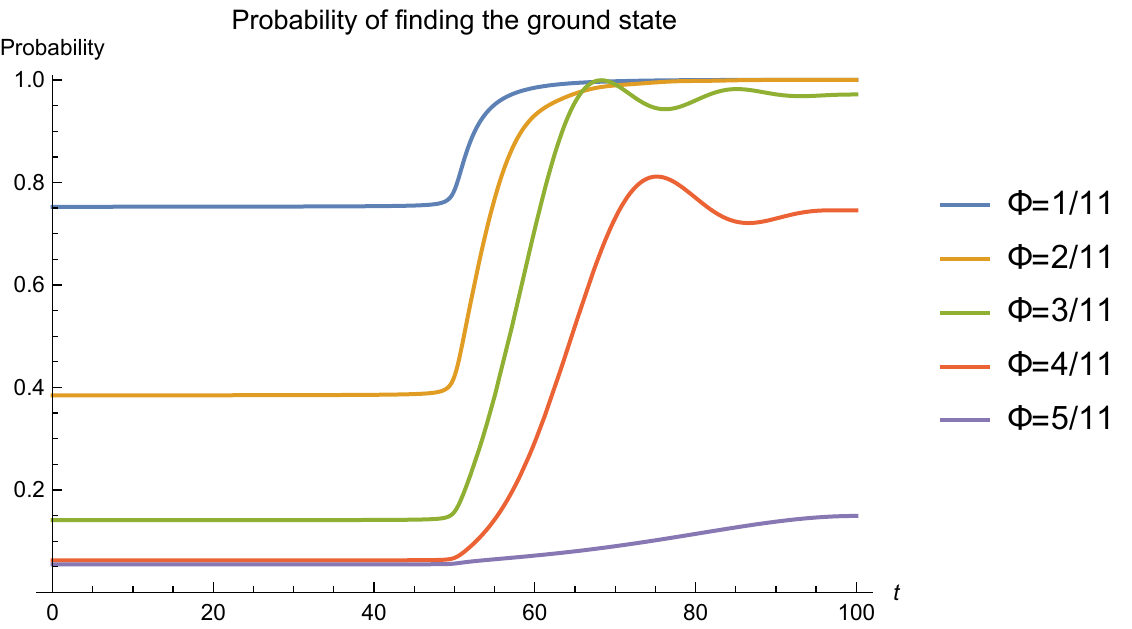}
\end{minipage}
\begin{minipage}{0.329\hsize}
\centering
    \includegraphics[width=\hsize]{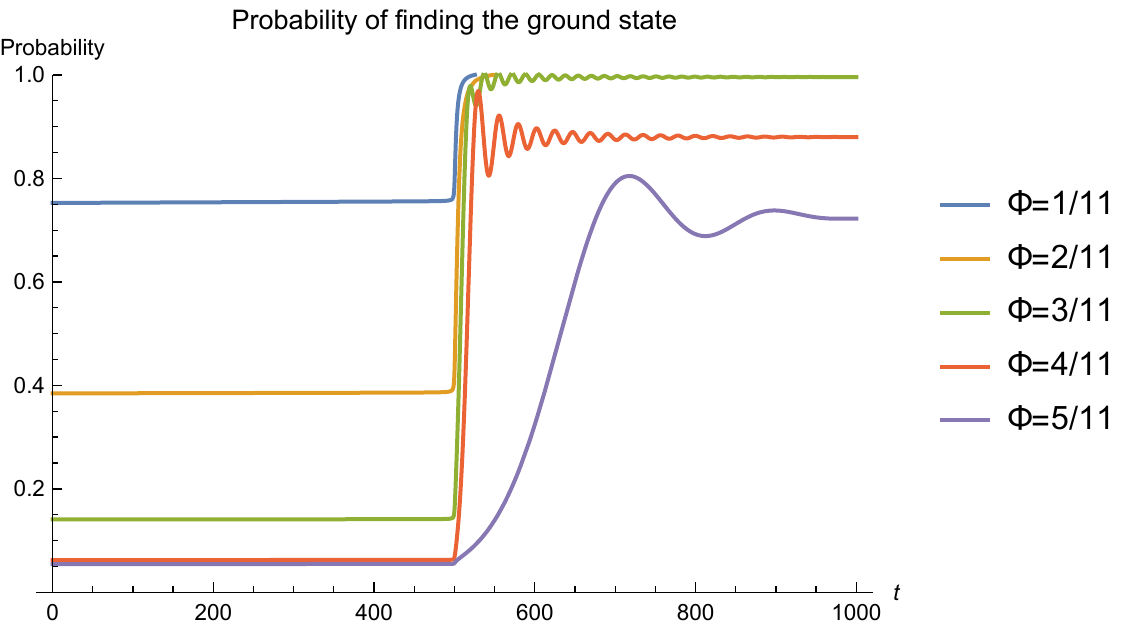}
\end{minipage}
\caption{Numerical results of probability of obtaining the ground states on a 3 by 3 square lattice under uniform magnetic flux $\phi$. The system obey the Hamiltonian $H(t)=(1-\Gamma(t))H_0+\Gamma(t)H_1$ whose time dependence is described by \eqref{eq:f}. [Left] $\tau=10$ [Middle] $\tau=100$ [Right] $\tau=1000$}
    \label{fig:probs3}
\end{figure}

\if{
\section{Machine Learning}
Let $x=\{x_i\}$ be a set of input data and $\{y_i\}$ be a set of output data. We denote by $P(x)$ the probability distribution and define $Q_W(x)$ by 
\begin{equation}
    Q_W(x)=\exp\left(\sum_i\left(\sum_{ab} J_{ab}{x_i}_a{x_i}_b+\sum_a h_a {x_i}_a\right)\right)/Z(J,h), 
\end{equation}
where $Z$ is the partition function. 
\begin{equation}
    P_0(x)=\frac{1}{D}\sum \delta(x-x_i)
\end{equation}

KL-divergence 
\begin{equation}
    D(P_0|Q_W)=\sum_{x}P_0(x)\log\left(\frac{P_0(x)}{Q_W(x)}\right)
\end{equation}

Taking derivative with respect to $h$, we obtain 
\begin{equation}
    -\frac{1}{D}\sum_i x_i+\sum_x xQ_W(x)
\end{equation}
the average of inputs and its expectation value with respect to $Q_W(x)$. 
}\fi

\section{Non-stoquastic Dynamics and Phase Transition}\label{sec:PT}
The general form of adiabatic quantum computation that we study here is given by the Hamiltonian 
\begin{equation}
    H(s)=sH_0+(1-s)H_1,~~~~~~~s\in[0,1] 
\end{equation}
where $H_0$ is a target Hamiltonian and $H_1$ is an initial Hamiltonian. They should not commute $[H_0, H_1]\neq 0$. The transverse magnetic field 
\begin{equation}
    H_1=-\sum_i^NX_i
\end{equation}
is widely used for the initial term \cite{PhysRevE.58.5355,2000quant.ph..1106F}. It is believed that adding a stoquastic Hamiltonian is not helpful for quantum speedup and there are some known examples of non-stoquastic terms that make problems efficiently solvable by adiabatic quantum computation \cite{2012PhRvE..85e1112S}. We add the following antiferromagnetic interactions
\begin{equation}
    H_{\text{AF}}=+N\left(\frac{1}{N}\sum_i^NX_i\right)^2
\end{equation}
as a non-stoquastic term in such a way that 
\begin{equation}
H(s,\lambda)=s(\lambda H_0+(1-\lambda)H_\text{AF})+(1-s)H_1.     
\end{equation}
The initial Hamiltonian should be $H(0,\lambda)=H_1$ with any $\lambda$ and the final Hamiltonian should be $H(1,1)=H_0$. In many cases, a phase transition occurs in the annealing process, some of which adversely affect the performance of annealing machines. Therefore it is crucial to clarify the properties of phase transitions. Since $X_i\ket{+}_i=\ket{+}_i$, at the initial stage of annealing $(s=0)$, the ground state $\ket{\psi_0}$ of $H_1$ is a super position of all possible $2^N$ states with the equal probability weight
\begin{align}
\begin{aligned}
\ket{\psi_0}&=\prod_i^N\ket{+}_i\\  
&=\frac{1}{\sqrt{2^N}}(\ket{\uparrow\uparrow\cdots\uparrow}+\ket{\uparrow\uparrow\cdots\uparrow\downarrow}+\cdots+\ket{\downarrow\downarrow\cdots\downarrow})
\end{aligned}
\end{align}
We call this phase as quantum paramagnetic (QP) phase. The ground state of $H_0$ is not necessary a PQ phase, hence a phase transition occurs in general. The first-order (second-order) phase transitions are defined by the discontinuity (continuity) of a given order parameter, respectively. 

To evaluate the required computational time, we refer to the adiabatic theorem. According to the adiabatic theorem, the computational time $t_*$ that is needed to efficiently obtain the ground state is proportional to the inverse square of the minimal energy gap $\Delta$ between the ground state and the first excited state ($t_*\sim\frac{1}{\Delta^2}$). For a large $N$, $\Delta$ is proportional to either $N^{-a}~(a>0)$ or $e^{-bN}~(b>0)$. (In the limit of $N\to \infty$, $\Delta$ goes to 0.) And by a lot of examples, it is known that $\Delta$ decays polynomially if a phase transition is second-order \cite{Damski_2013,PhysRevB.71.224420}, whereas $\Delta$ decays exponentially if it is first-order. Therefore, the problem on system with second-order phase transition is efficiently solved. It is known that when a non-stoquastic Hamiltonian is used, the first-order phase transitions can be avoided \cite{2012PhRvE..85e1112S,PhysRevA.95.042321}

Let $\ket{\theta,\phi}$ be the spin coherent state 
\begin{equation}
    \ket{\theta,\phi}=\bigotimes_{i}^N\ket{\theta,\phi}_i
\end{equation}
where $\ket{\theta,\phi}_i=\cos(\theta/2)\ket{0}_i+e^{i\phi}\sin(\theta/2)\ket{1}_i$ with $\theta\in[0,\pi], \phi\in[0,2\pi]$. Using ${}_i\bra{\theta,\phi}X_i\ket{\theta,\phi}_i=\sin\theta\cos\phi$, we find 
\begin{align}
    \begin{aligned}
         \bra{\theta,\phi}H_1\ket{\theta,\phi}&=-N\sin\theta\cos\phi\\
          \bra{\theta,\phi}H_{2}\ket{\theta,\phi}&=\frac{1}{N}\sum_{i}^Nh_{ii}+\frac{\sin^2\theta\cos^2\phi}{N}\sum_{i\neq j}h_{ij}\\
    \end{aligned}
\end{align}
The semi-classical potential $V(s,\lambda,\theta,\phi)$ is then defined by
\begin{equation}
    V(s,\lambda,\theta,\phi)=\lim_{N\to\infty}\frac{1}{N}\bra{\theta,\phi}H(s,\lambda)\ket{\theta,\phi}. 
\end{equation}
In what follows we address cases where $\bra{\theta,\phi}H_0\ket{\theta,\phi}$ is independent of $\phi$. Then it is easy to see that $V(s,\lambda,\theta,0)\le V(s,\lambda,\theta,\phi)$ for any $(s,\lambda,\theta,\phi)$. So $\phi=0$ gives a ground state. We define $\theta_{\min}$ by $V(s,\lambda,\theta_{\min},\phi) \le V(s,\lambda,\theta,\phi)$ for all $\theta$. 
The first-order phase transition occurs when $V$ is discontinuous with respect to $\theta_{\min}$. Starting $s=0,\lambda=1$, the ground state is initially located at $\theta=\pi/2$ and $\phi=0$, hence a second-order phase transition occurs when they satisfy
\begin{equation}\label{eq:2PT}
    \frac{\partial^2 V}{\partial \theta^2}\bigg|_{\theta=\frac{\pi}{2},\phi=0}=0
\end{equation}

\begin{figure}[H]
    \centering
    \includegraphics[width=10cm]{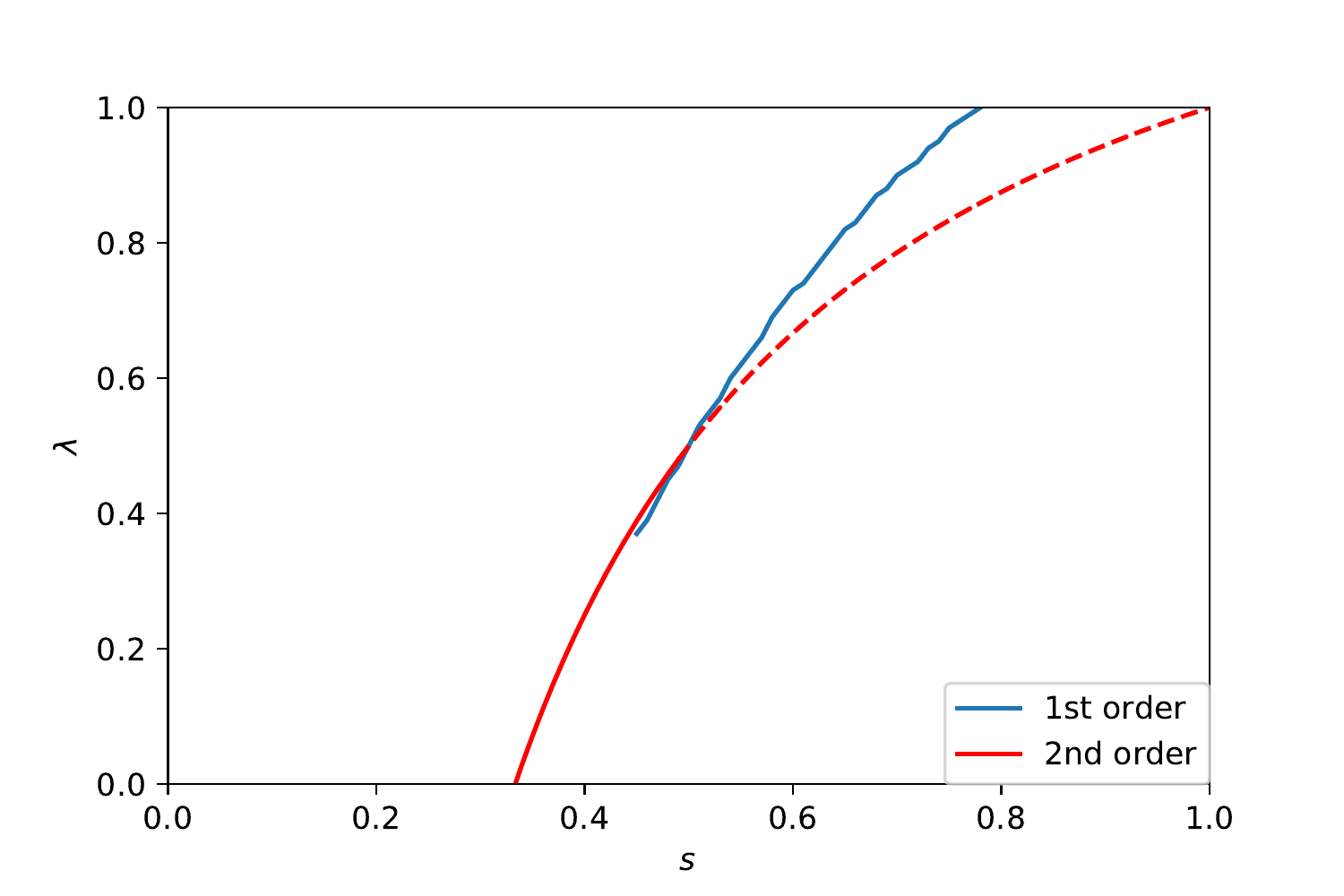}
    \caption{Phase diagram of the $p=6$ case.}
    \label{fig:PD2}
\end{figure}

\begin{figure}[H]
\begin{minipage}{0.329\hsize}
    \centering
    \includegraphics[width=\hsize]{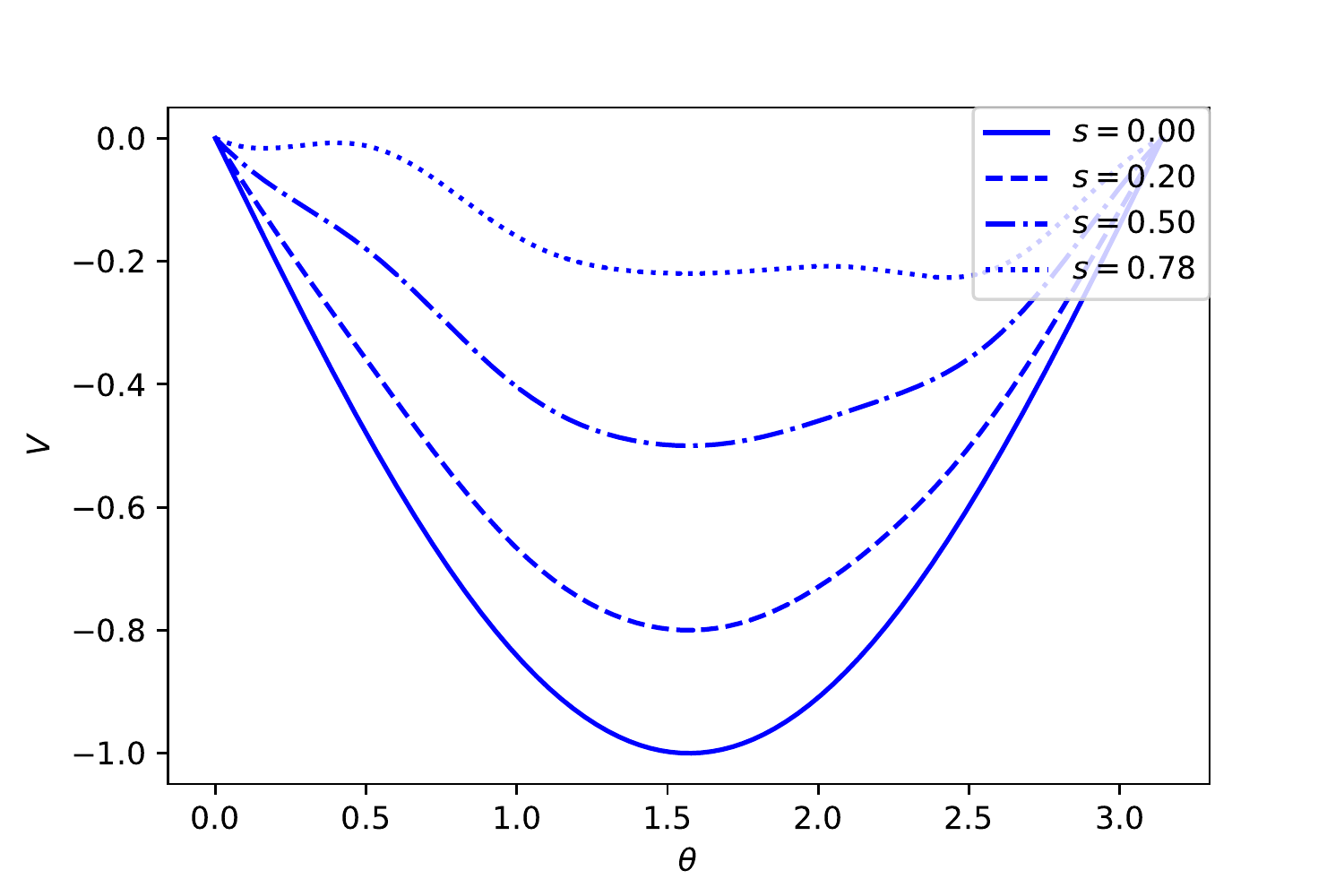}
\end{minipage}
\begin{minipage}{0.329\hsize}
    \centering
    \includegraphics[width=\hsize]{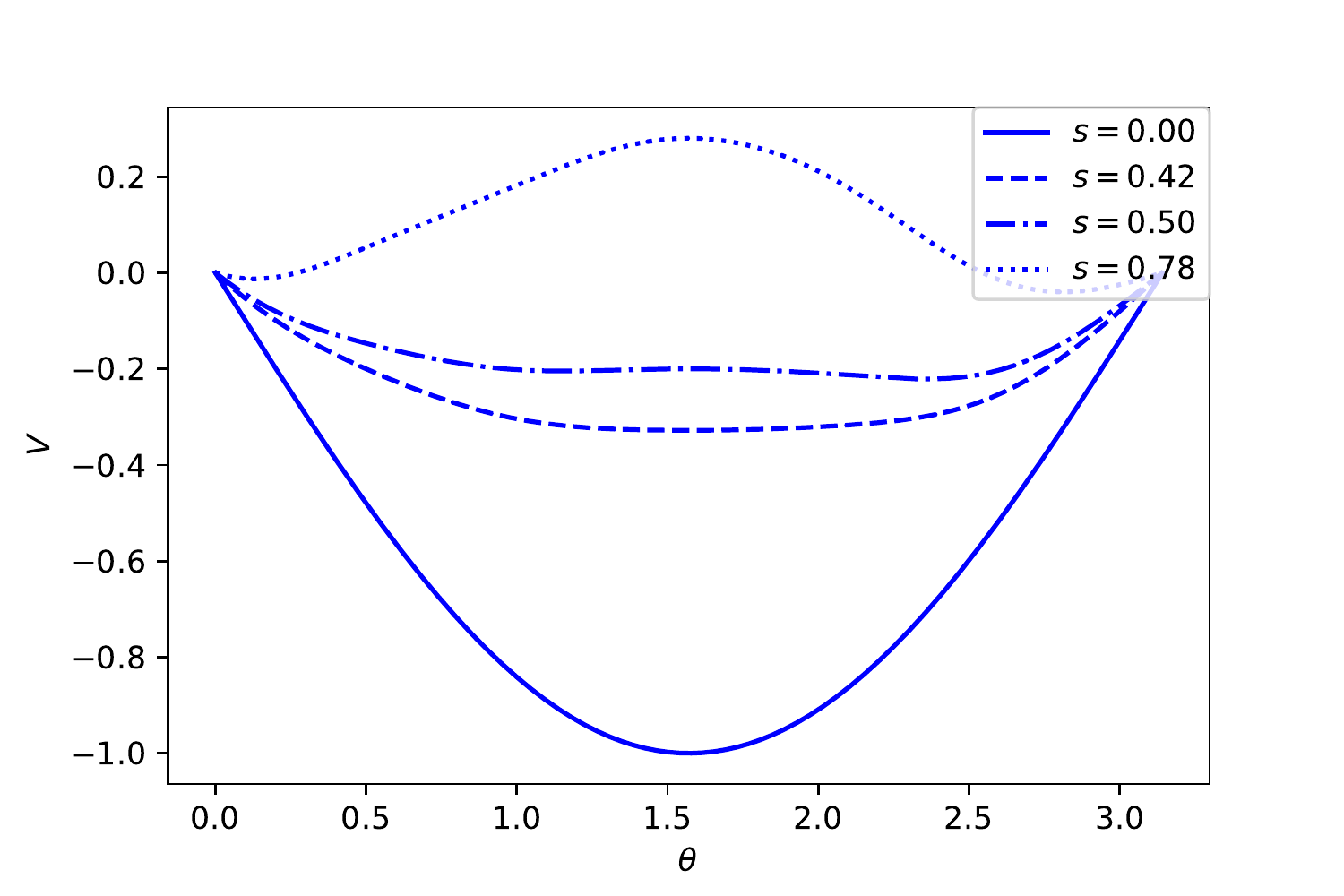}
\end{minipage}
\begin{minipage}{0.329\hsize}
    \centering
    \includegraphics[width=\hsize]{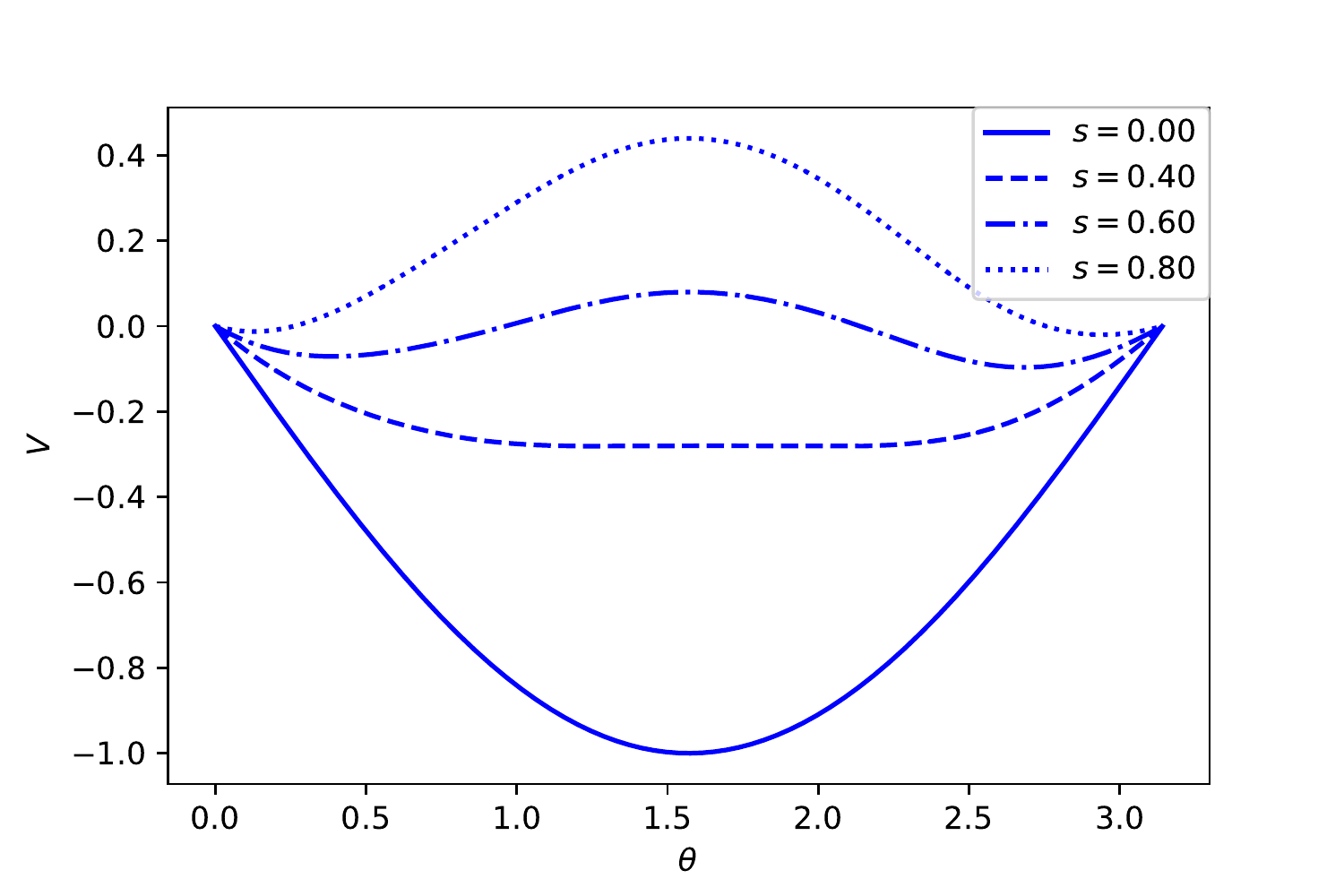}
\end{minipage}
\begin{minipage}{0.329\hsize}
    \centering
    \includegraphics[width=\hsize]{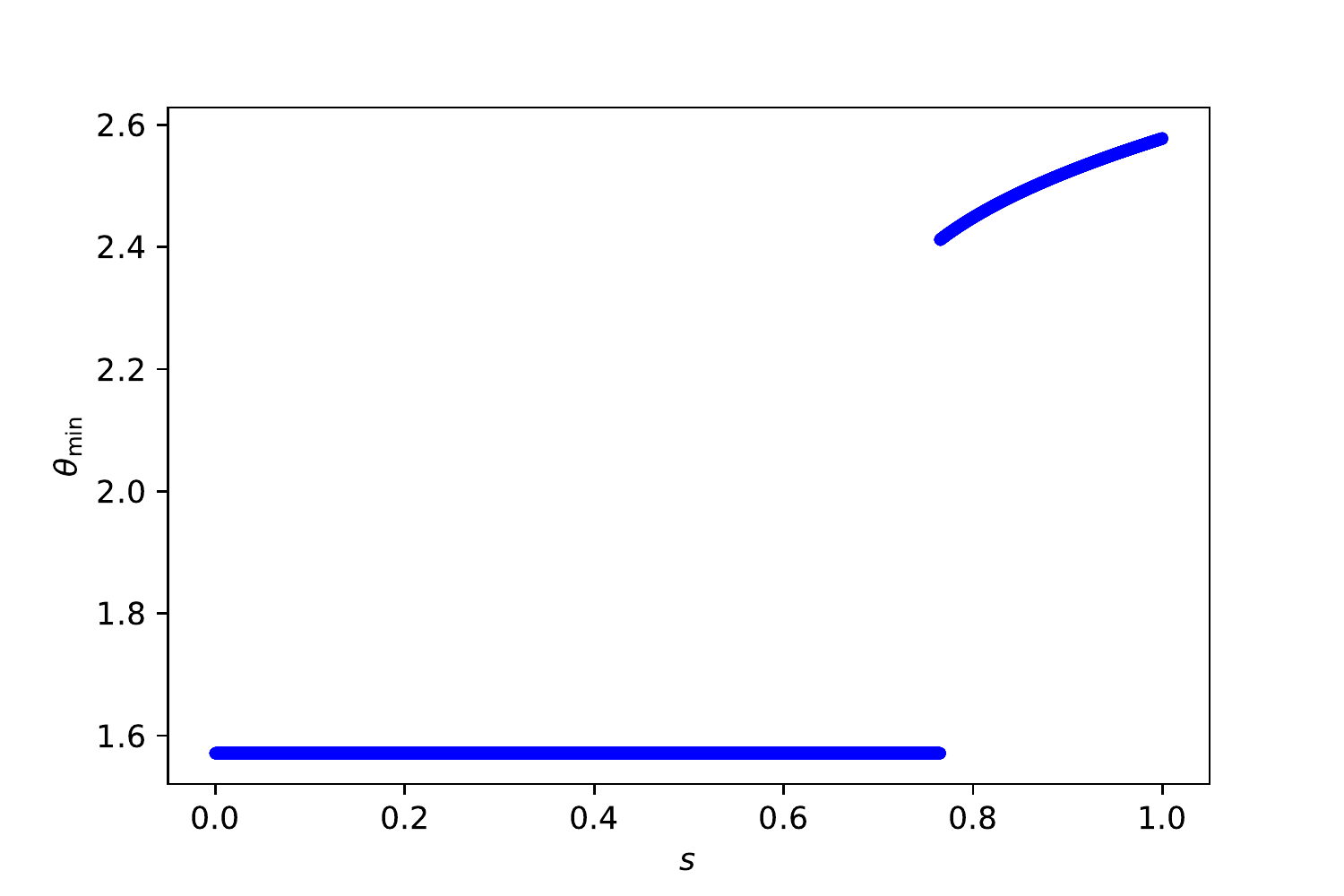}
\end{minipage}
\begin{minipage}{0.329\hsize}
    \centering
    \includegraphics[width=\hsize]{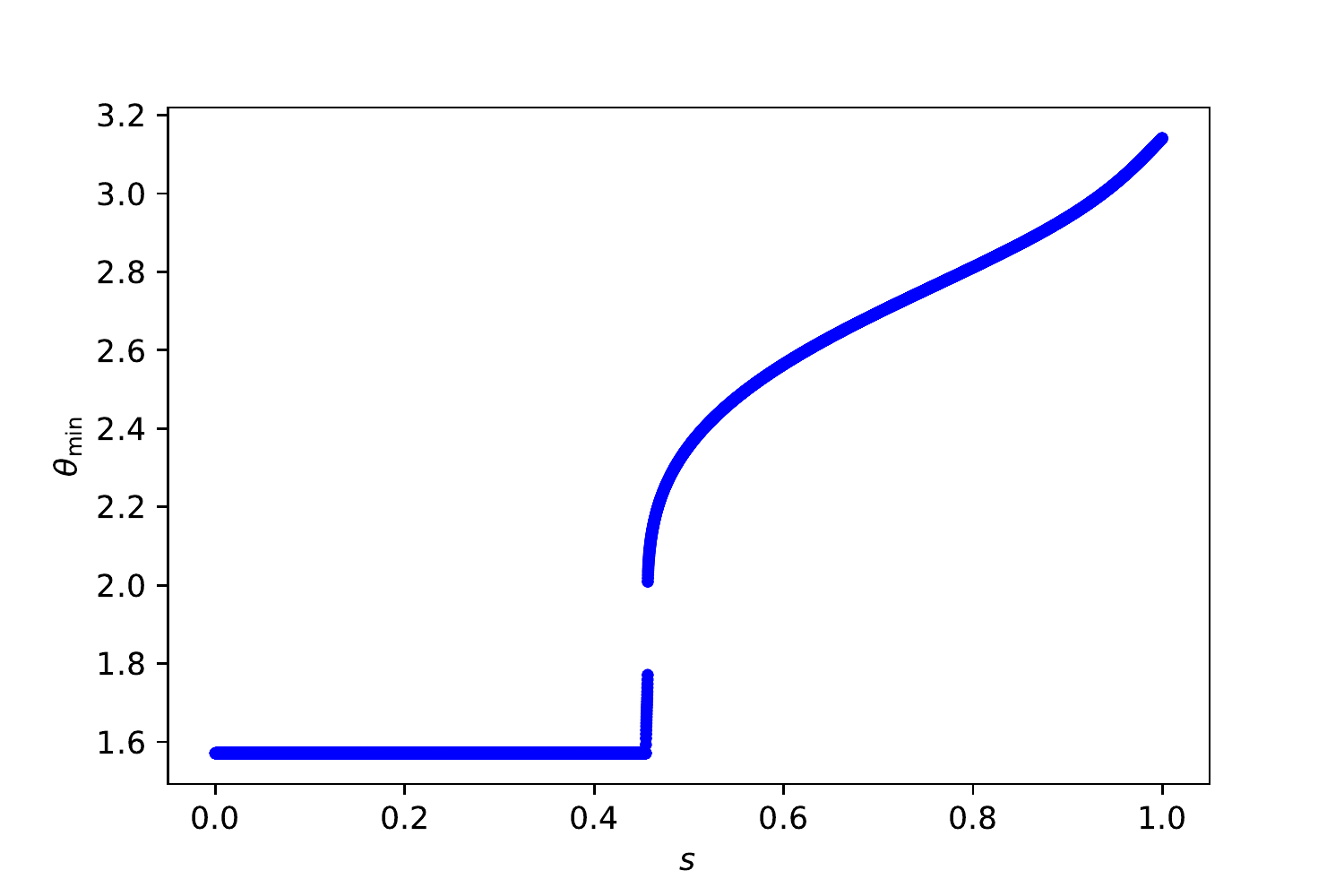}
\end{minipage}
\begin{minipage}{0.329\hsize}
    \centering
    \includegraphics[width=\hsize]{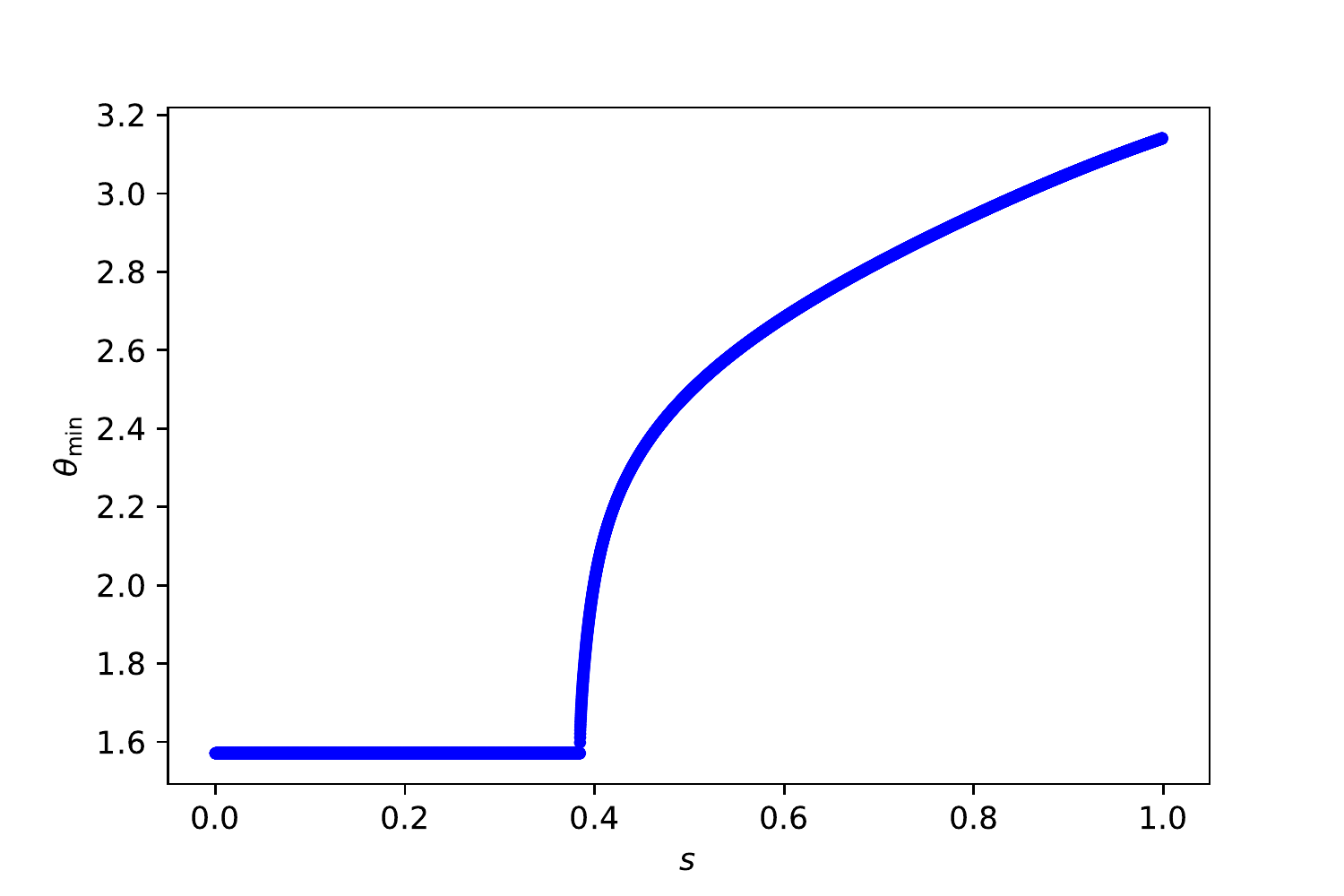}
\end{minipage}
    \caption{$p=6$ [Left] $\lambda=1.0$ [Middle] $\lambda=0.4$ [Right] $\lambda=0.2$}
    \label{fig:12}
\end{figure}

A model we are interested in has the Hamiltonian of Majorana fermions 
\begin{equation}\label{eq:HMF}
    H_0=i\sum_{k=1}^{N-p} c_{2(k+p)-1}c_{2k}+c_{2(k+p)}c_{2k-1}, 
\end{equation}
where $p$ is an integer and  $c_i$ is defined by the Jordan-Wigner formulation \eqref{eq:MJ}
\begin{align}
    \begin{aligned}
        c_{2k-1}&=Z_1\cdots Z_{k-1}X_k\\
        c_{2k}&=Z_1\cdots Z_{k-1}Y_k. 
    \end{aligned}
\end{align}
We find the potential 
\begin{equation}
    V=s\lambda\cos^{p-1}\theta\sin^2\theta-(1-s)\sin\theta\cos\phi+\frac{s(1-\lambda)\sin^2\theta\cos^2\phi}{N^2}\sum_{i\neq j}h_{ij},
\end{equation}
by using the forms 
\begin{align}
\begin{aligned}
    \bra{\theta,\phi}c_{2(k+p)-1}c_{2k}\ket{\theta,\phi}&=-i\cos^{p-1}\theta\sin^2\theta\cos^2\phi\\
    \bra{\theta,\phi}c_{2(k+p)}c_{2k-1}\ket{\theta,\phi}&=-i\cos^{p-1}\theta\sin^2\theta\sin^2\phi.
\end{aligned}
\end{align}
Then the condition of the second-order phase transitions is independent of $p$
\begin{equation}
    \lim_{N\to\infty}(1-s)-2s(1-\lambda)\frac{1}{N^2}\sum_{i\neq j}h_{ij}=0.
\end{equation}
In what follows we set $h_{ij}=1$. This model experiences various phase transitions (Fig. \ref{fig:PD2}). For a large $\lambda$, it is a first-order as shown in the left of Fig.\ref{fig:12}. There are no second-order phase transitions on the dashed line. For medium $\lambda\sim 0.4$, a first-order phase transition occurs after a second-order phase transition. For a small $\lambda$, a first-order phase transition is avoided. One can directly confirm some quantum effects by studying the trace distance between $\ket{\theta_{\min},0}$ and the ground state of $H$. So we can conclude that the non-stoquastic term plays a crucial role to avoid a first-order phase transition, which leads to quantum speedup. For a fist-order phase transition, even $p$ is important. One can confirm that a phase transition is second order if $p$ is odd. 
 
\begin{figure}[H]
\begin{minipage}{0.325\hsize}
    \centering
    \includegraphics[width=\hsize]{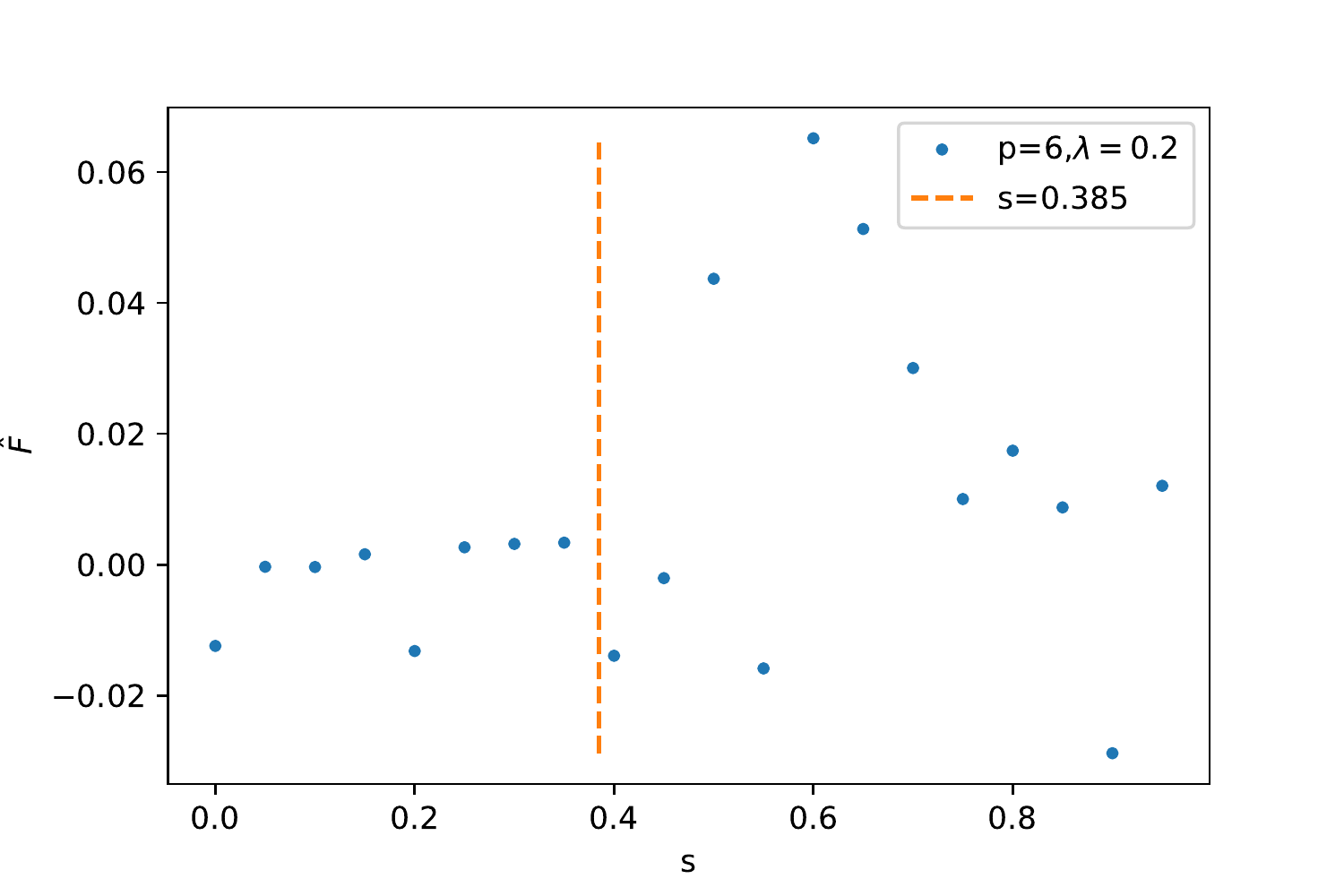}
\end{minipage}
\begin{minipage}{0.325\hsize}
    \centering
    \includegraphics[width=\hsize]{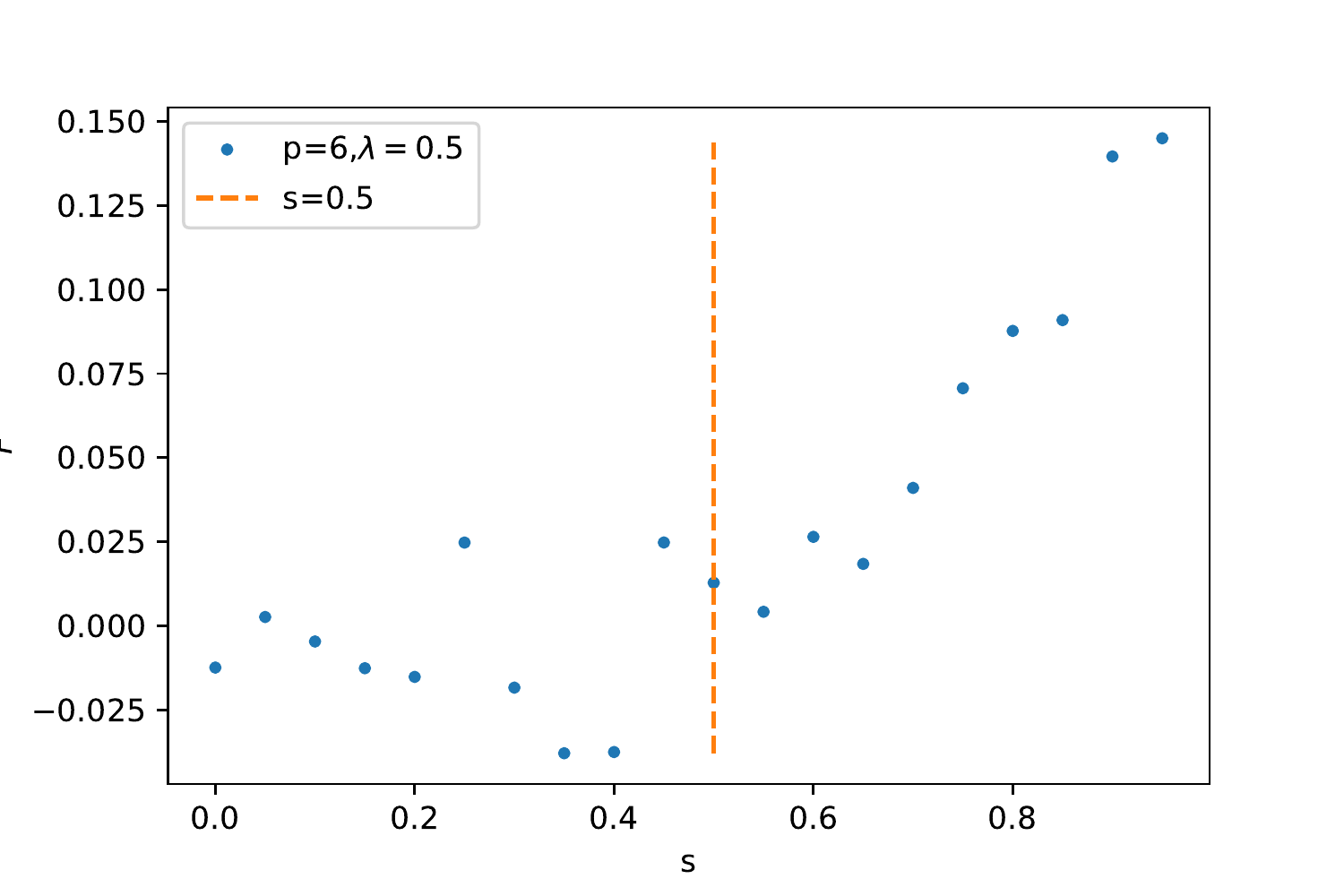}
\end{minipage}
\begin{minipage}{0.325\hsize}
    \centering
    \includegraphics[width=\hsize]{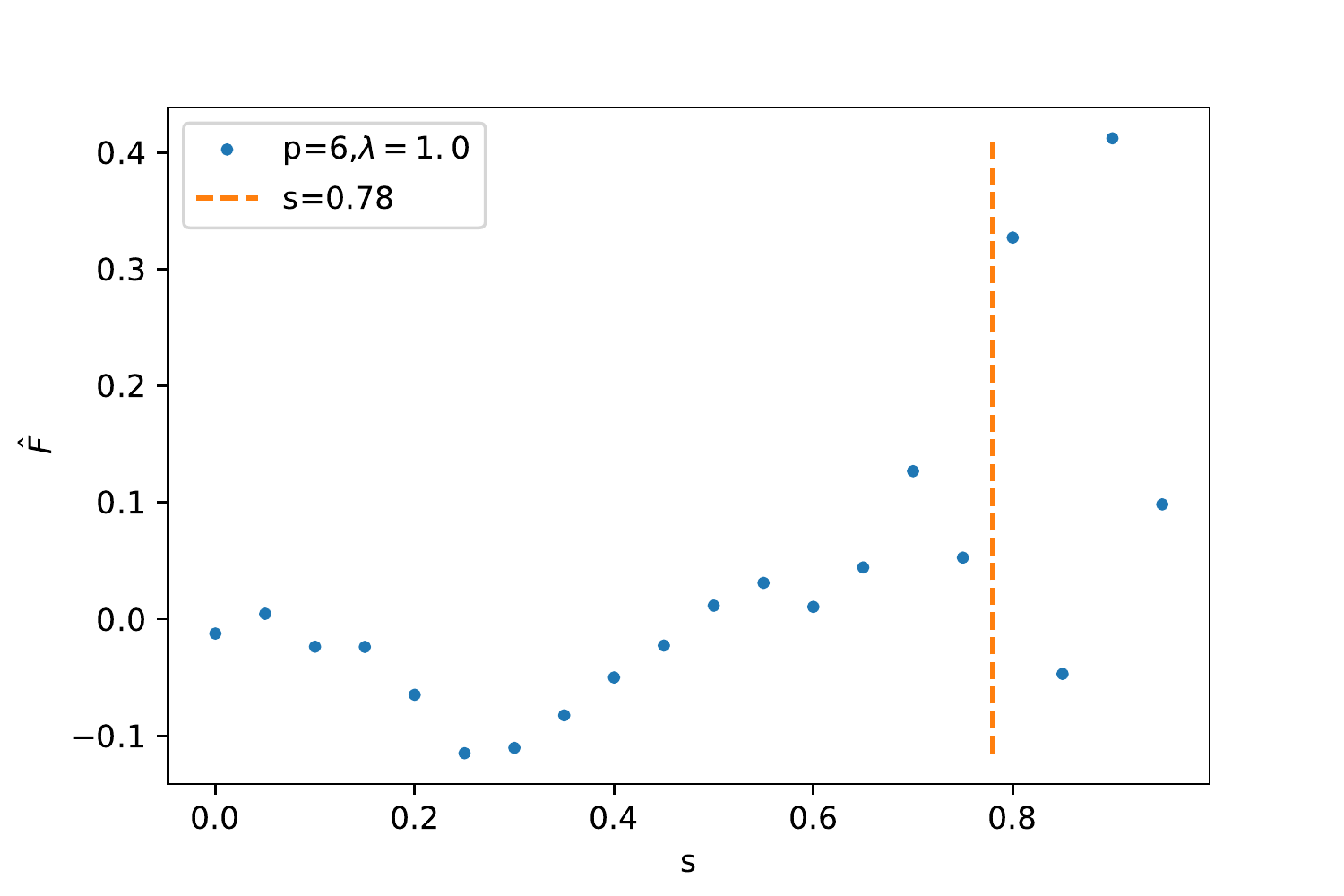}
\end{minipage}
    \caption{$N=8$}
    \label{fig:Ftilde}
\end{figure}

Now let us study a little bit more on the dynamics of our model from a viewpoint of quantum chaos. Roughly, chaos would trigger phase transition, which may affect the probability of obtaining states. This motivates us to study quantum chaos. There are various approaches to quantum chaos, but its formal definition has been illusive. The most standard way is to define a quantum counterpart of classical chaos. Generally non-zero Lyapunov exponent is a crucial factor of classical chaos, hence to find its quantum counterpart is a main interest. We study the phase transitions from a viewpoint of quantum chaos, especially we study the dynamics of OTOC \cite{larkin1969quasiclassical,Maldacena:2015waa,Kitaev14}.
\begin{equation}
    C(t,\psi,\varphi)=-\bra{\psi}[W(t), V(0)]^\dagger [W(t), V(0)]\ket{\varphi}\ge 0. 
\end{equation}
The OTOC contains the term $F(t)=\bra{\psi}W(t)VW(t)V\ket{\varphi}$.  It is believed that OTOC is a good measure of quantum chaos. A local operator $W=W(0)$ evolves to a complicated one $W(t)=e^{itH}W(0)e^{-it H}$, which can be written by sum of local operators 
\begin{equation}
    W(t)=W+it[H,W]+\frac{(it)^2}{2!}[H,[H,W]]+\frac{(it)^3}{3!}[H,[H,[H,W]]]+\cdots
\end{equation}
This implies that the OTOC is non-constant unless $[H,W]\neq 0$. In this work we use 
\begin{equation}
    F(t)=\bra{\psi_0}V(t)VV(t)V\ket{\psi_0}, 
\end{equation}
where $V$ is an operator and $\psi_0$ is the ground state. We find the time-average of $F(t)$ 
\begin{equation}
    \hat{F}=\frac{1}{T}\int_0^T F(t)dt
\end{equation}
can diagnose phase transition. The behavior of $\hat{F}$ drastically changes before and after the critical points (Fig. \ref{fig:Ftilde}). $\hat{F}$ would correspond to the steady value at large time, hence it is almost the same as $\hat{F}(t \rightarrow \infty)$, which is the contribution from the ground state and becomes dominant at large $t$. This is an intuitive explanation of $\hat{F}$'s behavior and its relation to the phase transition. The relation between the phase transition and the dynamics of $\hat{F}$ is not proven in general and should be confirmed by some other examples. In fact, the similar behavior of $\hat{F}$ is observed in some models \cite{Dag:2019yqu,2018arXiv181111191S,2018arXiv181201920W}.

\if{
As a different model, we consider the $p$-particle model 
\begin{equation}\label{eq:pp}
    H_0=-N\left(\frac{1}{N}\sum_{i}^Nn_i\right)^p
\end{equation}
Using ${}_i\bra{\theta,\phi}n_i\ket{\theta,\phi}_i=\sin^2(\theta/2)$ and $n^q_i=n_i$ for any positive integer $q$, we find 
\begin{align}
\begin{aligned}
    \bra{\theta,\phi}H_0\ket{\theta,\phi}&=-\frac{1}{N^{p-1}}\bra{\theta,\phi}\sum_{k_1+\cdots k_N=p}\frac{p!}{k_1!\cdots k_N!}n^{k_1}_1\cdots n^{k_N}_N\ket{\theta,\phi}\\
    &=-N\sin^{2p}(\theta/2)+O(1)
\end{aligned}
\end{align}
So the potential is 
\begin{equation}
    V=-s\lambda \sin^{2p}(\theta/2)-(1-s)\sin\theta\cos\phi+\frac{s(1-\lambda)\sin^2\theta\cos^2\phi}{N^2}\sum_{i\neq j}h_{ij}+O(1/N). 
\end{equation}
Unlike the previous case, the condition for the second-order phase transitions is dependent of $p$ in such a way that
\begin{equation}
    -2^{-p}s\lambda p(p-1)+(1-s)-2s(1-\lambda)\frac{1}{N^2}\sum_{i\neq j}h_{ij}=0.
\end{equation}
This model is resemble to the anti-ferromagnetic $p$-spin model, in which the condition of the second-order phase transition is independent of $p$. Taking a large $p$, they coincide.  Fig. \ref{fig:PD} is the phase diagram of the $h_{ij}=1$ case. In general, there are first-order and second-order phase transitions for $p>2$. The $p=2$ case essentially corresponds to the previous example we studied. Parameters on the dashed line in the figure satisfy $\partial^2V/\partial\theta^2|_{\theta=\pi/2,\phi=0}=0$ \eqref{eq:2PT}, but there are no phase transitions. Without the XX-interactions ($\lambda=1$), a phase transition is first-order and with some effect of the XX-interactions, the system experiences only a second-order phase transition (see also Fig. \ref{fig:11}). Therefore, in this case, the non-stoquastic Hamiltonian should be helpful to realize quantum speedup. One can directly confirm some quantum effects of the non-stoquastic term by studying the trace distance between $\ket{\theta_{\min},0}$ and the ground state of $H$. 
\begin{figure}[H]
    \centering
    \includegraphics[width=10cm]{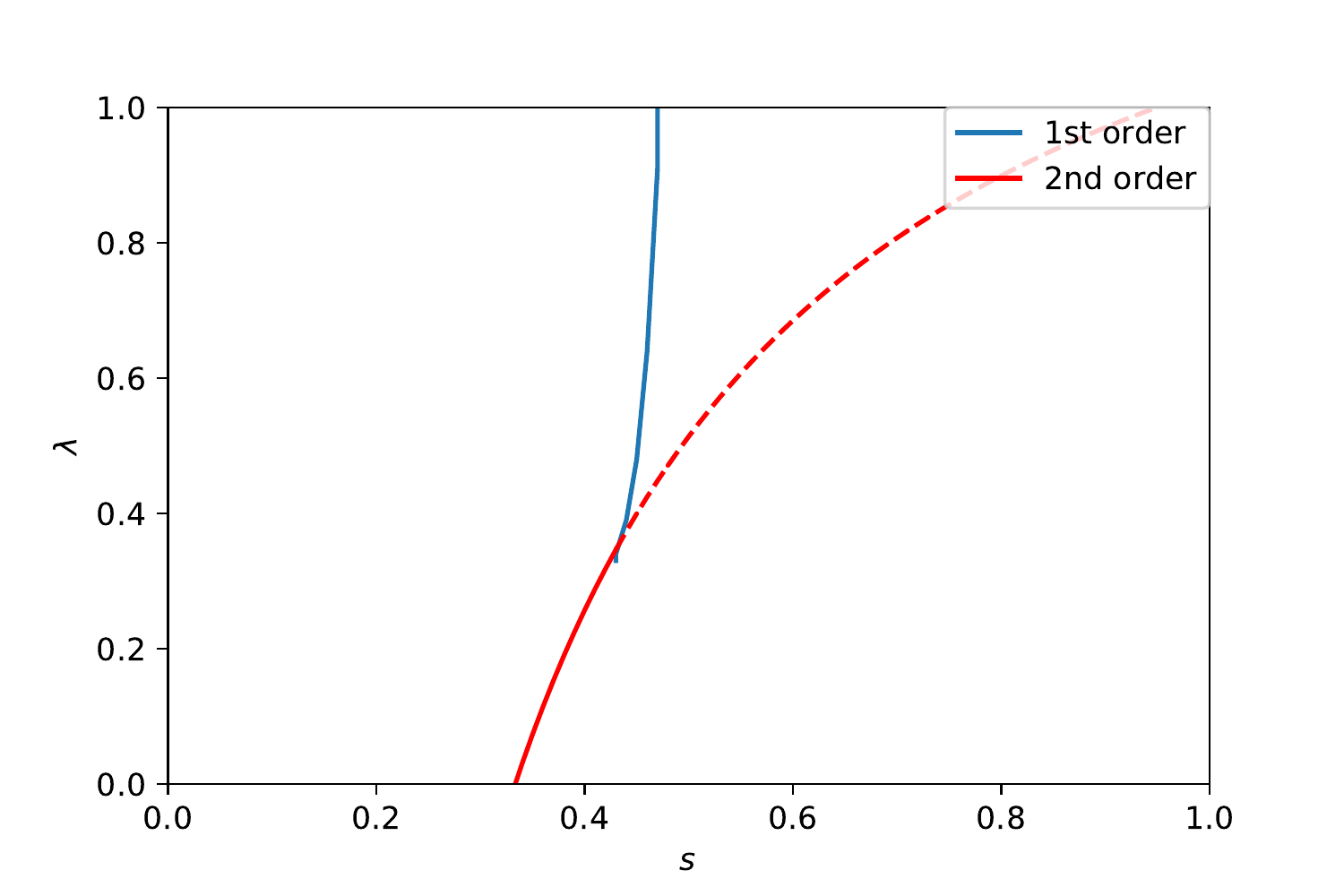}
    \caption{Phase diagram of the $p=11$ case.}
    \label{fig:PD}
\end{figure}

\begin{figure}[H]
\begin{minipage}{0.5\hsize}
    \centering
    \includegraphics[width=\hsize]{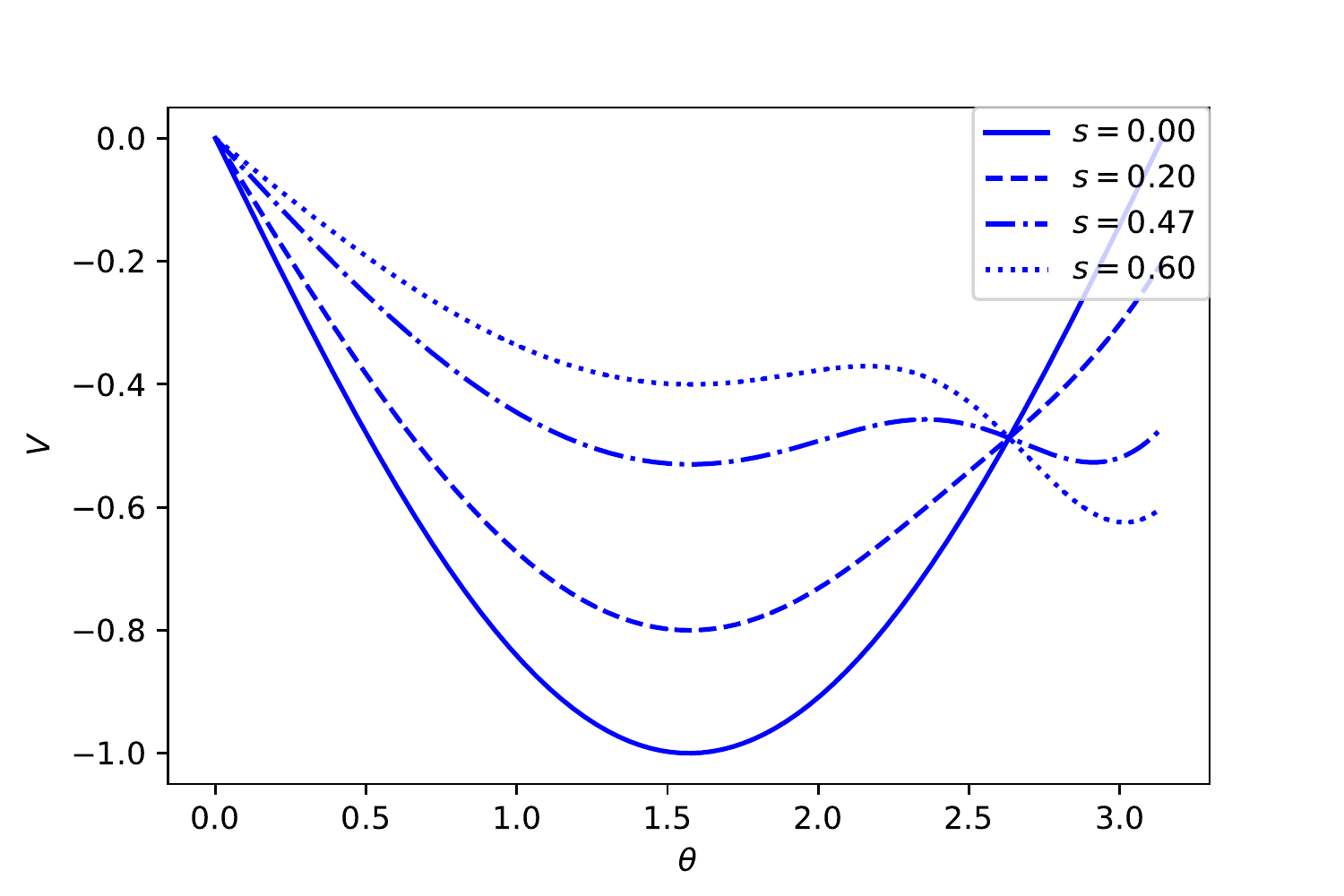}
\end{minipage}
\begin{minipage}{0.5\hsize}
    \centering
    \includegraphics[width=\hsize]{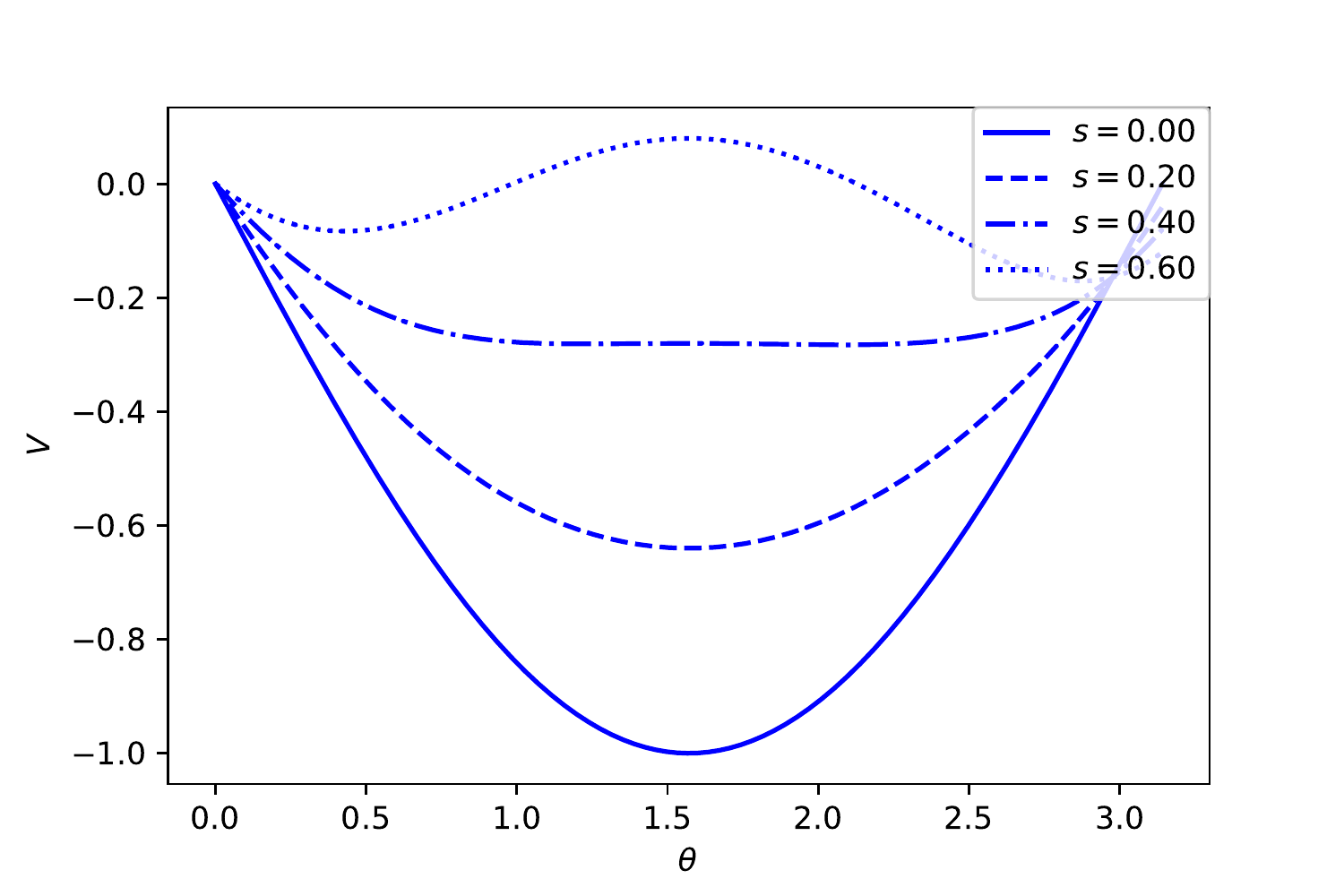}
\end{minipage}
\begin{minipage}{0.5\hsize}
    \centering
    \includegraphics[width=\hsize]{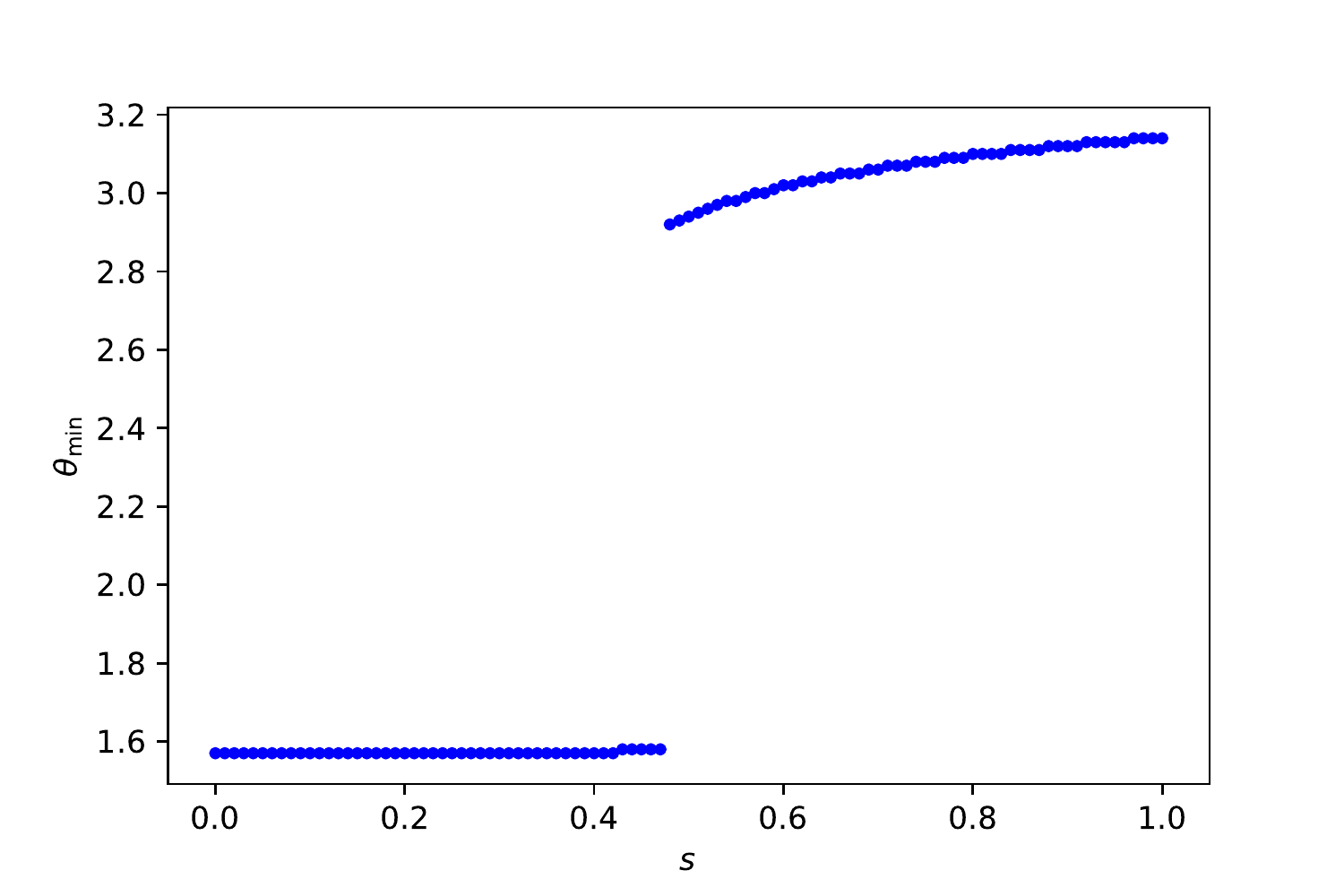}
\end{minipage}
\begin{minipage}{0.5\hsize}
    \centering
    \includegraphics[width=\hsize]{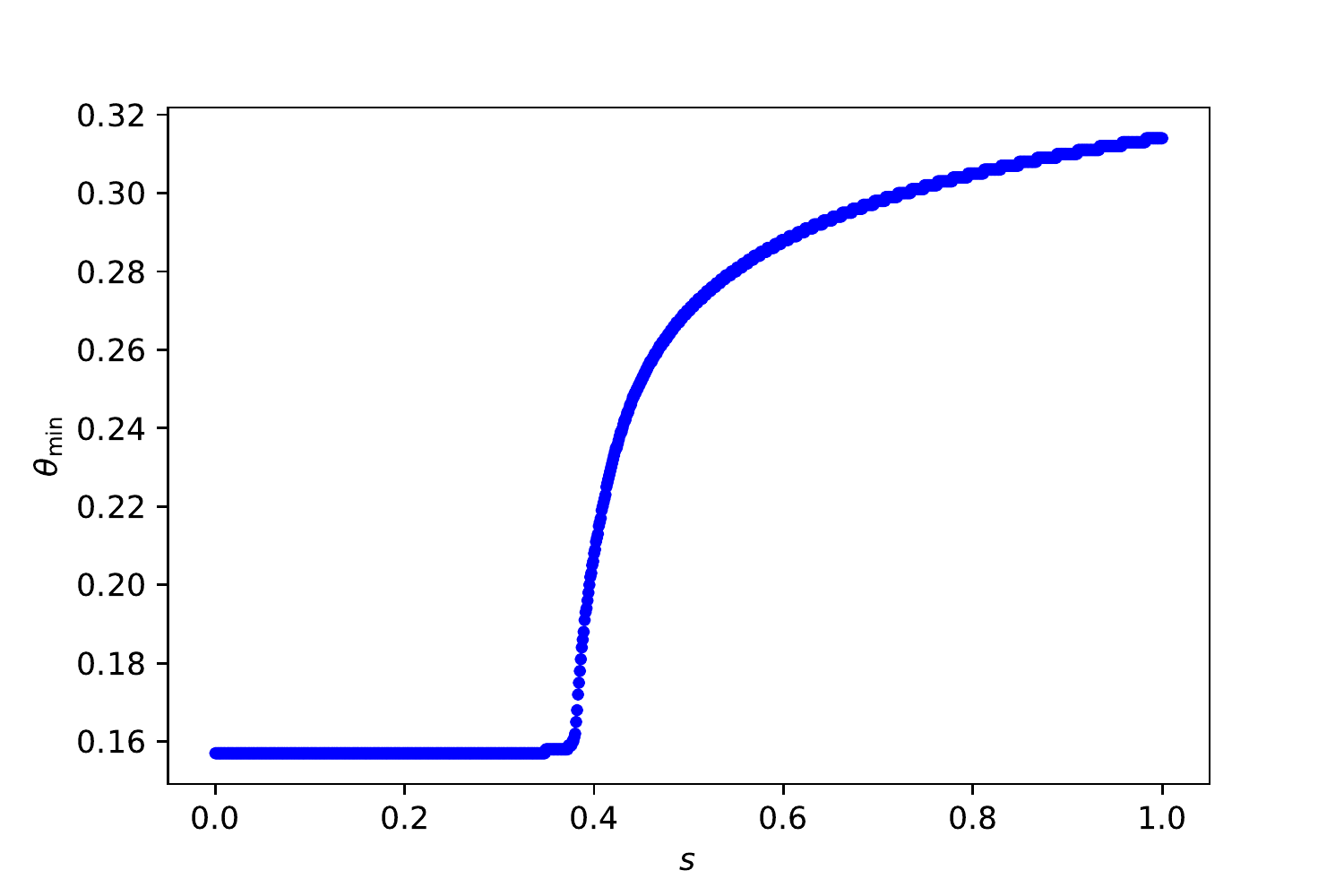}
\end{minipage}
    \caption{$p=11$ [Left] $\lambda=1.0$ [Right] $\lambda=0.2$}
    \label{fig:11}
\end{figure}
}\fi

\if{
\subsection{Second-order}
Second-order phase transitions make AQC more effective and it is also meaningful to know what kind of problems are efficiently solvable. As an example, we consider $H_0=\sum_{ij}J_{ij}(a^\dagger_ia_j+h.c.)+ \sum_{ij}K_{ij}n_in_j$.Putting $K_{ij}=0$ corresponds to a tight-binding model and $J_{ij}=0$ corresponds to a Hamiltonian of a binary optimization problem. Using ${}_i\bra{\theta,\phi}a_i\ket{\theta,\phi}_i=e^{i\phi}\cos(\theta/2)\sin(\theta/2)$ and ${}_i\bra{\theta,\phi}n_i\ket{\theta,\phi}_i=\sin^2(\theta/2)$, we find 
\begin{equation}
    \bra{\theta,\phi}H_0\ket{\theta,\phi}=\frac{1}{4}\sin^2\theta\sum_{ij}(J_{ij}+h.c.)+\frac{1}{4}(\cos\theta-1)^2\sum_{ij}K_{ij}. 
\end{equation}
For a large $N$, the potential $V$ is 
\begin{align}
\begin{aligned}
    V=& \frac{s\lambda}{4N}\sin^2\theta\left(\sum_{ij}J_{ij}+h.c.\right)+\frac{s\lambda}{4N}(\cos\theta-1)^2\sum_{ij}K_{ij}\\
    &-(1-s)\sin\theta\cos\phi+\frac{s(1-\lambda)\sin^2\theta\cos^2\phi}{N^2}\sum_{i\neq j}h_{ij}. 
\end{aligned}    
\end{align}
The condition of a second-order phase transition occurs when the parameters satisfy
\begin{equation}
    \Leftrightarrow -\frac{s\lambda}{2N}\left(\sum_{ij}(J_{ij}+h.c.)-K_{ij}\right)+(1-s)-2s(1-\lambda)\frac{1}{N^2}\sum_{i\neq j}h_{ij}=0. 
\end{equation}

We first put $K_{ij}=0$ and consider a simple one-dimensional chain model with a constant hopping parameter $J$ 
\begin{equation}J_{ij}=
\begin{cases}
J& |i-j|=1\\
0&|i-j|\neq 1
\end{cases}
\end{equation}
We also assume $h_{ij}$ is constant $h_{XX}$ everywhere. One can find a phase transition is second-order, if any. For $J,h_{XX}$ such that $J=2h_{XX}<0$, a phase transition does not occur. Several examples without or with XX interactions are shown in Fig. \ref{fig:pt1}. The left and middle figures show that the phase transitions are second-order, whereas the right figure shows there are no phase transitions.

\begin{figure}[H]
\begin{minipage}{0.329\hsize}
\centering
    \includegraphics[width=\hsize]{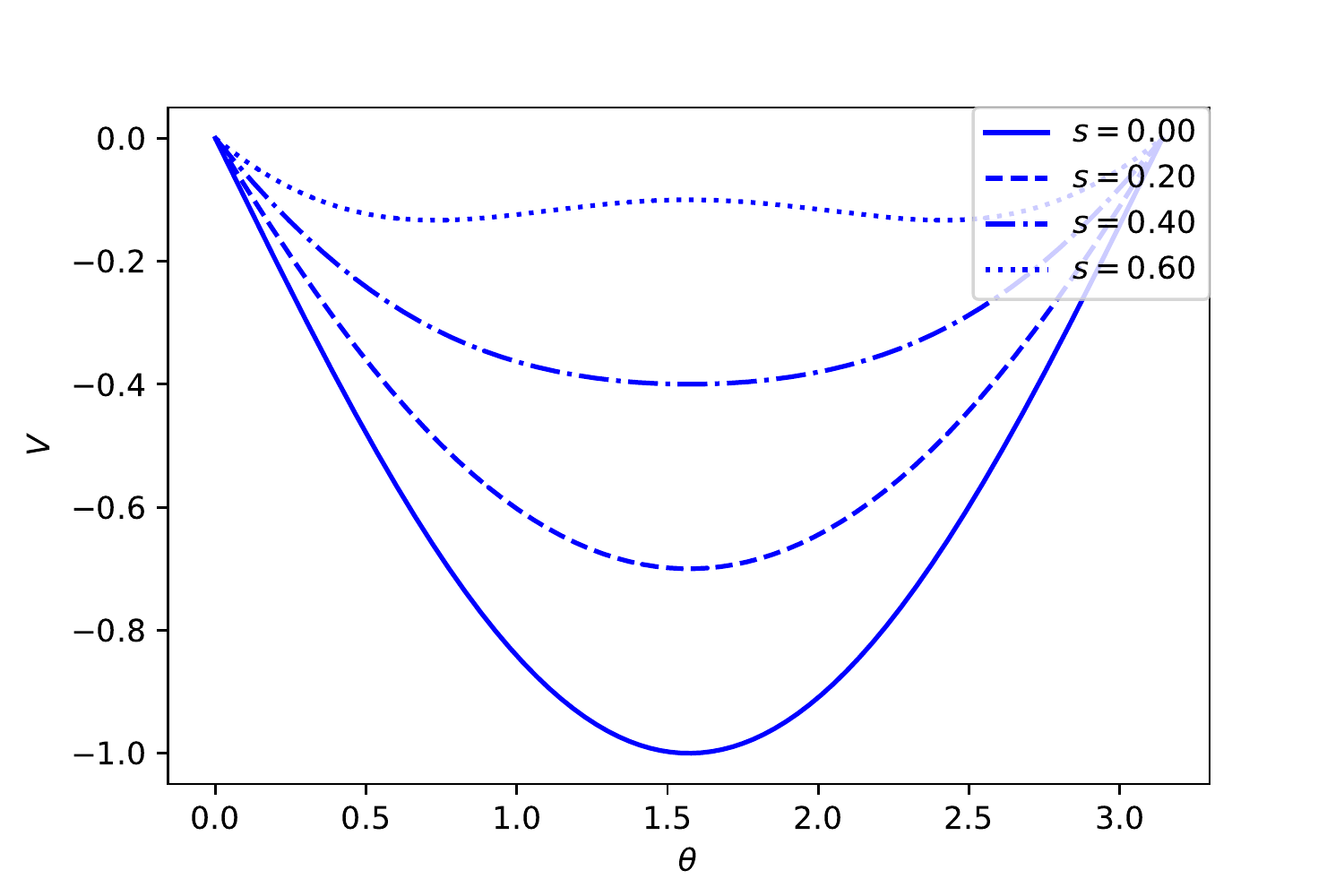}
\end{minipage}
\begin{minipage}{0.329\hsize}
\centering
    \includegraphics[width=\hsize]{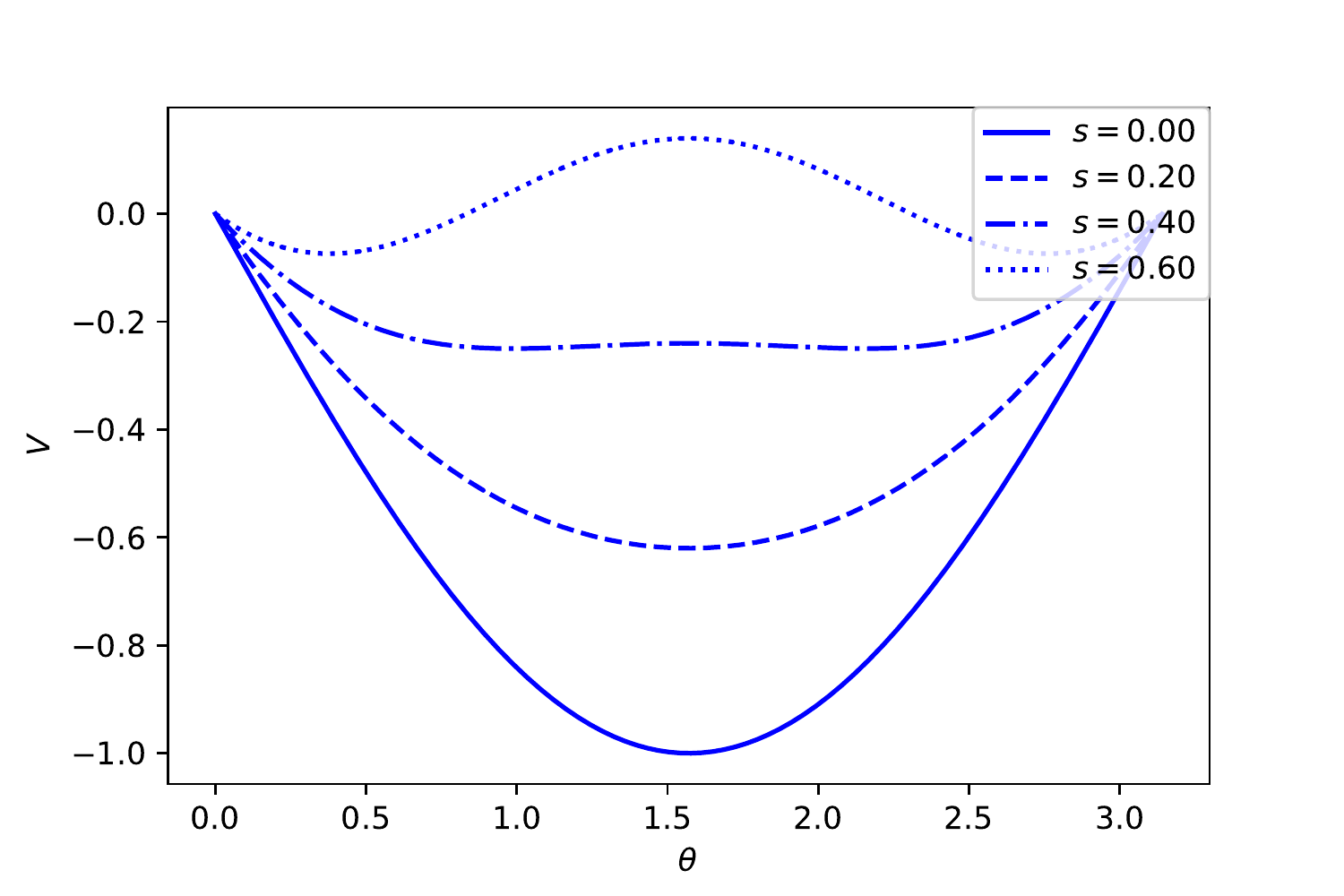}
\end{minipage}
\begin{minipage}{0.329\hsize}
\centering
    \includegraphics[width=\hsize]{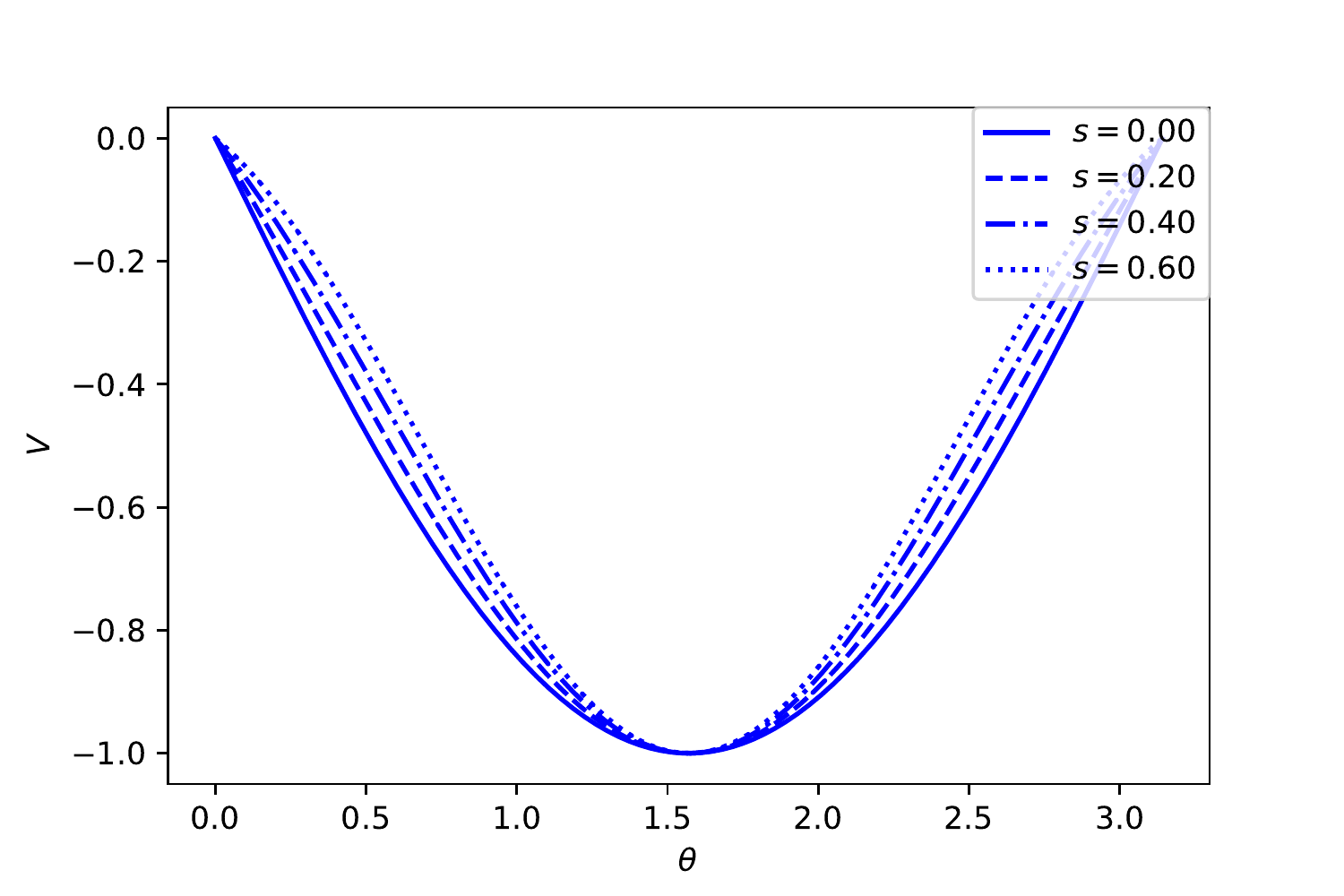}
\end{minipage}
\begin{minipage}{0.329\hsize}
\centering
    \includegraphics[width=\hsize]{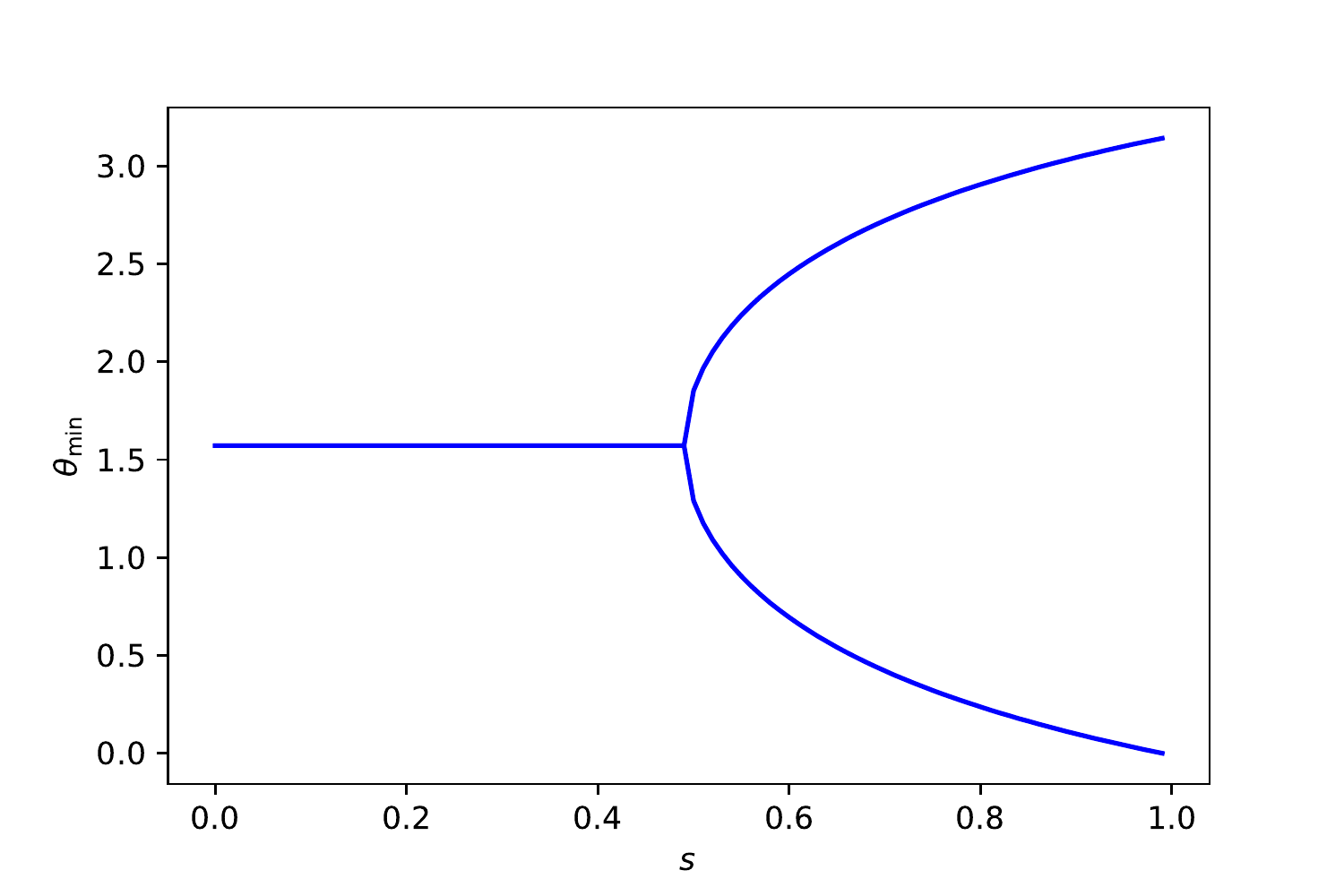}
\end{minipage}
\begin{minipage}{0.329\hsize}
\centering
    \includegraphics[width=\hsize]{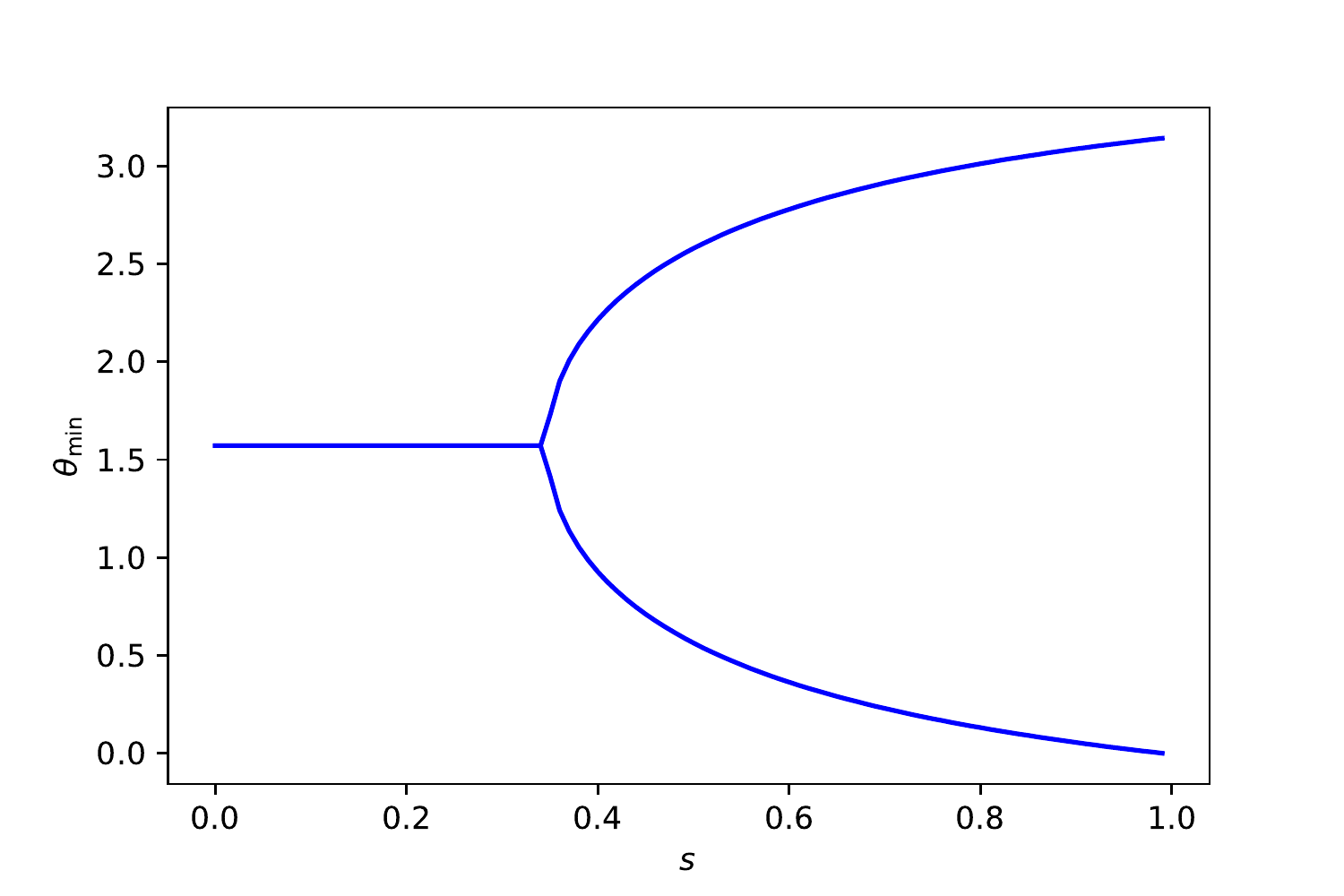}
\end{minipage}
\begin{minipage}{0.329\hsize}
\centering
    \includegraphics[width=\hsize]{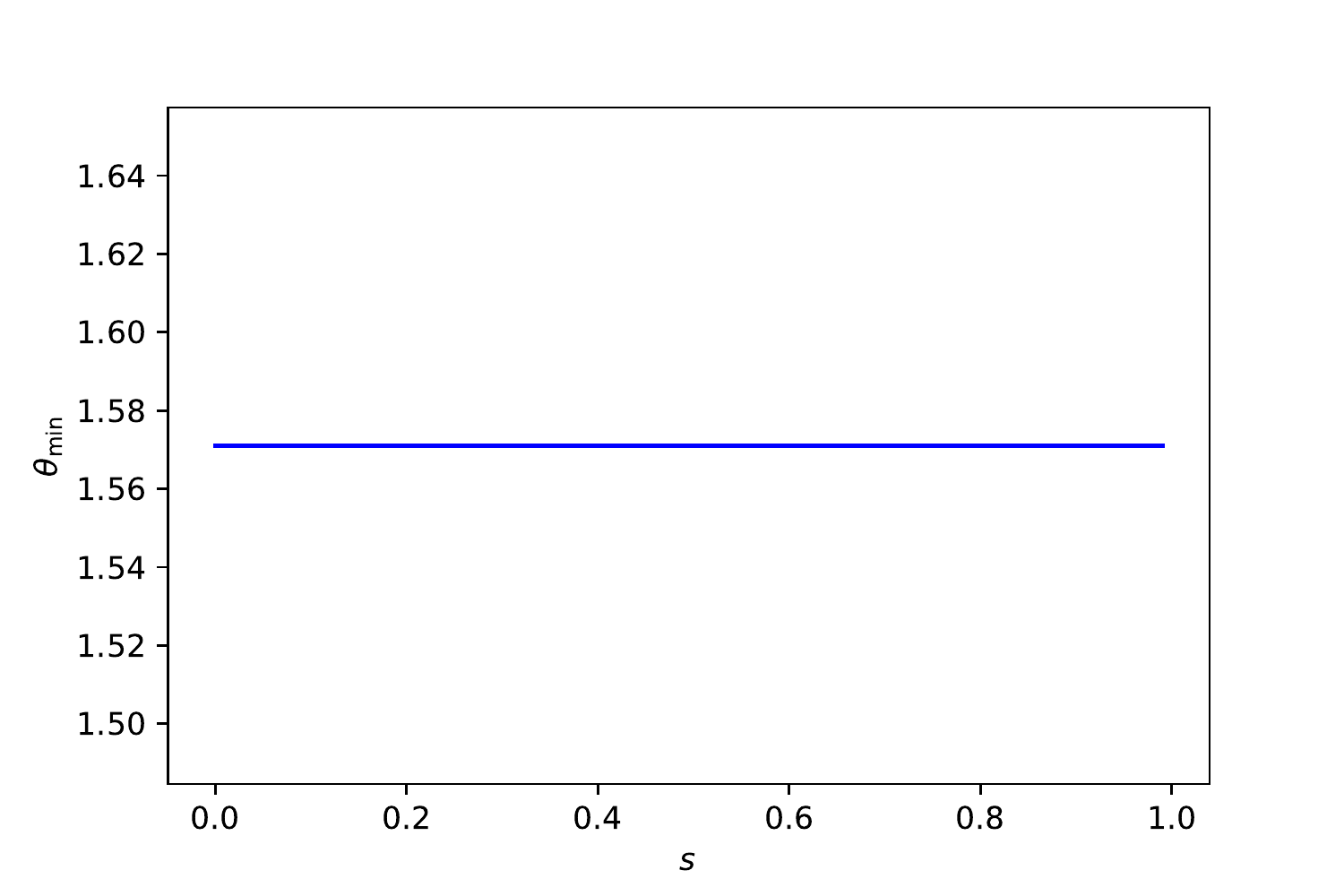}
\end{minipage}
\caption{[Left]$\lambda=1.0$, $h_{XX}=J=1.0$ [Middle] $\lambda=0.2$, $h_{XX}=J=1.0$ [Right] $\lambda=0.2$, $h_{XX}=2J=-1.0$. The upper figures show the $\theta$-dependence of the potential $V$ for several $s$. The lower figures show the $s$-dependence of $\theta_{\min}$. }
    \label{fig:pt1}
\end{figure}

We turn off $J$ and introduce $K_{ij}$:
\begin{equation}K_{ij}=
\begin{cases}
K& |i-j|=1\\
0&|i-j|\neq 1
\end{cases}
\end{equation}
In this case, phase transitions are second order as indicated by Fig. \ref{fig:pt2}. One can find similar results for $K_{ij}\neq 0$ and $J_{ij}\neq 0$. Therefore, those problems are efficiently solvable with our method.  
\begin{figure}[H]
\begin{minipage}{0.329\hsize}
\centering
    \includegraphics[width=\hsize]{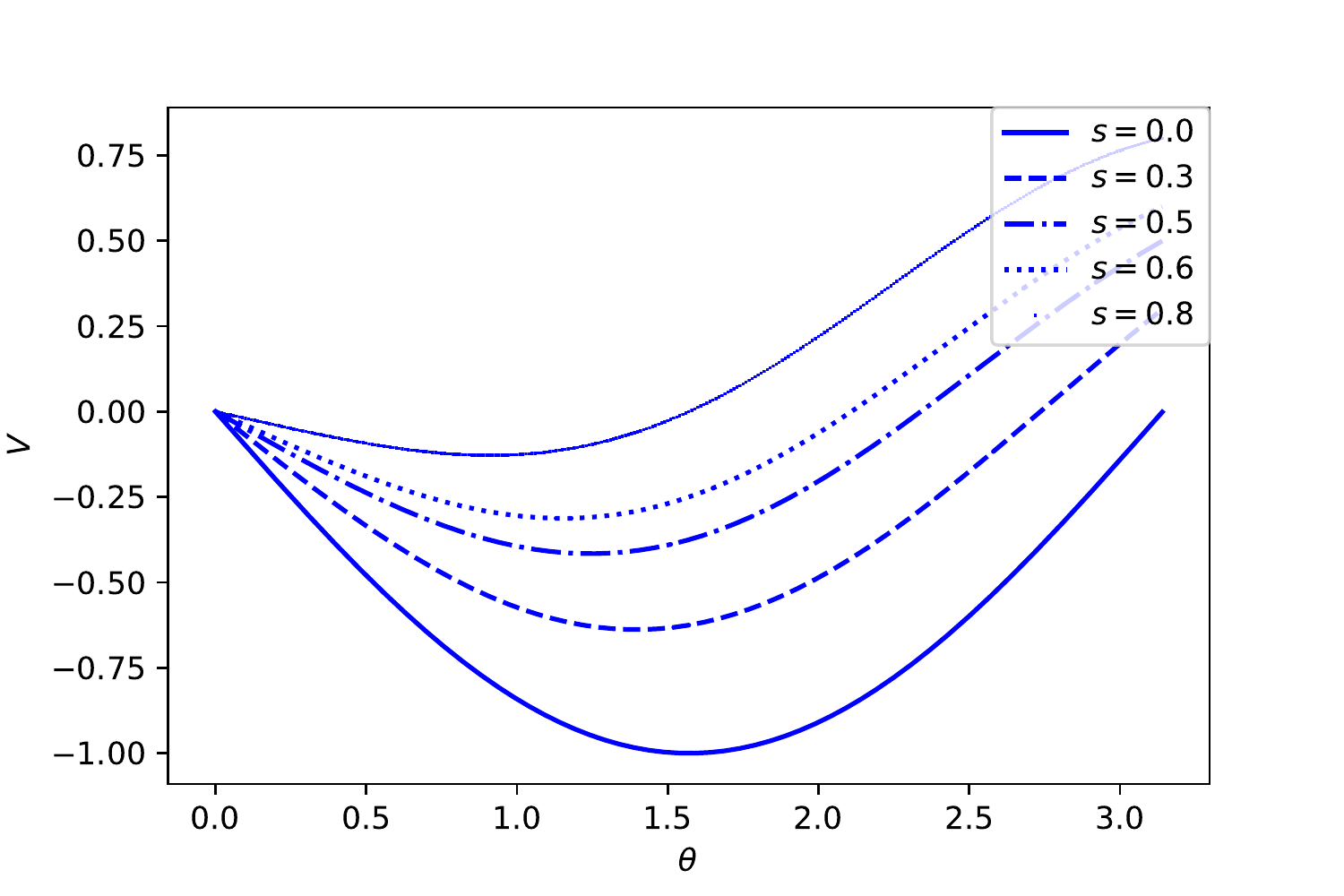}
\end{minipage}
\begin{minipage}{0.329\hsize}
\centering
    \includegraphics[width=\hsize]{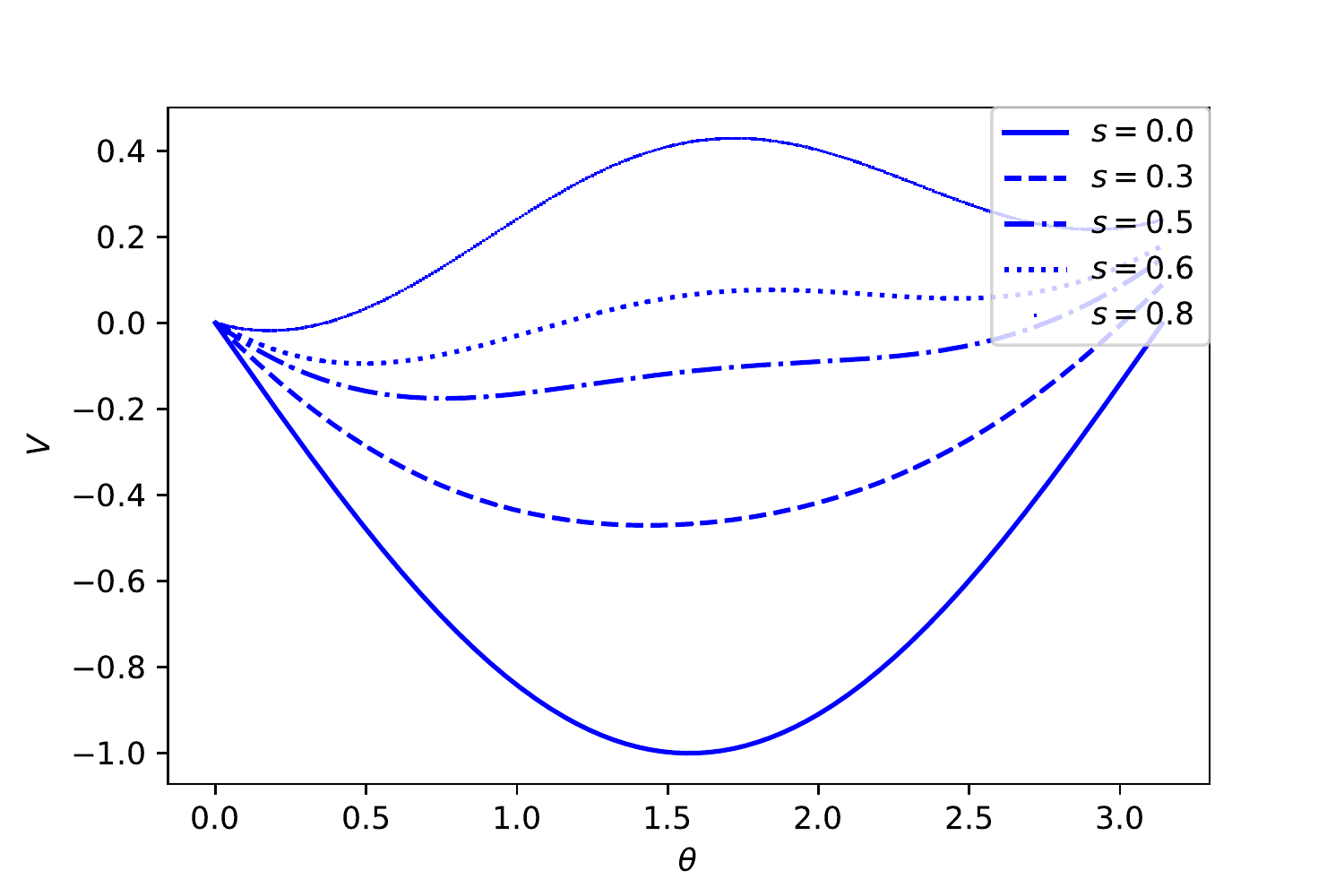}
\end{minipage}
\begin{minipage}{0.329\hsize}
\centering
    \includegraphics[width=\hsize]{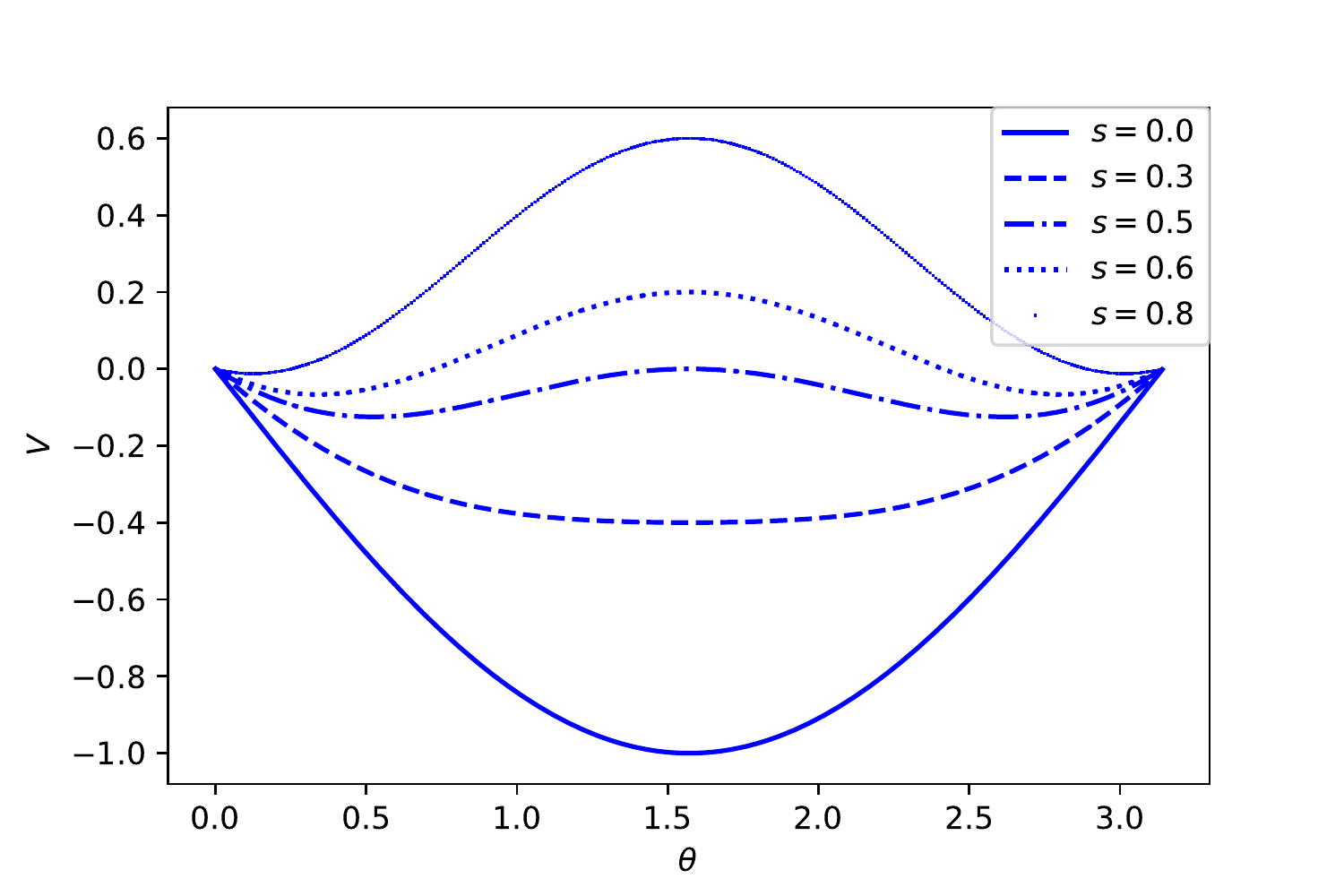}
\end{minipage}
\begin{minipage}{0.329\hsize}
\centering
    \includegraphics[width=\hsize]{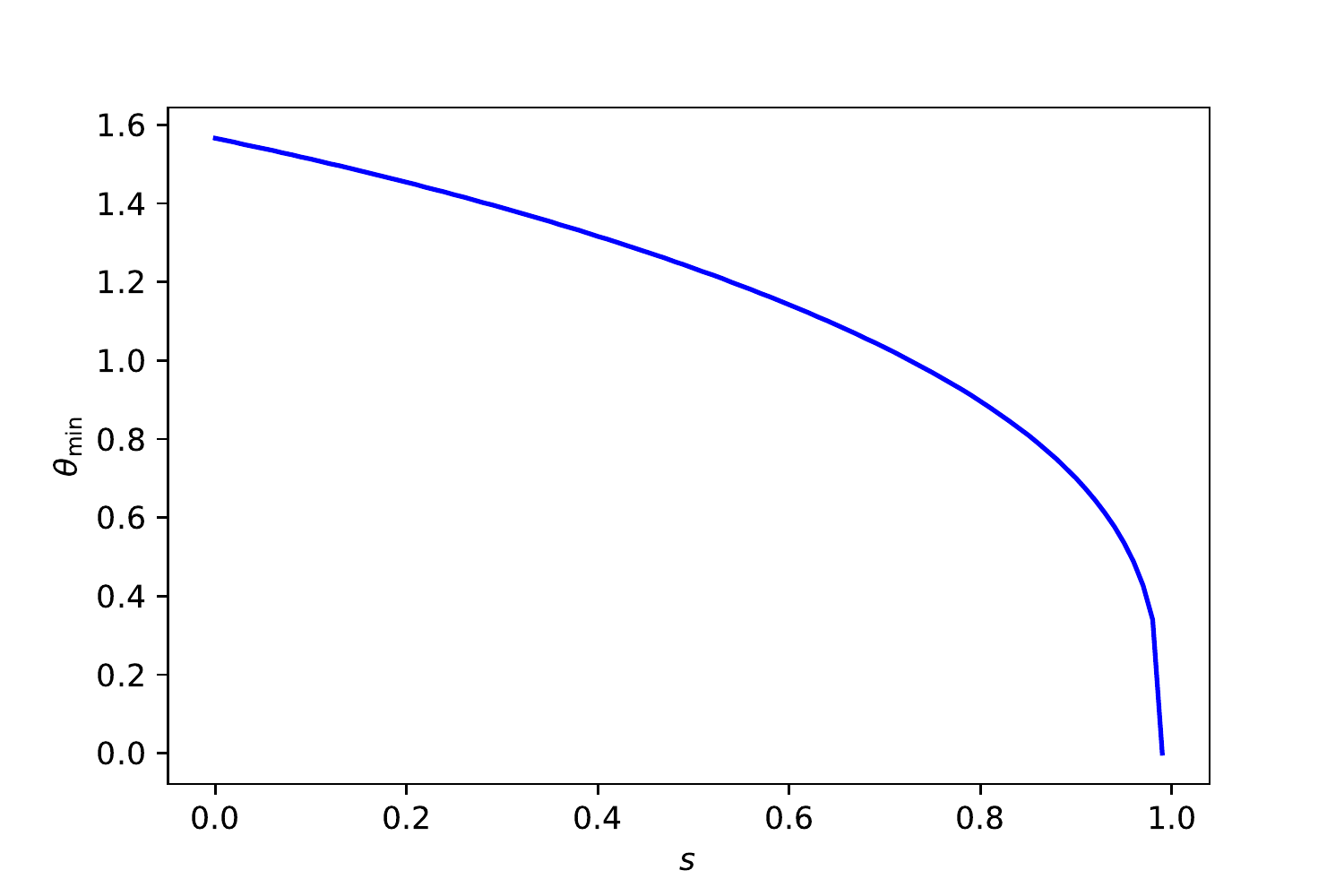}
\end{minipage}
\begin{minipage}{0.329\hsize}
\centering
    \includegraphics[width=\hsize]{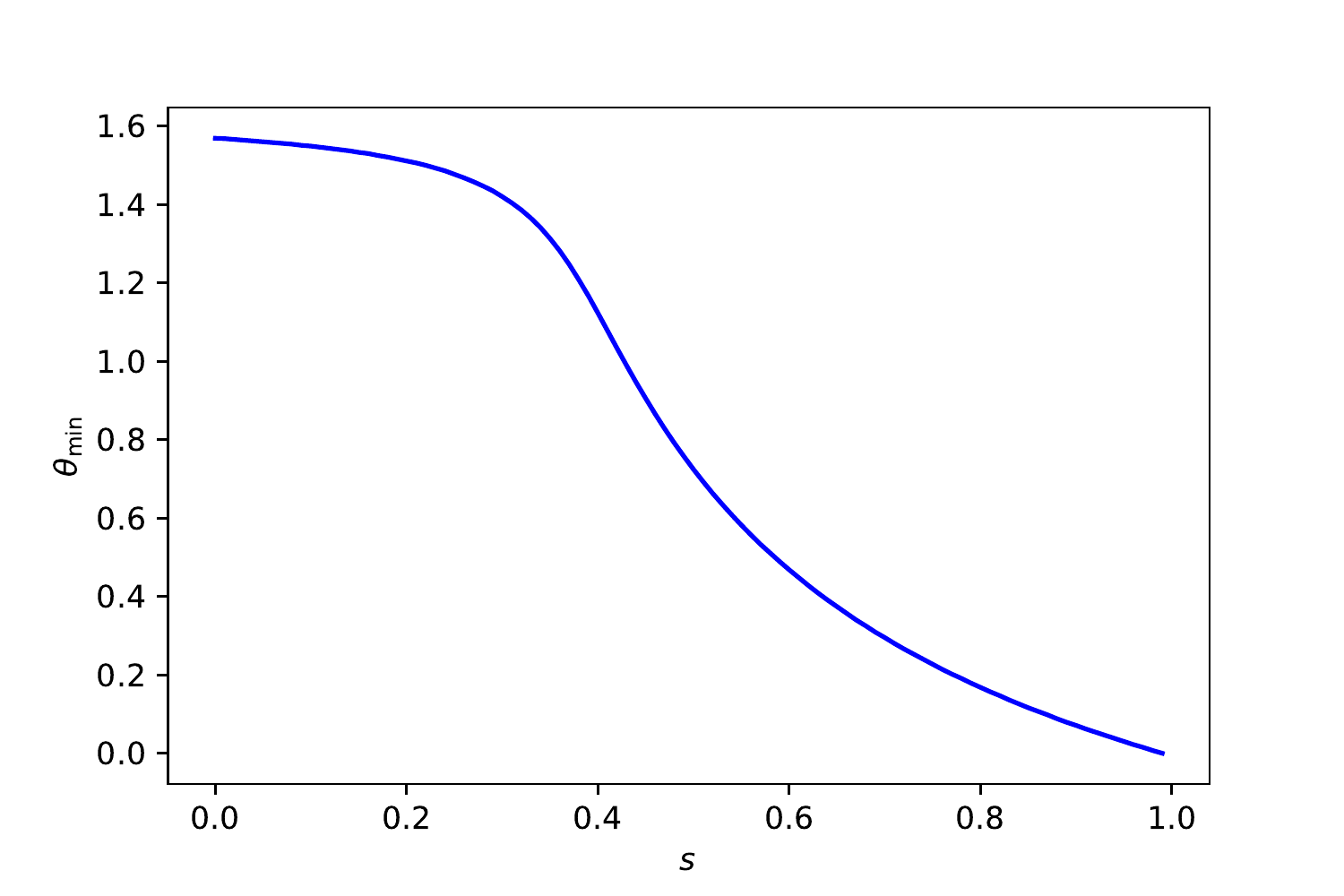}
\end{minipage}
\begin{minipage}{0.329\hsize}
\centering
    \includegraphics[width=\hsize]{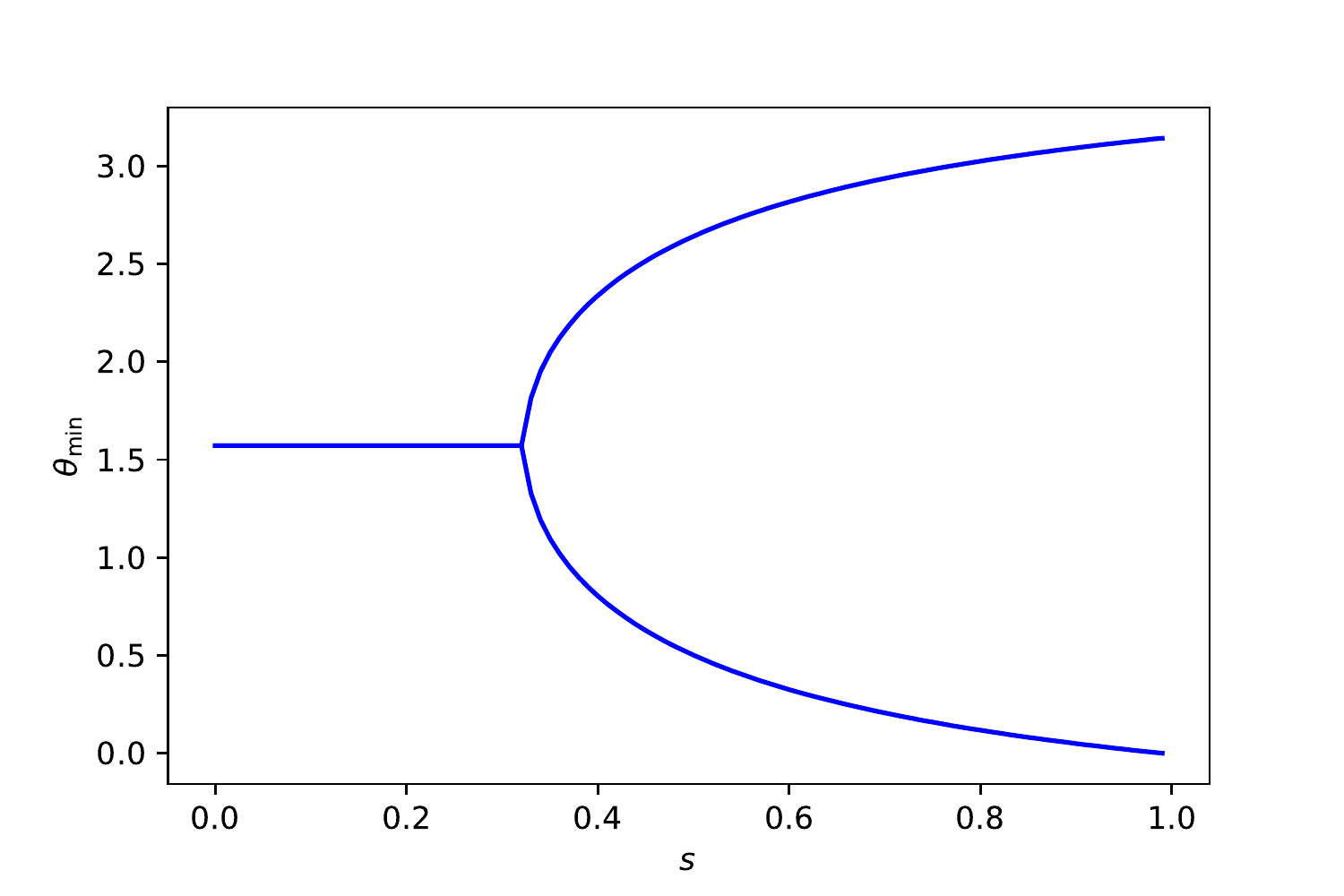}
\end{minipage}
\caption{[Left]$\lambda=1.0$, $h_{XX}=K=1.0$ [Middle] $\lambda=0.3$, $h_{XX}=K=1.0$ [Right] $\lambda=0.0$, $h_{XX}=K=1.0$. The upper figures show the $\theta$-dependence of the potential $V$ for several $s$. The lower figures show the $s$-dependence of $\theta_{\min}$.}
    \label{fig:pt2}
\end{figure}
}\fi

\if{
\section{Quantum Chaos}\label{sec:chaos}
Now let us study a little bit more on the dynamics of our model from a viewpoint of quantum chaos. Roughly, chaos would trigger phase transition, which may affect the probability of obtaining states. This motivates us to study quantum chaos. There are various approaches to quantum chaos, but its formal definition has been illusive. The most standard way is to define a quantum counterpart of classical chaos. Generally non-zero Lyapunov exponent is a crucial factor of classical chaos, hence to find its quantum counterpart is a main interest. Another characterization is level statistics. If a system is chaotic, level spacing distribution is approximated by a Wigner distribution $P(s)\approx s^\beta e^{-A s^2}$ \cite{PhysRevLett.52.1}, and if a system is classically integrable, it is a Poisson distribution $P(s)=e^{-s}$ \cite{doi:10.1098/rspa.1977.0140}. 

\if{
\subsection{$p$-spin ferromagnetic model}
Throughout this section, we work on the ferromagnetic $p$-spin model with X and XX interactions and employ the former perspective to analyze quantum chaos. More detailed analysis will be available elsewhere \cite{Ikeda2019_3}. Our Hamiltonian consists of 
\begin{align}\label{eq:SC}
\begin{aligned}
    H_0&=-N\left(\frac{1}{N}\sum_{i}^NZ_i\right)^p\\
    H_1&=-\sum_{i}^NX_i\\
    H_{2}&=+\frac{1}{N}\sum_{ij}^N h_{ij} X_iX_{j}
\end{aligned}
\end{align}
The Hamiltonian 
\begin{equation}
    H(s,\lambda)=s\lambda H_0+(1-s) H_1+s(1-\lambda)H_{2}
\end{equation}
is stoquastic if it is ferromagnetic $h_{ij}<0$ for all $i,j$ and non-stoquastic if $h_{ij}>0$ for some $i,j$. Let $\ket{\theta,\phi}$ be the spin coherent state 
\begin{equation}
    \ket{\theta,\phi}=\bigotimes_{i}^N\ket{\theta,\phi}_i
\end{equation}
where $\ket{\theta,\phi}_i=\cos(\theta/2)\ket{0}_i+e^{i\phi}\sin(\theta/2)\ket{1}_i$ with $\theta\in[0,\pi], \phi\in[0,2\pi]$. Using ${}_i\bra{\theta,\phi}X_i\ket{\theta,\phi}_i=\sin\theta\cos\phi$ and ${}_i\bra{\theta,\phi}Z_i\ket{\theta,\phi}_i=\cos\theta$, we find 
\begin{align}
    \begin{aligned}
        \bra{\theta,\phi}H_0\ket{\theta,\phi}&=-N\cos^p\theta+O(1)\\
         \bra{\theta,\phi}H_1\ket{\theta,\phi}&=-N\sin\theta\cos\phi\\
          \bra{\theta,\phi}H_{2}\ket{\theta,\phi}&=\frac{1}{N}\sum_{i}^Nh_{ii}+\frac{\sin^2\theta\cos^2\phi}{N}\sum_{i\neq j}h_{ij}\\
    \end{aligned}
\end{align}
The semi-classical potential $V(s,\lambda,\theta,\phi)$ is then defined by
\begin{equation}
    V(s,\lambda,\theta,\phi)=\frac{1}{N}\bra{\theta,\phi}H(\Gamma,\eta)\ket{\theta,\phi}. 
\end{equation}
For a large $N$, it is 
\begin{equation}
     V= -s\lambda \cos^p\theta-(1-s)\sin\theta\cos\phi+s(1-\lambda)\frac{\sin^2\theta\cos^2\phi}{N^2}\sum_{i\neq j}h_{ij}+O(1/N)
\end{equation}
It is easy to see that $V(\Gamma,\eta,\theta,0)\le V(\Gamma,\eta,\theta,\phi)$ for any $(\Gamma,\eta,\theta,\phi)$. So $\phi=0$ gives a ground state. We denote by $\theta_{\min}$ at which $V(\Gamma,\eta,\theta_{\min},\phi) \le V(\Gamma,\eta,\theta,\phi)$ for all $\theta$. 
The first-order phase transition occurs when $V$ is discontinuous with respect to $\theta_{\min}$. A second-oredr phase transition occurs when they satisfy
\begin{equation}
    \frac{\partial^2 V}{\partial \theta^2}\bigg|_{\theta=\frac{\pi}{2},\phi=0}=0\Leftrightarrow (1-s)-2s(1-\lambda)\frac{1}{N}\sum_{i\neq j}h_{ij}=0. 
\end{equation}

We will use $s(t)=1-\Gamma(t)$ and $\lambda(t)=s(t)^{q}$ with $q>0$. 
\begin{figure}
    \centering
    \includegraphics{SC1.pdf}
    \caption{Caption}
    \label{fig:my_label}
\end{figure}

}\fi

\subsection{Quantum Lyapunov Exponents}\label{sec:QLX}
We use the Hamiltonian 
\begin{equation}\label{eq:QA}
    H(t)=(1-\Gamma(t))H_0+\Gamma(t)H_1
\end{equation}
and the Schr\"{o}dinger equation 
\begin{equation}
    i\frac{d}{dt}U(t)=H(t)U(t). 
\end{equation}
of a unitary operator $U(t)$ that describes the time evolution of an operator $\mathcal{O}(t)=U^\dagger(t)\mathcal{O}U(t)$. We first consider time evolution of our operators $a_i(t),a_i^\dagger(t)~(i=1,\cdots,L)$ and define matrix by
\begin{equation}
    M_{ij}(t)=[a_i(t),a^\dagger_j(0)]. 
\end{equation}
At $t=0$, this corresponds to $Z_{i}\delta_{ij}$ by definition. Let $\ket{\phi}$ be a normalized state of a given Hamiltonian. Then with respect to $\ket{v_{ij}(t)}=M_{ij}(t)\ket{\phi}$, we define the $L^\phi$-matrix \cite{Gharibyan:2018fax} as follows:
\begin{equation}
    L^\phi_{ij}(t)=\sum_{k=1}^L\bra{v_{ki}(t)}v_{kj}(t)\rangle. 
\end{equation}
It is easy to see that $L^\phi_{ij}(0)=\delta_{ij}$ for any $\ket{\phi}$. The quantum Lyapunov exponents $\lambda_i$ are defined by
\begin{equation}\label{eq:L}
    \frac{1}{L}\text{Tr}L^\phi(t)=\sum_{i}^Le^{2\lambda_i(t)}, 
\end{equation}
which is 1 at $t=0$. If a system is chaotic, the function \eqref{eq:L} is expected to show exponential growth $e^{\lambda t}$ around $t=0$.  In this sense it is thought to be a quantum generalization of the spectrum of classical Lyapunov exponents.

In Fig. \ref{fig:L} we show $\frac{1}{L}\text{Tr}L^\phi(t)$ for a time-dependent case and for a constant case. All figures exhibit exponential growth around $t=0$. In both cases, the bigger $h_{XX}$ becomes, the more chaotic the systems get. The classical spin chain system with $H_0$ is an integrable model and the system gets chaotic when the quantum fluctuation term $H_1$ is introduced. Therefore as time passes, the time-dependent system experiences the chaotic/integrable transition.
  
\begin{figure}[H]
\begin{minipage}{0.329\hsize}
\centering
    \includegraphics[width=\hsize]{L_exp_05.pdf}
\end{minipage}
\begin{minipage}{0.329\hsize}
\centering
    \includegraphics[width=\hsize]{L_exp_03.pdf}
\end{minipage}
\begin{minipage}{0.329\hsize}
\centering
    \includegraphics[width=\hsize]{L_exp_01.pdf}
\end{minipage}
\caption{Plots of $\frac{1}{L}\text{Tr}L^\phi(t)$. The ground state of $H_0$ is used for $\ket{\phi}$ and $\Gamma(t)=\exp(-at)$ is used for the schedule. [Left] $a=0.5$ [Middle] $a=0.3$ [Right] $a=0.1$}
    \label{fig:L}
\end{figure}

\subsection{Out-of-Time-Ordered Correlator}\label{sec:OTOC}    
Another quantity which is expected as a good measure of quantum chaos is the OTOC (out-of-time-ordered correlator) \cite{larkin1969quasiclassical,Maldacena:2015waa,Kitaev14}. Here we consider the OTOC 
\begin{equation}
C(i,j,t)=\frac{1}{2^N}\text{Tr}(M_{ij}(t)^\dagger M_{ij}(t))
\end{equation}
to investigate quantum chaos of our model. The OTOC of a quantum chaotic system is expected to show exponential growth $e^{\lambda t}$ around $t=0$ and $\lambda$ would correspond to the classical Lyapunov exponent. We use $[a_i(t), a^\dagger_j(0)]$ and $[X_i(t), X_j(0)]$ for $M_{ij}(t)$. In what follows, we address the OTOC of the time-dependent Hamiltonian \eqref{eq:QA}. Some relations with quantum chaos and Ising model with constant $\Gamma(t)$ have been studied by many authors, however less is known for a system with a non-constant $\Gamma(t)$. Especially, this work is the first one which explores the OTOC with the quantum-annealing Hamiltonian \eqref{eq:QA}.

We show the parameter dependence of the OTOC with $M_{ij}(t)=[a_i(t),a^\dagger_j(0)]$ in Fig. \ref{fig:OTOC1}, and with $M_{ij}(t)=[X_i(t),X_j(0)]$ in Fig. \ref{fig:OTOC3}. As expected, exponential growth around $t=0$ is observed in each plot and some basic trends of the parameter dependence are consistent with the previous study on quantum Lyapunov exponents in Sec. \ref{sec:QLX}. 

\begin{figure}[H]
\begin{minipage}{0.5\hsize}
    \centering
    \includegraphics[width=\hsize]{OTOC_exp_01_lamda=1.pdf}
\end{minipage}
\begin{minipage}{0.5\hsize}
    \centering
    \includegraphics[width=\hsize]{OTOC_exp_01_lambda=06.pdf}
\end{minipage}
    \caption{OTOC of the Hamiltonian $H(t)$ with $M_{ij}=[X_i(t),X_j(0)]$ with $\Gamma(t)=\exp(-t/10)$. [Left] $\lambda=0.6$  [Right] non-stoquastic cases. }
    \label{fig:OTOC1}
\end{figure}

\begin{figure}[H]
\begin{minipage}{0.5\hsize}
    \centering
    \includegraphics[width=\hsize]{OTOC_c_exp_01.pdf}
\end{minipage}
\begin{minipage}{0.5\hsize}
    \centering
    \includegraphics[width=\hsize]{OTOC_c2_exp_01.pdf}
\end{minipage}
    \caption{Early-time growth in the OTOC of the Hamiltonian $H(t)$ with $M_{ij}=[a_i(t),a^\dagger_j(0)]$ and $\Gamma(t)=\exp(-t/10)$. [Left] stoquastic cases.  [Right] non-stoquastic cases.}
    \label{fig:OTOC3}
\end{figure}

\subsection{Relation to Performance of Computation}
Lastly we leave some short comments on how quantum chaos could affect the accuracy of computation. Let $\{\lambda_i,\ket{\phi_i(0)}\}$ be a normalized eigensystem at $t=0$. We consider the density matrix   
\begin{equation}
\rho(t)=\sum \lambda_i\ket{\phi_i(t)}\bra{\phi_i(t)},  
\end{equation}
where $\ket{\phi_i(t)}=U(t)\ket{\phi_i(0)}$. The probability of obtaining a state $\ket{\psi}$ at $t$ is given by 
\begin{equation}
    \text{Tr}(\ket{\psi}\bra{\psi}\rho(t))=\sum \lambda_i|\langle \psi\ket{\phi_i(t)}|^2. 
\end{equation}
With respect to the density matrix $\rho(t)$, we again consider the correlator 
\begin{align}
\begin{aligned}
    C(t)=&\bra{\psi}[\rho(t),\rho(0)]^\dagger [\rho(t),\rho(0)]\ket{\psi}\\
    =&2(1-\text{Re}\bra{\psi}\rho(t)\rho(0)\rho(t)\rho(0)\ket{\psi}). 
\end{aligned}
\end{align}
By definition $C(0)=0$ and $0\le C(t)\le 2$ for any $t$. 

We work on the spin chain Hamiltonian \eqref{eq:SC}. The probability of finding the ground state of $H_0$ is shown in Fig. \ref{fig:Prob_SC}. According to the figure, the smaller $h_{XX}$ becomes, the higher the probability gets overall. So comparing the results in Sec. \ref{sec:QLX} and \ref{sec:OTOC}, we find a strong correlation between quantum chaos and the probability of finding the ground states by adiabatic quantum computation. Namely, the ground state can be observed frequently in a less chaotic system. As previously mentioned in Sec. \ref{sec:ex}, the average probability approaches to 1 after a sufficiently long time.

\begin{figure}[H]
    \centering
    \includegraphics[width=\hsize]{Prob_SC_exp_01.pdf}
    \caption{The probability of finding the ground state of the spin chain model with $\Gamma(t)=\exp(-t/10)$.}
    \label{fig:Prob_SC}
\end{figure}
}\fi

\section{Conclusion and Future Directions}\label{sec:conc}
In this work we proposed several new techniques. We first introduced particles which cannot occupy the same position simultaneously and are symmetric under exchange of particle labels. It would request further investigations to clarify whether those particles do exist in nature. If they existed in a physical form, would they be fundamental particles? Even if they do not exist in nature or laboratories, for sure they are programmable and contribute to universal quantum computation as we show in Sec. \ref{sec:qfc}. Those programmable particles are relatively easy to understand and handle. One does not need to worry about particle labels and can write an algorithm just by creating or annihilating them. Therefore they could become a useful tool to develop some quantum programming language. In general, quantum computation does not always require knowledge of quantum physics, hence future quantum programming languages could be written in an unphysical manner. Functional usability of a language may get preference over rigorous theoretical aspects. In Sec. \ref{sec:ex} we addressed the two dimensional Bloch electron system without defects as an example. The conventional works on adiabatic quantum computation mostly address combinatorial optimization problems, but we showed it is also powerful enough to simulate quantum physics in our own way. There are many possible further research directions. For example, it would be a good exercise to simulate dynamics of systems with defects \cite{Matsuki_2019}. Moreover in principle, such dynamics can be simulated with a general Ising model with XX interactions. It would be interesting to implement and study it with a super conducting qubit system, though the current version of the quantum annealer \cite{Johnson2011,Troels14} allows us to tune only real number couplings and the transverse field $X$. In Sec. \ref{sec:PT} we studied phase transitions associated with AQC. With Majorana fermions and multiple particles we showed quantum speedup can be achieved by a non-stoquastic Hamiltonian. It will be also interesting to explore more on the novel Majorana fermion system \eqref{eq:HMF} we provided in this article.   

So far we have discussed computation on a discrete space. Now let us extend it to a theory on a connected space. Something unusual is the commutation relation of the creation and annihilation operators. Ours are neither bosonic nor fermionic. However, since the "particles" we have addressed do not have spins, they must obey the bononic commutation relation, otherwise causality should be broken. Indeed our formulation barely clears up this problem since they obey the boconic commutation relations almost everywhere: they do commute $[a_{x_i},a^\dagger_{x_j}]=0$ at different positions. Moreover, it is straightforward to generalize Theorem \ref{thm:universal} to a version on a connected space, hence any Hamiltonian can be reconstructed by $\{a_{x_i},a^\dagger_{x_i}\}_{i=1,2\cdots}$. As long as a Hamiltonian is Hermitian, any dynamical process is unitary, hence it does not cause any problem on the probability interpretation. Therefore apparently it could be possible to approximate the standard quantum field theory by such unusual creation and annihilation operators. They act on $\mathcal{H}_x=\{ \alpha\ket{1_x}+\beta\ket{0_x}:~|\alpha|^2+|\beta|^2=1, \alpha,\beta\in\mathbb{C}\}$ as $a^\dagger_x\ket{1_x}=0,a^\dagger_x\ket{0_x}=\ket{1_x},a_x\ket{0_x}=0,a_x\ket{1_x}=\ket{0_x}, \{a_x,a^\dagger_x\}=1_x$ and satisfy 
\begin{equation}
    [a_x,a^\dagger_y]=Z_x\delta(x-y),~[a_x,a_y]=[a^\dagger_x,a^\dagger_y]=0, 
\end{equation}
where $Z_x$ acts on states as $Z_x\ket{1_y}=-\delta(x-y)\ket{1_y}$ and $Z_x\ket{0_y}=\delta(x-y)\ket{0_y}$. The creation and annihilation operators describe particles which cannot occupy the same position simultaneously and particle labels are indistinguishable. It is an interesting open question to reconstruct QFT with those operators.

Furthermore, it is also a quite new and interesting direction to investigate quantum chaotic behavior of our system. Quantum chaos could be somehow related with quantum phase transition \cite{2018arXiv181111191S,2018arXiv181111191S}, hence it may have some effects on quantum computation. There are various approaches to quantum chaos, but its formal definition has been illusive. A characterization is done by level statistics. If a system is chaotic, level spacing distribution is approximated by a Wigner distribution $P(\epsilon)\approx \epsilon^\beta e^{-A \epsilon^2}$ \cite{PhysRevLett.52.1}, and if a system is classically integrable, it is a Poisson distribution $P(\epsilon)=e^{-\epsilon}$ \cite{doi:10.1098/rspa.1977.0140}. Another standard way is to define a quantum counterpart of classical chaos. Generally non-zero Lyapunov exponent is a crucial factor of classical chaos, hence to find its quantum counterpart is a main interest. A quantity which is expected as a good measure of quantum chaos is the OTOC (out-of-time-ordered correlator). In this work, we investigated the time average of the OTOC in order to diagnose quantum phase transition. Those phenomena should be checked with other models.

\if{Here we consider the OTOC 
\begin{equation}
C(i,j,t)=\frac{1}{2^N}\text{Tr}(M_{ij}(t)^\dagger M_{ij}(t))
\end{equation}
to investigate quantum chaos of our model. The OTOC of a quantum chaotic system is expected to show exponential growth $e^{\lambda_\text{OTOC} t}$ around $t=0$ and $\lambda_\text{OTOC}$ would correspond to the classical Lyapunov exponent. We use $[X_i(t), X_j(0)]$ for $M_{ij}(t)$. Fig. \ref{fig:OTOC} is the OTOC of the time-dependent Hamiltonian \eqref{eq:QA2} with the $p$-particle Hamiltonian \eqref{eq:pp} for $H_0$. We use $s(t)=t/100$ and $\lambda=s(t)^q$ for $t\in[0,100]$, hence both $s$ and $\lambda$ are defined over $[0,1]$. As shown in the figure, a first-order phase transition is avoided if $q=2.0$. $\lambda(t)$ controls the strength of the non-stoquastic term, hence growth rate in the OTOC would be somehow related to the $XX$-interactions. In this example, one may say that the larger the quantum Lyapunov exponent is, the higher the probability of obtaining the ground state becomes. It will be interesting to investigate more on quantum computation in terms of quantum chaos. For example, it is as open to investigate quantum chaos and probability of finding the ground state of the Hamiltonian. While there are many researches which associate first-order phase transitions with the performance of the adiabatic quantum computation, our results suggest that quantum chaos could be another factor to be taken into account. 

\begin{figure}[H]
\begin{minipage}{0.5\hsize}
    \centering
    \includegraphics[width=\hsize]{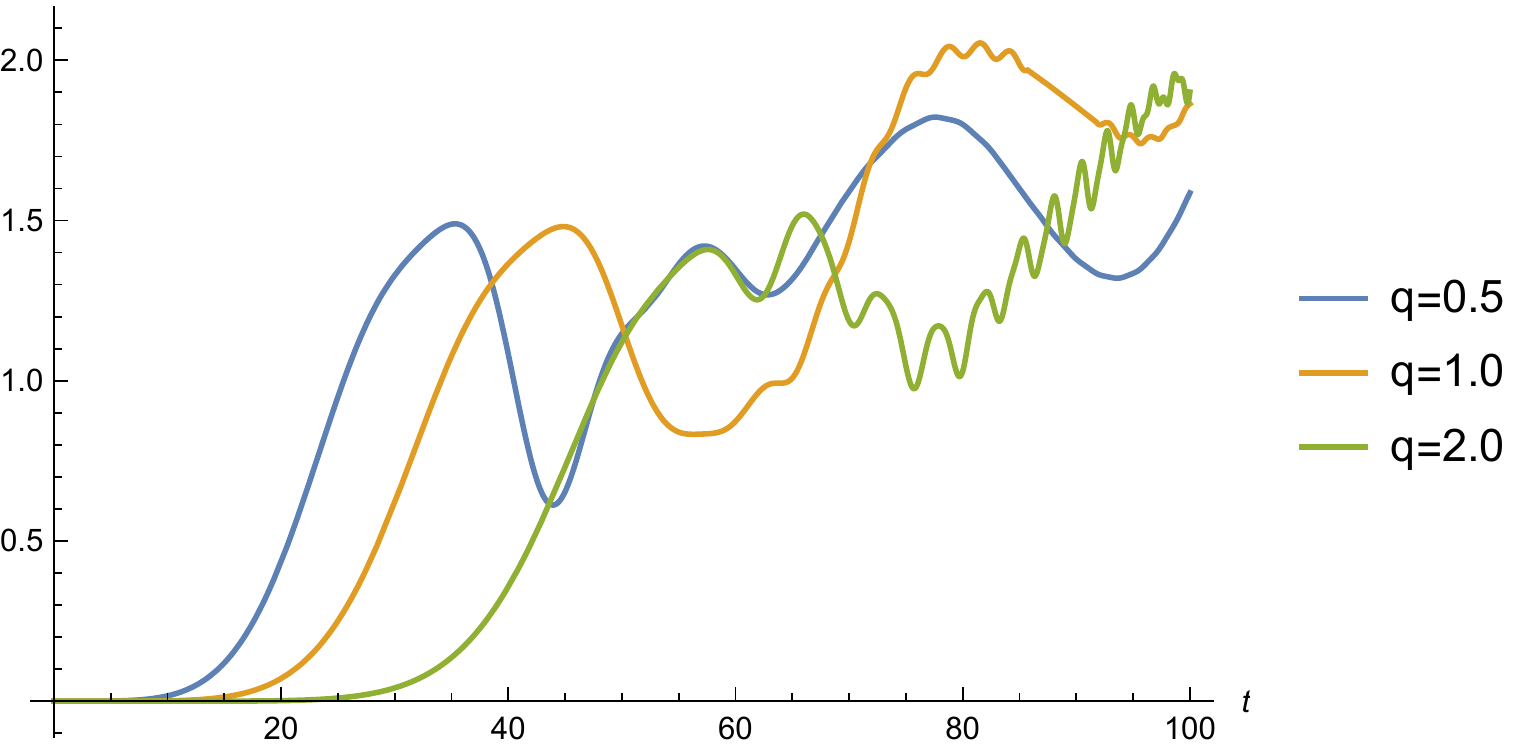}
\end{minipage}
\begin{minipage}{0.5\hsize}
    \centering
    \includegraphics[width=\hsize]{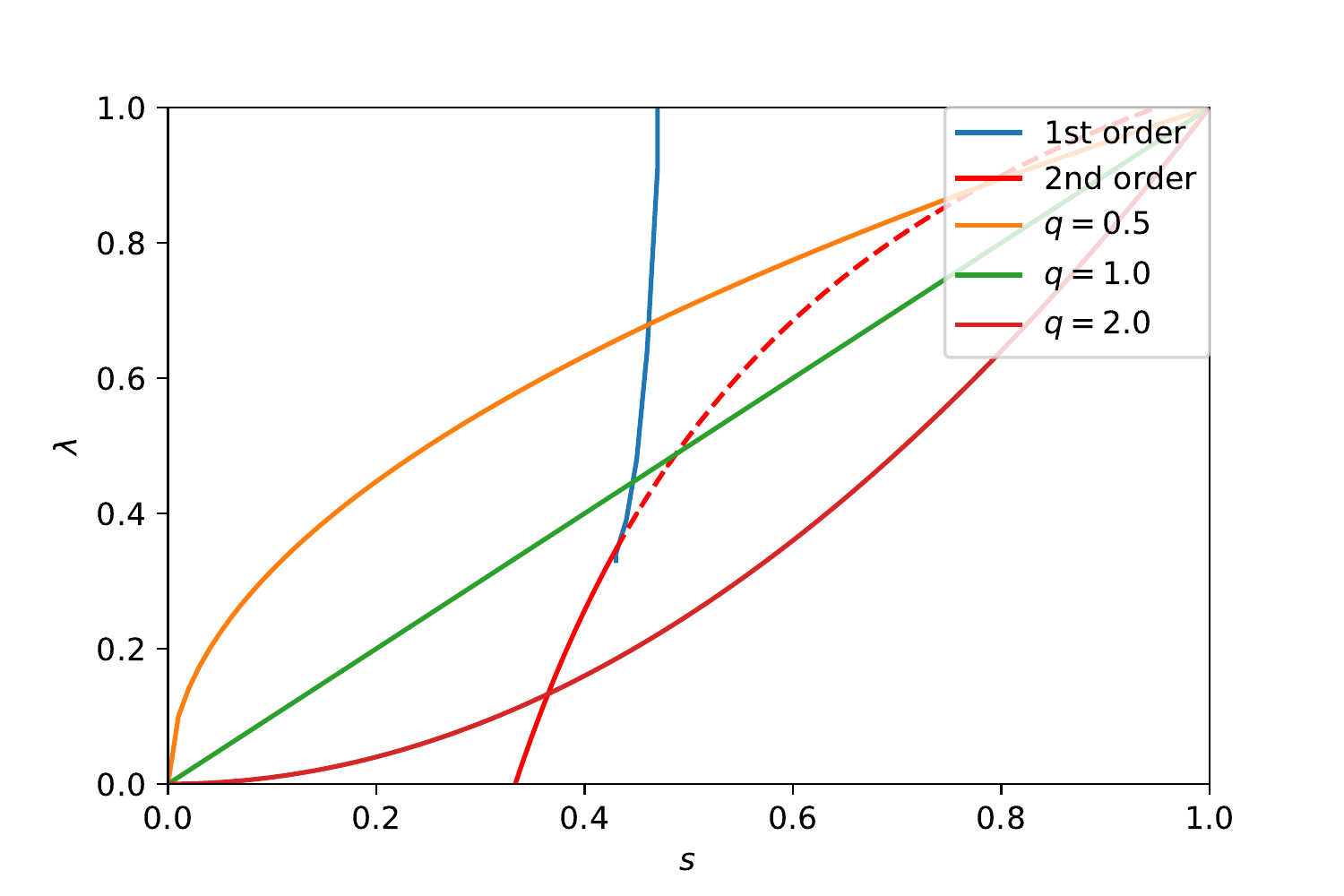}
\end{minipage}
    \caption{The average of OTOCs over all possible combinations of cites, computed with the non-stoqaustic Hamiltonian \eqref{eq:QA2} and \eqref{eq:pp}. We used the annealing schedule $s(t)=t/100$ and $\lambda(t)=s(t)^q$. }
    \label{fig:OTOC}
\end{figure}
}\fi

\section*{Acknowledgements}
I am grateful to Katsuya Hashino, Viktor Jahnke and Kin-ya Oda for stimulating discussion and useful comments on the draft. The author was partly supported by Grant-in-Aid for JSPS Research Fellow, No. 19J11073. 
\bibliographystyle{JHEP}
\bibliography{ref}

\providecommand{\href}[2]{#2}\begingroup\raggedright\begin{thebibliography}{10}

\bibitem{turing2009computing}
A.~M. Turing, \emph{Computing machinery and intelligence},  in \emph{Parsing
  the Turing Test}, pp.~23--65, Springer, (2009).

\bibitem{Feynman1986}
R.~P. Feynman, \emph{Quantum mechanical computers},
  \href{https://doi.org/10.1007/BF01886518}{\emph{Foundations of Physics}
  {\bfseries 16} (1986) 507}.

\bibitem{doi:10.1098/rspa.1985.0070}
D.~Deutsch, \emph{Quantum theory, the church-turing principle and the universal
  quantum computer},
  \href{https://doi.org/10.1098/rspa.1985.0070}{\emph{Proceedings of the Royal
  Society of London. A. Mathematical and Physical Sciences} {\bfseries 400}
  (1985) 97}
  [\href{https://arxiv.org/abs/https://royalsocietypublishing.org/doi/pdf/10.1098/rspa.1985.0070}{{\ttfamily
  https://royalsocietypublishing.org/doi/pdf/10.1098/rspa.1985.0070}}].

\bibitem{church1936unsolvable}
A.~Church, \emph{An unsolvable problem of elementary number theory},
  {\emph{American journal of mathematics} {\bfseries 58} (1936) 345}.

\bibitem{doi:10.1112/plms/s2-42.1.230}
A.~M. Turing, \emph{On computable numbers, with an application to the
  entscheidungsproblem},
  \href{https://doi.org/10.1112/plms/s2-42.1.230}{\emph{Proceedings of the
  London Mathematical Society} {\bfseries s2-42} (1937) 230}.

\bibitem{PhysRevLett.79.2586}
D.~S. Abrams and S.~Lloyd, \emph{Simulation of many-body fermi systems on a
  universal quantum computer},
  \href{https://doi.org/10.1103/PhysRevLett.79.2586}{\emph{Phys. Rev. Lett.}
  {\bfseries 79} (1997) 2586}.

\bibitem{berry2007efficient}
D.~W. Berry, G.~Ahokas, R.~Cleve and B.~C. Sanders, \emph{Efficient quantum
  algorithms for simulating sparse hamiltonians}, {\emph{Communications in
  Mathematical Physics} {\bfseries 270} (2007) 359}.

\bibitem{1998RSPSA.454..313Z}
C.~{Zalka}, \emph{{Simulating quantum systems on a quantum computer}},
  \href{https://doi.org/10.1098/rspa.1998.0162}{\emph{Proceedings of the Royal
  Society of London Series A} {\bfseries 454} (1998) 313}
  [\href{https://arxiv.org/abs/quant-ph/9603026}{{\ttfamily
  quant-ph/9603026}}].

\bibitem{Jordan:2011ne}
S.~P. Jordan, K.~S.~M. Lee and J.~Preskill, \emph{{Quantum Algorithms for
  Quantum Field Theories}},
  \href{https://doi.org/10.1126/science.1217069}{\emph{Science} {\bfseries 336}
  (2012) 1130} [\href{https://arxiv.org/abs/1111.3633}{{\ttfamily 1111.3633}}].

\bibitem{Jordan2018bqpcompletenessof}
S.~P. Jordan, H.~Krovi, K.~S.~M. Lee and J.~Preskill, \emph{{BQP}-completeness
  of scattering in scalar quantum field theory},
  \href{https://doi.org/10.22331/q-2018-01-08-44}{\emph{{Quantum}} {\bfseries
  2} (2018) 44}.

\bibitem{2000quant.ph..1106F}
E.~{Farhi}, J.~{Goldstone}, S.~{Gutmann} and M.~{Sipser}, \emph{{Quantum
  Computation by Adiabatic Evolution}}, {\emph{arXiv e-prints} (2000) quant}
  [\href{https://arxiv.org/abs/quant-ph/0001106}{{\ttfamily
  quant-ph/0001106}}].

\bibitem{RevModPhys.90.015002}
T.~Albash and D.~A. Lidar, \emph{Adiabatic quantum computation},
  \href{https://doi.org/10.1103/RevModPhys.90.015002}{\emph{Rev. Mod. Phys.}
  {\bfseries 90} (2018) 015002}.

\bibitem{PhysRevE.58.5355}
T.~Kadowaki and H.~Nishimori, \emph{Quantum annealing in the transverse ising
  model}, \href{https://doi.org/10.1103/PhysRevE.58.5355}{\emph{Phys. Rev. E}
  {\bfseries 58} (1998) 5355}.

\bibitem{Johnson2011}
M.~W. Johnson, M.~H.~S. Amin, S.~Gildert, T.~Lanting, F.~Hamze, N.~Dickson
  et~al., \emph{Quantum annealing with manufactured spins}, {\emph{Nature}
  {\bfseries 473} (2011) 194 EP }.

\bibitem{Troels14}
T.~F. R{\o}nnow, Z.~Wang, J.~Job, S.~Boixo, S.~V. Isakov, D.~Wecker et~al.,
  \emph{Defining and detecting quantum speedup},
  \href{https://doi.org/10.1126/science.1252319}{\emph{Science} {\bfseries 345}
  (2014) 420}.

\bibitem{10.3389/fphy.2014.00005}
A.~Lucas, \emph{Ising formulations of many np problems},
  \href{https://doi.org/10.3389/fphy.2014.00005}{\emph{Frontiers in Physics}
  {\bfseries 2} (2014) 5}.

\bibitem{ikeda2019NSP}
K.~Ikeda, Y.~Nakamura and T.~S. Humble, \emph{Application of quantum annealing
  to nurse scheduling problem},
  \href{https://doi.org/10.1038/s41598-019-49172-3}{\emph{Scientific Reports}
  {\bfseries 9} (2019) 12837}.

\bibitem{Feynman:85}
R.~P. Feynman, \emph{Quantum mechanical computers},
  \href{https://doi.org/10.1364/ON.11.2.000011}{\emph{Optics News} {\bfseries
  11} (1985) 11}.

\bibitem{PhysRevA.78.012352}
J.~D. Biamonte and P.~J. Love, \emph{Realizable hamiltonians for universal
  adiabatic quantum computers},
  \href{https://doi.org/10.1103/PhysRevA.78.012352}{\emph{Phys. Rev. A}
  {\bfseries 78} (2008) 012352}.

\bibitem{2008RSPSA.464.3089J}
R.~{Jozsa} and A.~{Miyake}, \emph{{Matchgates and classical simulation of
  quantum circuits}},
  \href{https://doi.org/10.1098/rspa.2008.0189}{\emph{Proceedings of the Royal
  Society of London Series A} {\bfseries 464} (2008) 3089}
  [\href{https://arxiv.org/abs/0804.4050}{{\ttfamily 0804.4050}}].

\bibitem{Jordan1928}
P.~Jordan and E.~Wigner, \emph{{\"U}ber das paulische {\"a}quivalenzverbot},
  \href{https://doi.org/10.1007/BF01331938}{\emph{Zeitschrift f{\"u}r Physik}
  {\bfseries 47} (1928) 631}.

\bibitem{doi:10.1063/1.4998635}
K.~Ikeda, \emph{Hofstadter's butterfly and langlands duality},
  \href{https://doi.org/10.1063/1.4998635}{\emph{Journal of Mathematical
  Physics} {\bfseries 59} (2018) 061704}
  [\href{https://arxiv.org/abs/https://doi.org/10.1063/1.4998635}{{\ttfamily
  https://doi.org/10.1063/1.4998635}}].

\bibitem{Ikeda:2017uce}
K.~Ikeda, \emph{{Quantum Hall Effect and Langlands Program}},
  \href{https://doi.org/10.1016/j.aop.2018.08.002}{\emph{Annals Phys.}
  {\bfseries 397} (2018) 136}
  [\href{https://arxiv.org/abs/1708.00419}{{\ttfamily 1708.00419}}].

\bibitem{Ikeda:2018tlz}
K.~Ikeda, \emph{{Topological Aspects of Matters and Langlands Program}},
  \href{https://arxiv.org/abs/1812.11879}{{\ttfamily 1812.11879}}.

\bibitem{PhysRevB.14.2239}
D.~R. Hofstadter, \emph{Energy levels and wave functions of bloch electrons in
  rational and irrational magnetic fields},
  \href{https://doi.org/10.1103/PhysRevB.14.2239}{\emph{Phys. Rev. B}
  {\bfseries 14} (1976) 2239}.

\bibitem{Hatsuda:2016mdw}
Y.~Hatsuda, H.~Katsura and Y.~Tachikawa, \emph{{Hofstadter's butterfly in
  quantum geometry}},
  \href{https://doi.org/10.1088/1367-2630/18/10/103023}{\emph{New J. Phys.}
  {\bfseries 18} (2016) 103023}
  [\href{https://arxiv.org/abs/1606.01894}{{\ttfamily 1606.01894}}].

\bibitem{2012PhRvE..85e1112S}
Y.~{Seki} and H.~{Nishimori}, \emph{{Quantum annealing with antiferromagnetic
  fluctuations}}, \href{https://doi.org/10.1103/PhysRevE.85.051112}{\emph{Phys.
  Rev. E} {\bfseries 85} (2012) 051112}
  [\href{https://arxiv.org/abs/1203.2418}{{\ttfamily 1203.2418}}].

\bibitem{Damski_2013}
B.~Damski and M.~M. Rams, \emph{Exact results for fidelity susceptibility of
  the quantum ising model: the interplay between parity, system size, and
  magnetic field},
  \href{https://doi.org/10.1088/1751-8113/47/2/025303}{\emph{Journal of Physics
  A: Mathematical and Theoretical} {\bfseries 47} (2013) 025303}.

\bibitem{PhysRevB.71.224420}
S.~Dusuel and J.~Vidal, \emph{Continuous unitary transformations and
  finite-size scaling exponents in the lipkin-meshkov-glick model},
  \href{https://doi.org/10.1103/PhysRevB.71.224420}{\emph{Phys. Rev. B}
  {\bfseries 71} (2005) 224420}.

\bibitem{PhysRevA.95.042321}
Y.~Susa, J.~F. Jadebeck and H.~Nishimori, \emph{Relation between quantum
  fluctuations and the performance enhancement of quantum annealing in a
  nonstoquastic hamiltonian},
  \href{https://doi.org/10.1103/PhysRevA.95.042321}{\emph{Phys. Rev. A}
  {\bfseries 95} (2017) 042321}.

\bibitem{larkin1969quasiclassical}
A.~Larkin and Y.~N. Ovchinnikov, \emph{Quasiclassical method in the theory of
  superconductivity}, {\emph{Sov Phys JETP} {\bfseries 28} (1969) 1200}.

\bibitem{Maldacena:2015waa}
J.~Maldacena, S.~H. Shenker and D.~Stanford, \emph{{A bound on chaos}},
  \href{https://doi.org/10.1007/JHEP08(2016)106}{\emph{JHEP} {\bfseries 08}
  (2016) 106} [\href{https://arxiv.org/abs/1503.01409}{{\ttfamily
  1503.01409}}].

\bibitem{Kitaev14}
A.~{Kitaev}, ``Hidden correlations in the hawking radiation and thermal
  noise.''

\bibitem{Dag:2019yqu}
C.~B. Dağ, K.~Sun and L.~M. Duan, \emph{{Detection of Quantum Phases via
  Out-of-Time-Order Correlators}},
  \href{https://doi.org/10.1103/PhysRevLett.123.140602}{\emph{Phys. Rev. Lett.}
  {\bfseries 123} (2019) 140602}
  [\href{https://arxiv.org/abs/1902.05041}{{\ttfamily 1902.05041}}].

\bibitem{2018arXiv181111191S}
Z.-H. {Sun}, J.-Q. {Cai}, Q.-C. {Tang}, Y.~{Hu} and H.~{Fan},
  \emph{{Out-of-time-order correlators and quantum phase transitions in the
  Rabi and Dicke model}}, {\emph{arXiv e-prints} (2018) arXiv:1811.11191}
  [\href{https://arxiv.org/abs/1811.11191}{{\ttfamily 1811.11191}}].

\bibitem{2018arXiv181201920W}
Q.~{Wang} and F.~{P{\'e}rez-Bernal}, \emph{{Probing excited-state quantum phase
  transition in a quantum many body system via out-of-time-ordered
  correlator}}, {\emph{arXiv e-prints} (2018) arXiv:1812.01920}
  [\href{https://arxiv.org/abs/1812.01920}{{\ttfamily 1812.01920}}].

\bibitem{Matsuki_2019}
Y.~Matsuki and K.~Ikeda, \emph{Comments on the fractal energy spectrum of
  honeycomb lattice with defects},
  \href{https://doi.org/10.1088/2399-6528/ab18de}{\emph{Journal of Physics
  Communications} {\bfseries 3} (2019) 055003}.

\bibitem{PhysRevLett.52.1}
O.~Bohigas, M.~J. Giannoni and C.~Schmit, \emph{Characterization of chaotic
  quantum spectra and universality of level fluctuation laws},
  \href{https://doi.org/10.1103/PhysRevLett.52.1}{\emph{Phys. Rev. Lett.}
  {\bfseries 52} (1984) 1}.

\bibitem{doi:10.1098/rspa.1977.0140}
M.~V. Berry, M.~Tabor and J.~M. Ziman, \emph{Level clustering in the regular
  spectrum}, \href{https://doi.org/10.1098/rspa.1977.0140}{\emph{Proceedings of
  the Royal Society of London. A. Mathematical and Physical Sciences}
  {\bfseries 356} (1977) 375}
  [\href{https://arxiv.org/abs/https://royalsocietypublishing.org/doi/pdf/10.1098/rspa.1977.0140}{{\ttfamily
  https://royalsocietypublishing.org/doi/pdf/10.1098/rspa.1977.0140}}].

\end{thebibliography}\endgroup

\end{document}